\documentclass[prb,preprint,superscriptaddress,floatfix]{revtex4-1}
\usepackage{amsmath}
\usepackage{subfigure}
\usepackage{graphicx}

\begin{document}
\bibliographystyle{prbrev}

\title{ 
Density Functional Theory of doped superfluid liquid helium and nanodroplets}

\author{Francesco Ancilotto}
\affiliation{Dipartimento di Fisica e Astronomia ``Galileo Galilei''
and CNISM, Universit\`a di Padova, via Marzolo 8, 35122 Padova, Italy}
\affiliation{ CNR-IOM Democritos, via Bonomea, 265 - 34136 Trieste, Italy }

\author{Manuel Barranco}
\affiliation{
Universit\'e Toulouse 3 and CNRS, Laboratoire des Collisions, Agr\'egats et R\'eactivit\'e,
IRSAMC, 118 route de Narbonne, F-31062 Toulouse Cedex 09, France
}
\affiliation{Departament FQA, Facultat de F\'{\i}sica,
Universitat de Barcelona. Diagonal 645,
08028 Barcelona, Spain}
\affiliation{Institute of Nanoscience and Nanotechnology (IN2UB),
Universitat de Barcelona.}

\author{Fran\c{c}ois Coppens}
\affiliation{
Universit\'e Toulouse 3 and CNRS, Laboratoire des Collisions, Agr\'egats et R\'eactivit\'e,
IRSAMC, 118 route de Narbonne, F-31062 Toulouse Cedex 09, France
}

\author{Jussi Eloranta}
\affiliation{
Department of Chemistry and Biochemistry, California State University at Northridge, California 91330, USA
}

\author{Nadine Halberstadt}
\affiliation{
Universit\'e Toulouse 3 and CNRS, Laboratoire des Collisions, Agr\'egats et R\'eactivit\'e,
IRSAMC, 118 route de Narbonne, F-31062 Toulouse Cedex 09, France
}

\author{Alberto Hernando}
\affiliation{
Social Thermodynamics Applied Research (SThAR) EPFL Innovation Park B\^{a}timent C, CH-1015 Lausanne, Switzerland
}

\author{David Mateo}
\affiliation{Applied Complexity Group, Singapore University of Technology and Design. 8 Somapah Road, Singapore 487372
}

\author{Mart\'{\i} Pi}
\affiliation{Departament FQA, Facultat de F\'{\i}sica,
Universitat de Barcelona. Diagonal 645,
08028 Barcelona, Spain}
\affiliation{Institute of Nanoscience and Nanotechnology (IN2UB),
Universitat de Barcelona.}

\begin{abstract}
During the last decade, density function theory (DFT) in its static and dynamic time dependent 
forms, has emerged as a powerful tool to describe the structure and dynamics of doped liquid helium 
and droplets. In this review, we summarize the activity carried out in this field within the DFT framework since the publication of the previous review article 
on this subject [M. Barranco {\it et al.}, J. Low Temp. Phys. {\bf 142}, 1 (2006)]. Furthermore, a comprehensive presentation of the actual implementations of helium DFT is given, 
which have not been discussed in the individual articles or are scattered in the existing literature.
This is an Accepted Manuscript of an article published on August 2, 2017 by Taylor \& Francis  Group in 
 Int. Rev. Phys. Chem. {\bf 36}, 621 (2017), available online:  
http://dx.doi.org/10.1080/0144235X.2017.1351672
\end{abstract}

\date{\today}

\maketitle

\centerline{\bfseries Contents}\vspace{-14pt}\hfill {\sc{page}}\medskip

\hspace*{-13pt} {\bf{{I.}    Introduction}}\hfill {\pageref{intro}}\\
{\bf{{II.}   Density functional theory of liquid $^4$He at zero temperature}}\hfill {\pageref{2}}\\
\hspace*{10pt}{II.A.}  Theoretical basis of density functional theory \hfill {\pageref{2.1}}\\
\hspace*{10pt}{II.B.} The Orsay-Trento density functional  \hfill {\pageref{2.2}}\\
\hspace*{10pt}{II.C.}  Recent improvements of the OT-DFT functional  \hfill {\pageref{2.3}}\\
\hspace*{24pt} {II.C.1.} The `solid' density functional \hfill {\pageref{2.3.1}}\\
\hspace*{24pt} {II.C.2.} Instability of the backflow term \hfill {\pageref{2.3.2}}\\
{\bf{{III.}   Time-independent calculations}}\hfill {\pageref{3}}\\
\hspace*{10pt}{III.A.}  General considerations \hfill {\pageref{3.1}}\\
\hspace*{10pt}{III.B.}  Introduction of vorticity \hfill {\pageref{3.2}}\\
{\bf{{IV.}   Dynamics }}\hfill {\pageref{4}}\\
\hspace*{10pt}{IV.A.} Heavy impurities \hfill {\pageref{4.1}}\\
\hspace*{10pt}{IV.B.}  Test particle method for light impurities \hfill {\pageref{4.2}}\\
\hspace*{10pt}{IV.C.}  Simulation of absorption and emission spectra using the density\\
\hspace*{30pt} fluctuation method \hfill {\pageref{4.3}}\\
{\bf{{V.}  Recent applications of  DFT for impurity doped superfluid helium }}\hfill {\pageref{5}}\\
\hspace*{10pt}{V.A.} Alkali metal doped helium droplets: solvation and absorption spectra  \hfill {\pageref{5.1}}\\
\hspace*{10pt}{V.B.}  Alkaline earth metal doped helium droplets:  solvation and absorption \\
\hspace*{30pt}  spectra  \hfill {\pageref{5.2}}\\
\hspace*{10pt}{V.C.} Droplets doped with more than one species  \hfill {\pageref{5.3}}\\
\hspace*{10pt}{V.D.} Cluster-doped helium droplets \hfill {\pageref{5.4}}\\
\hspace*{10pt}{V.E.}  Doped  mixed $^3$He-$^4$He and $^3$He droplets \hfill {\pageref{5.5}}\\
\hspace*{10pt}{V.F.} Electrons in liquid helium \hfill {\pageref{5.6}}\\
\hspace*{10pt}{V.G.} Cations in liquid helium and droplets \hfill {\pageref{5.7}}\\
\hspace*{10pt}{V.H.} Intrinsic helium impurities \hfill {\pageref{5.8}}\\
\hspace*{10pt}{V.I.} Translational motion of ions below the Landau critical velocity \hfill {\pageref{5.9}}\\
\hspace*{10pt}{V.J.} Critical Landau velocity in small $^4$He droplets \hfill {\pageref{5.10}}\\
\hspace*{10pt}{V.K}.  Rotational superfluidity \hfill {\pageref{5.11}}\\
\hspace*{10pt}{V.L.} Interaction of impurities with vortex lines \hfill {\pageref{5.12}}\\
\hspace*{24pt} {V.L.1.} Electrons  \hfill {\pageref{5.12.1}}\\
\hspace*{24pt} {V.L.2.} Atomic and molecular impurities  \hfill {\pageref{5.12.2}}\\
\hspace*{10pt}{V.M.} Vortex arrays in $^4$He droplets \hfill {\pageref{5.13}}\\
\hspace*{10pt}{V.N.} Dynamics of alkali atoms excited on the surface of $^4$He droplets \hfill {\pageref{5.14}}\\
\hspace*{10pt}{V.O.} Capture of impurities by $^4$He droplets \hfill {\pageref{5.15}}\\
\hspace*{24pt} {V.P.1.} Pure droplets  \hfill {\pageref{5.15.1}}\\
\hspace*{24pt} {V.P.2.} Droplets hosting vortices \hfill {\pageref{5.15.2}}\\
\hspace*{10pt}{V.Q.} Liquid helium on nanostructured surfaces \hfill {\pageref{5.16}}\\
\hspace*{10pt}{V.R.}  Soft-landing of helium droplets \hfill {\pageref{5.17}}\\
{\bf{{VI}   Summary and outlook }}\hfill {\pageref{6.0}}\\\\

\section{Introduction}
\label{intro}

Liquid helium-4 becomes superfluid below the lambda transition at 2.17 K due to partial Bose-Einstein condensation (BEC).
It exhibits unusual macroscopic 
behavior such as e.g. vanishing viscosity and the thermo-mechanical effect.\cite{Til74} On the atomic scale, the response of this fascinating quantum liquid has
 been studied experimentally by using solvated atomic and molecular species as probes.\cite{Bor07} The early experiments employed bulk liquid helium samples 
 where only ionic species and intrinsic helium excimers could be introduced. A breakthrough in this area has been  the development of the helium droplet technique, 
 which made it possible to embed neutral atomic and molecular species in superfluid helium droplets at 0.37 K.\cite{Har95,Har96}
 In addition to their intrinsic interest as a superfluid object of finite size, helium droplets provide an ideal matrix for
  spectroscopic experiments due to their low temperature and weak interaction with the solvated species.\cite{Leh98,Toe04,Bar06,Sza06,Sti06,Cho06,Tig07,Sle08,Cal11a,Yan13,
Mud14,Toe90,Wha94,Wha98,Toe98,Nor01,Kro02} 

From the theoretical point of view, superfluid helium must be considered as a high dimensional quantum system. Quantum Monte Carlo (QMC)\cite{Kro02} 
and direct quantum mechanical\cite{deL06,deL10,Agu13} calculations are the most accurate methods, but their computational demand quickly exceeds currently
 available computer resources when the number of helium atoms increases. Furthermore, QMC cannot describe dynamic evolution of superfluid helium in real time. 
  To address these limitations, approximate methods based on  density functional theory (DFT) formalism have been introduced.\cite{Str87a,Str87b,Dal95}  
 DFT can be applied to much larger systems than QMC and allows for time-dependent formulation. As such, it offers a good compromise between accuracy and
  computational feasibility. The main drawback of DFT is that the exact energy functional is not known and must therefore be constructed in a semi-empirical manner. 
  Nevertheless, DFT is the only method to date that can successfully reproduce results from a wide range of time-resolved experiments in superfluid helium on the atomic scale.

Application of recently developed femtosecond laser techniques  to study helium droplets\cite{Mud14,Zie15}  highlights the importance of time-dependent DFT (TDDFT). For example, TDDFT can
 be used to analyze experiments that employ free electron laser pulses to visualize vortex arrays in helium droplets,\cite{Gom14} or  the  dynamics following
 optical excitation of guest atoms or molecules embedded in helium droplets.\cite{Bra13,Van17} It is the only method that allows for such a close interplay between theory 
 and time-resolved helium droplet experiments. In fact, many of the results presented in this review were obtained as joint experimental-theoretical collaborative work. 
 
 Despite
  the wide success of both DFT and TDDFT, they have known limitations, especially when the interaction between the guest species and helium is strong.\cite{Lea14b,Fie12} 
  New strategies for resolving with such problems are also summarised in this review.  In addition, applications of DFT and time-dependent DFT will be reviewed with a focus 
  on the new developments that have appeared after the previous review article on this topic.\cite{Bar06} 

We provide a comprehensive presentation of the most recent DFT models and their applications to superfluid helium droplets and bulk liquid.
Selected topics dealing with DFT of non-superfluid $^3$He are also briefly discussed;
some practical details of the DFT implementation  are given in Ref. \onlinecite{dft-guide}. 
As stated by Frank Stienkemeier and Kevin Lehmann  in their 2006 topical review,\cite{Sti06} a  truly comprehensive review of the activity carried out recently in this field would require a monograph
 instead of a review article;
the reader is thus referred to the appropriate literature, in particular to some recent reviews\cite{Toe04,Bar06,Sza06,Sti06,Cho06,Tig07,Sle08,Cal11a,Yan13,
Mud14,Zie15}  for the subjects not considered in detail here.

\begin{figure}[t]
\vspace{42pt}
\begin{center}
\resizebox*{10cm}{!}{\includegraphics{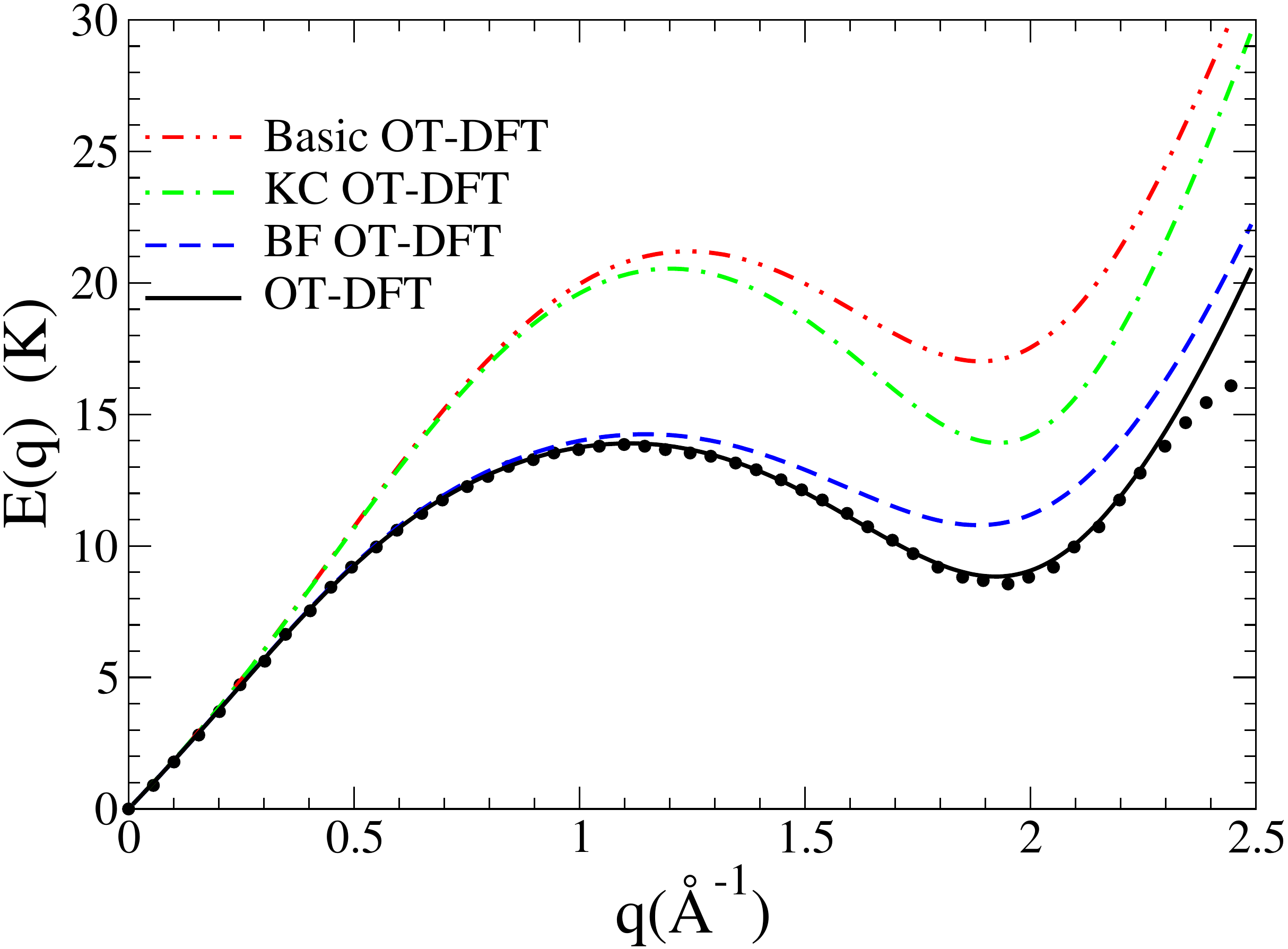}}
\caption{\label{fig1}
Dispersion relation for elementary excitations in liquid $^4$He calculated as in  Ref. \onlinecite{Mat10a}.  
`Basic' indicates the OT-DFT without the non-local kinetic energy correlation (KC) nor the backflow (BF) terms; 
KC OT-DFT adds  to the basic OT-DFT the KC term; BF OT-DFT adds to the basic
OT-DFT the BF term.
The dots are the experimental data from Ref. \onlinecite{Don81}.
The Landau velocity $v_L = E(q)/(\hbar\,q)|_{min}$ obtained for each functional is  60.3 m/s (OT-DFT); 75.1 m/s (BF OT-DFT); 94.4 m/s (KC OT-DFT);
118 m/s (basic OT-DFT); and 57.5 (experiment).
}
\end{center}
\end{figure}

\section{Density functional theory of liquid $^4$He at zero temperature}
\label{2} 

\subsection{Theoretical basis of density functional theory}
\label{2.1}

The starting point is the Hohenberg-Kohn (HK) theorem,\cite{Hoh64} which
states that the total energy $E$ of a many-body quantum system at $T=0$ is
a functional of the one-particle density 
$\rho ({\mathbf r})=\langle\Phi |\sum _i \delta ({\mathbf r}-{\mathbf r}_i)|\Phi \rangle$ 
($\Phi $  being the many-body wave function):
\begin{equation}
E[\rho] = {\cal T}[\rho] +\int d{\mathbf r}\, {\cal E}[\rho]
\label{eq1}
\end{equation}
where the kinetic energy functional has been separated from the interaction part.

The Kohn-Sham formulation\cite{Koh65} of the HK theorem allows to write the above functional in the form
\begin{equation}
E[\rho] = T[\rho] + \int d{\mathbf r} \,{\cal E}_c[\rho]
\label{eq2}
\end{equation}
where $T[\rho]$ is the kinetic energy of
a fictitious system of {\it non-interacting} particles, with the same density
of the original one,
described by single-particle orbitals
$\{ \phi _i({\mathbf r}) \}$
\begin{equation}
T= -\frac{\hbar^2}{2m_4} \sum_i \int d{\mathbf r}\, \phi _i^\ast ({\mathbf r})
\nabla^2  \phi _i ({\mathbf r})
\label{eq3}
\end{equation}
The sum extends to the $N_4$ particles  of mass $m_4$ in the system.
The difference ${\cal T}[\rho]-T[\rho]$ has been buried in the
interaction term ${\cal E}_c$.
The density of such non-interacting system is thus
$\rho ({\mathbf r})=\sum _i |\phi _i ({\mathbf r})|^2 $.
We conform here to the common notation 
used for DFT studies of helium systems,\cite{Dal95}
which defines as `correlation energy density' the
functional ${\cal E}_c$, even if it includes
also He-He interactions at the mean-field level
(first term in Eq. (\ref{eq8}) below).

Assuming complete Bose-Einstein condensation at $T=0$
(i.e. all the $^4$He atoms are in the same single-particle orbital $\phi_0$),
the many-body wave function is simply
\begin{equation}
\Phi({\mathbf r}_1, {\mathbf r}_2, \cdots  {\mathbf r}_{N_4}) = \prod_{i=1}^{i=N_4} \phi_0({\mathbf r}_i)
\label{eq4}
\end{equation}
while $\rho ({\mathbf r})=N_4 |\phi _0({\mathbf r})|^2$.
 Although the actual condensate fraction of superfluid $^4$He is less than 10\%, the available helium density functionals have been devised such that, by starting
  from a fully condensed state, the interaction term ${\cal E}_c$ allows to reproduce the relevant physical properties of liquid helium at $T = 0$.

It is customary to define an 
\emph{order parameter} $\Psi$ (also called \emph{effective wave function}) 
as $\Psi(\textbf{r}) = \sqrt{N_4} \phi_0(\textbf{r})$. 
The kinetic energy of the condensate is thus
\begin{equation}
T[\rho] =
-\frac{\hbar^2}{2m_4} N_4 \langle \phi_0 | \nabla^2 |\phi_0\rangle =
\frac{\hbar^2}{2m_4} \int d {\mathbf r} |\nabla \Psi|^2
\label{eq5}
\end{equation}
The Runge-Gross theorem extends DFT to describe the time evolution of the system through the time-dependent DFT (TDDFT) formalism.\cite{Run84} 
In this case, functional variation of the associated Lagrangian leads to a time-dependent Euler-Lagrange (EL) equation
\begin{equation}
\imath \hbar\frac{\partial}{\partial t} \Psi({\mathbf r},t) = \left\{-\frac{\hbar^2}{2m_4}\nabla^2 + \frac{\delta{\cal E}_{c}}{\delta\rho}\right\}\Psi(\textbf{r},t)  \equiv
{\cal H}\left[\rho\right] \,\Psi(\textbf{r},t) 
\label{eq6}
\end{equation}
Given the initial state $\Psi(\textbf{r},0)$, solution of this non-linear equation yields 
$\Psi\left(\textbf{r},t\right)$ which, in the hydrodynamic picture,\cite{Mad27} can be decomposed into liquid density and the associated velocity potential field. 
For stationary states, $\Psi(\textbf{r},t)=\Psi_0(\textbf{r})e^{-\i\mu t/\hbar}$ and Eq. (\ref{eq6}) can be cast into a non-linear time-independent EL equation
\begin{equation}
\left\{-\frac{\hbar^2}{2m_4} \nabla^2 + \frac{\delta {\cal E}_{c}}{\delta \rho}\right\}\Psi_0(\textbf{r}) = \mu \Psi_0(\textbf{r})
\label{eq7}
\end{equation}
where $\mu$ is the chemical potential. Iterative solution of Eq. (\ref{eq7}) determines the particle density $\rho(\textbf{r})=|\Psi_0(\textbf{r})|^2$ 
(a similar relationship holds in the time-dependent situation)
and hence the total energy of the system. 
 
In its time-independent formulation, 
DFT is a ground-state theory.
However, within the HK theorem, the variational principle is
applicable to the lowest state of a given symmetry, which may be different from the true ground state of the system.
For example, this can be employed to obtain stationary vortex solutions in helium droplets by DFT.
Similarly, minimization of the energy functional in the presence of additional constraints (e.g.,
fixed total angular momentum) will
provide the correct density for the associated excited state. In particular, this technique can be used to produce vortex arrays in helium droplets.
However, for general excited states, there is no equivalent HK
theorem and TDDFT must be used to model them.
 
In the case of phenomenological helium DFT,
the quality of the results depends on the functional form used.
As an example, TDDFT calculation of the dispersion relation for uniform liquid helium
(an excited state property) is shown in Fig. \ref{fig1}. The OT-DFT introduced below gives results in 
agreement, by construction, with the experimental (`exact') results.
This obviously does not guarantee that the same functional
would also give reliable results for inhomogeneous systems. However,
based on our experience, these functionals are highly `transferable'
to such situations and provide results that are generally in good agreement
with experiments.

Approximate representations for the interaction energy density functional ${\cal E}_c$, which are capable of describing inhomogeneous 
$^4$He systems quantitatively, are discussed in the following Section.

\subsection{The Orsay-Trento density functional}
\label{2.2}

\begin{table}[t]
\vspace{0.1cm}
{\begin{tabular}{cccccc}
\hline\hline
$\epsilon_{LJ}$  (K)    & $\sigma$ (\AA)& $h$ (\AA) & $c_2$ (K \AA$^6$)  & $c_3$ (K \AA$^9$) & $\alpha_s$ (\AA$^3$) \\
 10.22   & 2.556 &2.190323 & -2.41186 $\times 10^4$ & 1.85850 $\times 10^6$ & 54.31 \\
\colrule
 $\rho_{0s}$ (\AA$^{-3}$)& $l$ (\AA)&$C$ (Hartree) &$\beta$ (\AA$^3$) & $\rho_m$ (\AA$^{-3}$) & $\gamma_{11}$ \\  
  0.04& 1. & 0.1 &40.   & 0.37 & -19.7544 \\
  \colrule
  $\gamma_{12}$ (\AA$^{-2}$)& $\alpha_1$  (\AA$^{-2}$) & $\gamma_{21}$  & $\gamma_{22}$ (\AA$^{-2}$) & $\alpha_2$ (\AA$^{-2}$) &      \\  
   12.5616 &1.023 &  -0.2395 & 0.0312 & 0.14912  & \\
\hline\hline
\end{tabular}}
\caption{\label{table1}
Model parameters for the OT-DFT and solid functionals.
}
\end{table}

The  first and simplest DFT model for superfluid $^4$He was developed by Stringari and coworkers.\cite{Str87a,Str87b} In this approach, ${\cal E}_c[\rho] $ consists of a sum of terms that only depend
 on the local density $\rho({\bf r})$. More recent models include  also finite-range and non-local terms, which greatly improve the accuracy of the method,
  especially when applied to highly inhomogeneous systems. 
 
The most successful approach to date is the finite range, non-local Orsay-Trento DFT model (OT-DFT),\cite{Dal95} which has been calibrated to reproduce bulk liquid properties such as
 the energy per atom, the equilibrium density, the dispersion relation, and the compressibility at $P=T=0$. The OT-DFT energy functional is written as 
\begin{eqnarray}
{\cal E}_c[\rho ,\mathbf{v}] &=&  
\frac{1}{2} \int {\rm d}{\bf r'} \rho({\bf r}) V_{LJ}(|{\bf r}-{\bf r'}|) \rho({\bf r'}) 
\nonumber \\
&& + \frac{1}{2} c_2\, \rho({\bf r}) \left[{\bar \rho}({\bf r}) \right]^2 
+ \frac{1}{3} c_3 \, \rho({\bf r}) \left[ {\bar \rho}({\bf r}) \right]^3 
\nonumber \\
&&- \frac{\hbar^2}{4m_4} \alpha_s \int {\rm d}{\bf r'} F(|{\bf r}-{\bf r'}|) 
\left[ 1- \frac{{\tilde \rho}({\bf r})}{\rho_{0s}} \right]
\nabla \rho({\bf r}) \cdot \nabla' \rho({\bf r'})
\left[ 1- \frac{{\tilde \rho}({\bf r'})}{\rho_{0s}} \right] 
\nonumber
\\
&& - \frac{m_4}{4} \int d\mathbf{r}' \,V_J(| \mathbf{r} - \mathbf{r}'|)\, \rho(\mathbf{r}) \, \rho(\mathbf{r}')\,  [\mathbf{v}(\mathbf{r}) - \mathbf{v}(\mathbf{r}')]^2
\label{eq8}
\end{eqnarray} 
The first term corresponds to a classical Lennard-Jones interaction between helium atoms, which is truncated at short distances where 
the correlation effects become important
\begin{eqnarray}
V_{LJ}(r) &=& 4 \epsilon_{LJ} \left[ \left(\frac{\sigma}{r} \right)^{12} - 
\left(\frac{\sigma}{r} \right)^{6} \right] \quad {\rm if} \quad r  > h 
\nonumber \\
&=& 0 \quad {\rm otherwise}
\label{eq9}
\end{eqnarray} 
The second line in Eq. (\ref{eq8}) accounts for short-range correlation effects.
The third line (`$\alpha_s$ term') is a non-local kinetic energy correction (KC)
-- which partially accounts for the difference 
${\cal T}[\rho]-T[\rho]$ in the
interaction term ${\cal E}_c$ --
 and the last term is the \emph{backflow} (BF) contribution that affects the 
dynamic response of the functional. Note that the BF term only contributes when the order parameter is a complex valued function 
(e.g. time-dependent problem or vortex state). The velocity $\mathbf{v}(\mathbf{r}) $ is determined from the current 
\begin{equation}
\mathbf{j}(\mathbf{r}) =  - \frac{\imath \hbar}{2 m_4} [\Psi^*(\mathbf{r}) \nabla \Psi(\mathbf{r}) - \Psi(\mathbf{r}) \nabla \Psi^*(\mathbf{r})]
\label{eq10}
\end{equation}
as $\mathbf{v}(\mathbf{r}) = \mathbf{j}(\mathbf{r}) / \rho(\mathbf{r}) = \hbar/m_4 \times {\rm Im}\{\nabla \Psi(\mathbf{r})/\Psi(\mathbf{r})\}$.
The two coarse-grained averages of the liquid density, $\bar{\rho}$ and $\tilde{\rho}$, entering into the short-range correlation terms in Eq. (\ref{eq8}), are given by
\begin{equation}
{\bar \rho}({\bf r}) = \int {\rm d}{\bf r'} \rho({\bf r'}) w(|{\bf r}-{\bf r'}|)
\label{eq11}
\end{equation}
where
\begin{eqnarray}
w(r) &=& \frac{3}{4 \pi h^3} \quad {\rm if} \quad r <h \nonumber \\
&=& 0 \quad {\rm otherwise.} 
\label{eq12}
\end{eqnarray} 
and 
\begin{equation}
{\tilde \rho}({\bf r}) = \int {\rm d}{\bf r'} \rho({\bf r'}) F(|{\bf r}-{\bf r'}|)
\label{eq13}
\end{equation}
where $F(r)$ is a Gaussian kernel
\begin{equation}
F(r)= \frac{1}{\pi^{3/2}l^3} {\rm e}^{-r^2/l^2}
\label{eq14}
\end{equation}
The function $V_J(r)$ presents in the backflow term is defined as 
\begin{equation}
V_J(r) = (\gamma_{11} +\gamma_{12} \, r^2) e^{-\alpha_1 r^2}+(\gamma_{21} +\gamma_{22} \, r^2) e^{-\alpha_2 r^2}
\label{eq15}
\end{equation}
The various parameters entering the OT-DFT functional are specified in Table \ref{table1}.

While OT-DFT can model the response of superfluid helium very accurately, it is seldom applied to inhomogeneous systems due to its complexity.\cite{Anc10} Furthermore, in most time-dependent applications, both the kinetic energy correlation and backflow terms are often neglected because their evaluation is time consuming and they tend to exhibit numerical instabilities, especially for highly inhomogeneous systems. Strategies for overcoming  these instabilities are presented in the next  section.

The backflow and non-local kinetic energy correlation terms in OT-DFT are required for a quantitative description of the elementary excitation spectrum of superfluid helium. 
While both terms influence the energetics of the roton minimum, the backflow term has the most important contribution of the two as demonstrated in Fig. \ref{fig1}. Note that the 
Landau critical velocity predicted by the functional, which determines the onset of bulk dissipative behavior in time-dependent applications, is the slope of a straight line passing through  
the origin and tangent to the dispersion curve near the roton minimum.\cite{Dal95,Pop15}
 The influence of these terms to the description of a vortex line structure is discussed in Ref. \onlinecite{Mat15a}, see also Sec. \ref{2.3.2} below.

\begin{figure}[t]
\vspace{42pt}
\begin{center}
\resizebox*{10cm}{!}{\includegraphics{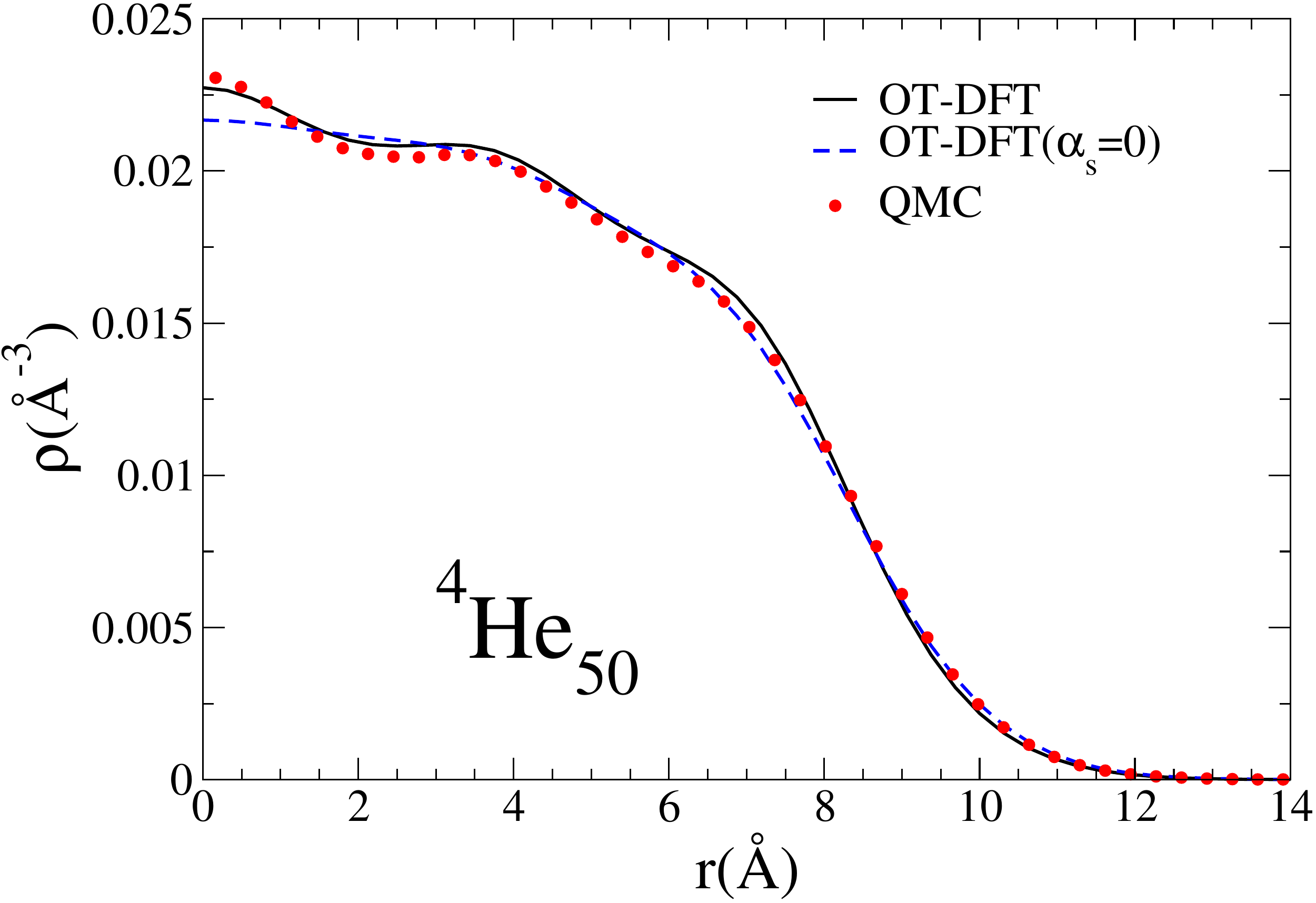}}
\caption{\label{fig2}
Comparison between OT-DFT and QMC calculations for the density profile 
of the $^4$He$_{50}$ droplet. 
The QMC calculations have been carried out by M. Rossi, University of Padova.}
\end{center}
\end{figure}

The accuracy of OT-DFT can be further assessed by comparing the obtained density profiles of pure helium droplets against QMC calculations. By way of an example, such a comparison
 is shown in Fig. \ref{fig2} for a droplet with $N_4 = 50$. Since DFT should generally work better when the number of particles increases, OT-DFT will retain its accuracy 
 for the typical droplet sizes produced in experiments (a few thousand $^4$He atoms). Even with the kinetic energy correlation term omitted (i.e. $\alpha_s = 0$), the agreement 
 with QMC remains rather good as demonstrated in Fig. \ref{fig2}.

In contrast to DFT employing local functionals, the performance of finite-range functionals is superior when processes such as atomic/molecular impurity solvation or their spectroscopy is considered (see e.g.
 Fig. 1 of Ref. \onlinecite{Mat13b}). Any process that requires the correct liquid response on the \AA{}ngstr\"om-scale must employ a finite range, non-local model.
 However, in some applications the non-local terms are not very important and it is possible to use the much simpler local functionals. Local density functionals of different complexity have been used to describe static and dynamic properties of pure and doped superfluid helium.\cite{Ber01,Mat11a,Jin10a,Jin10b,Jin10c} Very recently, a zero-range reduction of the OT functional has also been applied to study inelastic scattering of Xe atoms by quantised vortices in superfluid helium.\cite{Psh16}

The original  OT-DFT  formulation only applies 
to superfluid $^4$He at $T=0$.  
It has been extended up to $T=3$ K by considering the wetting properties of various metals,\cite{Anc00} see also Ref. \onlinecite{Bib02,Jin12}. 
A non-local extension of the functional has also been introduced for mixed $^3$He-$^4$He systems.\cite{Bar97}
 The latter model has been used recently to study elementary excitations of superfluid $^3$He-$^4$He mixtures\cite{Mat10a} and to study  the solvation of OCS in  mixed $^3$He-$^4$He droplets.\cite{Lea13} Various functionals have also been developed for pure $^3$He, see e.g. Refs. \onlinecite{Her07,Her02} and  references therein. Finally, we note that a method  similar to the one used for superfluid $^4$He has also been used to describe cold dipolar Bose gases\cite{Aba10} and para-hydrogen clusters, for which a DFT-based approach is also available.\cite{Anc16}

\subsection{Recent improvements of  the OT-DFT functional}
\label{2.3}
\subsubsection{The `solid' density functional}
\label{2.3.1}

The OT-DFT functional becomes unstable in the presence of highly inhomogeneous liquid density distributions, like those occurring e.g. for the solvation of cations inside $^4$He. 
To overcome this problem, an additional cutoff term, which was originally developed to account for the liquid-solid phase transition of $^4$He,\cite{Anc05a,Cau07} can be employed to it.
This is essentially a penalty term that prevents excessive liquid density accumulation
\begin{equation}
{\cal E}_{\rm sol}[\rho(\mathbf{r})] = C\,\rho(\mathbf{r}) \{1 + \tanh (\beta [\rho(\mathbf{r})-\rho_{\rm m}])\}
\label{eq16}
\end{equation}
Since ${\cal E}_{\rm sol}[\rho]$ is only significant when the liquid density is comparable to $\rho_{\rm m}$ or larger, it does not alter the original OT-DFT functional at densities lower than the
(large) cutoff value $\rho_{\rm m}$. 
For instance, the total energy of pure $^4$He$_{1000}$ droplet is $-$5400.34 K where the contribution of the penalty term is only 4.2 $\times 10^{-5}$ K.
The model parameters used are specified in Table \ref{table1}.
 
Inclusion of the `solid' term in the OT-DFT model has made it possible to use it in complex situations where the impurity-helium interaction is  strongly attractive. However,  
 while ${\cal E}_{\textnormal{sol}}[\rho] $ in Eq. (\ref{eq16}) can prevent the unphysical density pile-up,
it cannot eliminate the often observed spontaneous symmetry 
 breaking of the $^4$He order parameter in the presence of strongly attractive external potentials.
 For instance, the numerical solution can become non-spherical even when the external potential is strictly spherically symmetric.
 Taking a spherical average of the symmetry broken solution appears however to yield results very close to QMC 
 calculations.\cite{Anc07,Fie12} It is not clear at the moment how to preserve the desired symmetry during the calculations.
Note  that a  spontaneous symmetry breaking is expected to occur  around very attractive impurities, which form `snowball' structures 
  with a solid-like first solvation layer. The solid OT-DFT functional, consisting of ${\cal E}_{\rm sol}$ and the first three terms of Eq. (\ref{eq8}), 
  has often been used in the static and dynamic applications discussed in the next sections. 

\begin{figure}[t]
\vspace{42pt}
\begin{center}
\resizebox*{10cm}{!}{\includegraphics{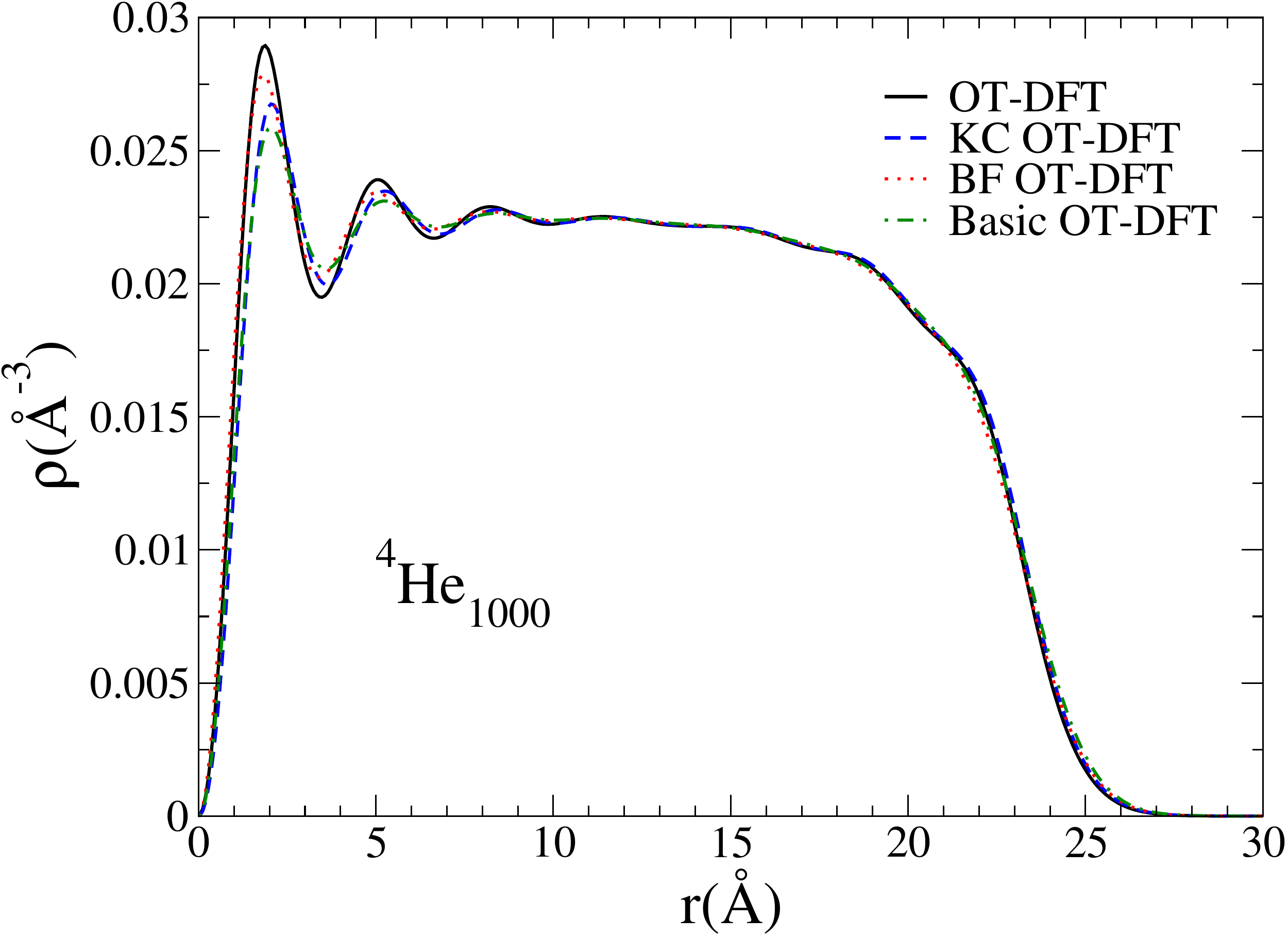}}
\caption{\label{fig3}
Density profile of the $^4$He$_{1000}$ droplet hosting a linear vortex along its diameter. As in Fig. \ref{fig1}, the calculations have been carried out using the full OT-DFT or including
only some of its terms. 
} 
\end{center}
\end{figure}

\subsubsection{Instability of the backflow term}
\label{2.3.2}

\begin{table}[t]
\vspace{0.1cm}
{\begin{tabular}{cccccc}
\hline\hline
$P$       & $\rho$ &   $q_R$ (exp)   & $\Delta$ (exp)   & $q_R$ (OT)  & $\Delta$ (OT)  \\
  (bar)   & (10$^{-2}$ \AA$^{-3})$  & (\AA$^{-1})$ &  (K) &   (\AA$^{-1})$ &   (K)  \\
\hline
0     & 2.1836  & 1.93    & 8.62  & 1.92  & 8.84\\
5     & 2.2994  & 1.97    & 8.33 &  1.95  &8.52 \\ 
10   & 2.3916  & 2.01    & 8.03 &  1.97  &8.24\\
15   & 2.4694  & 2.03    & 7.75 &  1.99         &7.98 \\
20   & 2.5374  & 2.06    & 7.44  &  2.01         &7.73 \\
24   & 2.5865  & 2.05    & 7.30 &  2.01 & 7.54\\
 73.2    & 3.0  &   & &   2.10 & 5.48\\
 181.4    & 3.5    &  &  &2.21 & 1.50\\
 195.9    & 3.55  &    &  &   2.22 & 0.74\\
 211.2    & 3.6    & &   &   -  & -\\
\hline\hline
\end{tabular}}
\caption{\label{table2}
OT-DFT zero temperature equation of state $P(\rho)$ of liquid $^4$He  and roton minimum parameters calculated by the method of  Ref. \onlinecite{Mat10a}.
The experimental results are from Ref. \onlinecite{Gib99}.
}
\end{table}

The dispersion relation of elementary excitations in liquid $^4$He is shown in Fig \ref{fig1}. The low wavenumber ($q$) region exhibits linear behaviour and corresponds to phonons (sound), followed by a maximum (maxon), and a high-$q$ region that corresponds to collective excitations called rotons. The latter region exhibits a distinct minimum around $q\approx 2$ \AA$^{-1}$ (roton minimum). Within the previously developed microscopic variational approach of Feynman and Cohen,\cite{Fey56} a quantitative description of the roton minimum required the introduction of specific corrections
to describe the correlated motion around each atom in the superfluid (backflow).

The formulation of the BF term in OT-DFT,\cite{Dal95} which is shown on the fourth line of Eq. (\ref{eq8}), was inspired by a  previous 
work of Thouless.\cite{Tho69} With this term included, OT-DFT can accurately reproduce the $T=0$ experimental dispersion relation up to the solidification 
pressure\cite{Gib99} with the exception of the turn-over region at high momenta beyond the roton region.\cite{Don81}

 The roton minimum can be charaterised by two parameters, $\mu_R$ and $q_R$, by fitting 
the experimental dispersion relation close to the minimum with the following function at $T = 0.5$ K\cite{Gib99}
\[
E(q) = \Delta + \frac{\hbar^2}{2 \mu_R} (q - q_R)^2 
\]
where $q$ is the wavenumber, $\Delta$ is the roton energy, and $\mu_R$ defines the curvature at the roton minimum. A comparison of the experimental $\Delta$ and $q_R$  values
with those obtained with the OT-DFT functional  in shown in Table \ref{table2}.

The BF term becomes numerically unstable when $\rho\rightarrow 0$ and $\left|\textbf{v}\right| \ne 0$. This instability is present in the 
energy functional  as well as in the corresponding functional derivative yielding the effective potential in Eq. (\ref{eq8}).\cite{Gia03,Leh04} Since 
the contribution of the BF term should be negligible at low densities, this problem can be eliminated by introducing a  density cutoff for 
evaluating the velocity field from the probability current: $\textbf{v} = \textbf{j}/\left(\rho+\epsilon_v\right)$ where $\epsilon_v$ is the density cutoff 
value. Typical values of $\epsilon_v$ applied in recent work\cite{Mat15a} are in the order of $7\times 10^{-5}$ \AA{}$^{-3}$, which can
 be compared with the bulk liquid density $\rho_0 = 2.1836\times 10^{-2}$ \AA{}$^{-3}$ at $P=T=0$. An alternative approach is to neglect the BF 
 term when the density becomes smaller than a given threshold value (\textit{ca.} $10^{-6}$ \AA$^{-3}$).\cite{Anc10} 

Figure \ref{fig3} shows the density profile for a $^4$He$_{1000}$ droplet hosting a vortex line, which was calculated by the full OT-DFT 
or including only some of its term. The  BF term reduces the vortex line energy $E_V$ by about 20 K.
 Indeed, by using Eq. (\ref{eq30}) and the definition of $E_V$ given  in Sec. \ref{3.2}, one finds 127.1 K (basic OT-DFT);
 124.2 K (KC OT-DFT); 107.5 (BF OT-DFT),   and 105.5 K (OT-DFT).

With the sole exception of electrons,\cite{Anc10} all attempts made so far to include the BF term in calculations modelling impurity dynamics in superfluid 
helium droplets or in bulk liquid have  failed. One possible reason for such a  failure is 
  the appearance of a dynamic instability: the rationale for this being that the OT-DFT roton minimum energy collapses to zero around densities between 
  0.0355 and 0.036 \AA$^{-3}$ as shown in Table \ref{table2}.
 Indeed, local liquid densities around  impurities which  exhibit strong binding towards helium may reach densities much higher than solid helium,  leading 
 to an unphysical behavior of the BF term  and break down of the OT-DFT model. 
 
 It is likely that the BF term should only be applied in the liquid phase  
 for which it was originally intended for. For example, the first solvation layers around snowball structures should be excluded from the BF interaction.
 A similar remark applies to the KC term, although it has not been found  to be unstable.
  On the other hand, the solvation 
 structures of electrons and vortices are free from such  huge density pile-ups and the OT-DFT functional can be employed.\cite{Anc10}

 The instability of the backflow term appearing at high densities  calls for improvements. We present here a modified BF term that is numerically stable, only acts in the liquid phase
 and, by construction, yields  a functional of the same quality as  the original one\cite{Dal95} in that  physical region.

 Consider a BF term of the following form
\begin{equation}
{\cal H}_{BF} =  - \frac{m_4}{4} \int \int d\mathbf{r} \,d\mathbf{r}' \,V_J(| \mathbf{r} - \mathbf{r}'|)\, G(\rho(\mathbf{r}))  \, G(\rho(\mathbf{r}'))\,
 \rho(\mathbf{r}) \, \rho(\mathbf{r}')\,  [\mathbf{v}(\mathbf{r}) - \mathbf{v}(\mathbf{r}')]^2
\label{eq17}
\end{equation}
If one takes 
$$ G(\rho(\mathbf{r})) = [1-\Theta(\rho(\mathbf{r})-\rho_{BF})]$$
with $\rho_{BF}= 0.033$ \AA$^{-3}$ and $\Theta(x)= 1$ if $x\geq 0$ and zero otherwise,
this will make the BF contribution effective only when helium is a `true' liquid.
The function $G(\rho(\mathbf{r}))$ is difficult to handle numerically. In practice, it has been substituted by 
\begin{equation}
G( \rho(\mathbf{r})) = \frac{1}{2} \big\{ 1 - \tanh [\xi( \rho(\mathbf{r}) -  \rho_{BF})] \big\}
\label{eq18}
\end{equation}
with a  sufficiently large value of $\xi$  to make it steep at $\rho(\mathbf{r}) \sim \rho_{BF}$.
Values used in the calculations are $\xi= 10^4$ \AA$^3$ and $\rho_{BF} = 0.033$ \AA$^{-3}$.

The contribution of ${\cal H}_{BF}$ to the mean field, 
applied to the effective wave function $\Psi(\mathbf{r})$, is  
\begin{eqnarray}
&& - \frac{m_4}{2} \left[ \left.\rho(\mathbf{r}) \frac{d G}{d \rho} \right|_{\rho(\mathbf{r})}     +        G( \rho(\mathbf{r}))                             \right]
  \left\{ \int d\mathbf{r}' \,V_J(| \mathbf{r} - \mathbf{r}'|)\, \rho(\mathbf{r}') \,  G( \rho(\mathbf{r}')) [\mathbf{v}(\mathbf{r}) - \mathbf{v}(\mathbf{r}')]^2 \right\} \Psi(\mathbf{r})
\nonumber
\\
&& + \frac{\imath \hbar}{2} \frac{1}{\rho(\mathbf{r})} \nabla_{\mathbf{r}} \cdot  
  \left\{ \int d\mathbf{r}' \,V_J(| \mathbf{r} - \mathbf{r}'|) \, \rho(\mathbf{r}) \,  G( \rho(\mathbf{r})) \, \rho(\mathbf{r}') \,  G( \rho(\mathbf{r}')) [\mathbf{v}(\mathbf{r}) - \mathbf{v}(\mathbf{r}')] \right\} \Psi(\mathbf{r})
\label{eq19}
\end{eqnarray}
Putting $G( \rho(\mathbf{r})) =1$  reduces it  to  the OT-DFT expression.\cite{Gia03,Leh04} Eq. (\ref{eq19})  is as complex to use as the original OT-DFT form.  
 The modified OT-DFT functional  (MOT-DFT), that includes the solid and modified BF terms, has been tested in dynamic calculations where OT-DFT was unstable; at variance,   
 MOT-DFT  has been found to be stable. 
 
Having solved the instability problem in practice, let us mention that it is unclear  what is the actual relevance of the BF term -- and KC term --  
 for most items addressed in this review. 
 Processes such as photoexcitation and photoionisation of impurities usually involve high liquid densities and velocities in their first stages, 
 which can lead to the production of shock waves, cavitation, and vorticity. 
None of them are very sensitive to the accurate description of the roton minimum
and therefore the contribution of the BF term should be minimal
 (except for vorticity, for which we have already estimated the error if it is  not included in the calculations). It is only after most of the excess energy has been dispersed into 
 the fluid that the proper description of the elementary excitations becomes important, and the OT-DFT model can be applied at that point. 
 
 Last but not least,
 even if its systematic use might be computationally prohibitive,  the stable MOT-DFT may be useful  to carry out test calculations to calibrate simpler DFT approaches.

\section{Time-independent calculations}
\label{3}

In this section, we describe how to solve the time-independent EL equation, Eq. (\ref{eq7}),
to obtain the energetics and structure of solvated impurities in superfluid helium. For example,
it can be used to determine absorption and emission spectra of atoms/molecules embedded in helium droplets.
It also provides a starting point for subsequent time-dependent calculations. 

\subsection{General considerations}
\label{3.1}

The ground state liquid density of a system can be obtained by solving Eq. (\ref{eq7}). This is most often achieved by employing the imaginary time-method  (ITM).\cite{Leh07b} Normalization of the solution to a fixed number of helium atoms in the droplet, $N_4$, determines the corresponding chemical potential $\mu$. On the other hand, for the bulk liquid $\mu$ is dictated by the liquid equation of state, which can be obtained from ${\cal E}_c$.\cite{Pi07} Provided that a sufficiently large volume of liquid is considered, the value of  the bulk chemical potential also applies to systems with solvated impurities or free surfaces.

Impurities much heavier than helium can be described classically as point-like particles, providing an external field for the helium density.
In contrast, light impurities have to be modelled quantum mechanically based on the Schr\"odinger equation. In both cases, the impurity-He atom interaction, $V_X$, must be known: it is used to construct the impurity-liquid interaction using the pairwise sum approximation. For classical impurities, this interaction is included as an external field in the energy functional, $E\left[\rho\right]$, by integrating over the liquid density
\begin{equation}
E[\rho] \rightarrow E[\rho] +  \int d {\mathbf r} \rho({\mathbf r}) V_X(|{\mathbf r} - {\mathbf r}_I|)  
\label{eq20}
\end{equation}
where ${\mathbf r}_I$ is the location of the impurity. 
Eq. (\ref{eq7}) is then written as
\begin{equation}
\left\{-\frac{\hbar^2}{2m_4} \nabla^2 + \frac{\delta {\cal E}_c}{\delta \rho}  + V_X(|{\mathbf r} - {\mathbf r}_I|) \right\}\Psi({\mathbf r})  
= \mu \Psi({\mathbf r})
\label{eq21}
\end{equation}

For impurities requiring quantum mechanical treatment, $E[\rho]$ must also take  into account their zero point motion
\begin{equation}
E[\rho] \rightarrow  E[\rho] + \frac{\hbar^2}{2m_I} \int  d {\mathbf r}_I |\nabla_I \phi({\mathbf r}_I)|^2
+  
\int \int d {\mathbf r} d {\mathbf r}_I \rho({\mathbf r}) V_X(|{\mathbf r} - {\mathbf r}_I|)  |\phi({\mathbf r}_I)|^2
\label{eq22}
\end{equation}
where $\phi({\mathbf r}_I)$ is the impurity wave function and  $m_I$ its mass. 
This yields two coupled equations, one for the liquid and another for the impurity
\begin{eqnarray}
\left\{-\frac{\hbar^2}{2m_4} \nabla^2 + \frac{\delta {\cal E}_c}{\delta \rho}  + 
\int d {\mathbf r}_I V_X(|{\mathbf r} - {\mathbf r}_I|) |\phi({\mathbf r}_I)|^2 \right\}\Psi({\mathbf r})  
&=& \mu \Psi({\mathbf r})
\nonumber
\\
\left\{-\frac{\hbar^2}{2m_I} \nabla^2 _I + 
\int d {\mathbf r}\,V_X(|{\mathbf r} - {\mathbf r}_I|) \rho({\mathbf r}) \right\}\phi({\mathbf r}_I)  
&=& \varepsilon \phi({\mathbf r}_I)
\label{eq23}
\end{eqnarray}

In some applications, it may be necessary to fix the distance between the impurity and the centre of mass (COM) of the
droplet.
This is the case, e.g., when calculating  the energy as a function of that distance allows
to determine possible energy barriers hindering the motion of the impurity.
This can be achieved by including a constraint 
term in the energy functional. Assuming  that the  classical impurity  lies along the $z$-axis, the constraint can be introduced as
\begin{equation}
E[\rho] +  \int d {\mathbf r} \rho({\mathbf r}) V_X(|{\mathbf r} - {\mathbf r}_I|)  +  \frac{\lambda_C}{2}  \left({\cal Z} - {\cal Z}_0\right)^2
\label{eq24} 
\end{equation}
where ${\cal Z}$ is the instantaneous distance between the impurity and the COM of the droplet, ${\cal Z}_0$ is the corresponding preset constrained distance, 
and $\lambda_C$ is a constant determining the strength of the penalty  term. 
Typical values of $\lambda_C$ to ensure that the desired distance is retained within 0.1\% accuracy are  in the $1000-3000$ K \AA$^{-2}$ range. 

To illustrate how this constraint influences the EL equations when the impurity is treated as a quantum particle, we first define the droplet COM position and the expectation 
value for the quantum impurity position along the $z$-axis
\begin{eqnarray}
Z_{CM}&= &\frac{1}{N_4}   \int d {\mathbf r}\, z \, \rho({\mathbf r})  
\nonumber
\\
z_{imp}& = &\int d {\mathbf r}_I  \, z_I \, |\phi({\mathbf r}_I)|^2  
\label{eq25}
\end{eqnarray}
With these definitions, Eqs (\ref{eq23}) become
\begin{eqnarray}
&&\left\{-\frac{\hbar^2}{2m_4} \nabla^2 + \frac{\delta {\cal E}_c}{\delta \rho}   +
\int d {\mathbf r}_I V_X(|{\mathbf r} - {\mathbf r}_I|) |\phi({\mathbf r}_I)|^2  
+ \lambda_C [{\cal Z}- {\cal Z}_0] \left(\frac{-z}{N_4}\right) \right\}\Psi({\mathbf r}) 
= \mu \Psi({\mathbf r})
\nonumber
\\
&&\left\{-\frac{\hbar^2}{2m_I} \nabla^2 _I + 
\int d {\mathbf r}\,V_X(|{\mathbf r} - {\mathbf r}_I|) \rho({\mathbf r})
+ \lambda_C [{\cal Z}- {\cal Z}_0] \,z_I\,\right\} \phi({\mathbf r}_I)  
= \varepsilon \phi({\mathbf r}_I)
\label{eq26}
\end{eqnarray}
where ${\cal Z}=z_{imp}- Z_{CM}$. For a classical impurity, the term $\lambda_C [{\cal Z}- {\cal Z}_0](- z/N_4)$ has to be added to the left hand side of 
Eq. (\ref{eq21}); in this case  $z_{imp}$ is the impurity position.

The DFT equations (\ref{eq21}) or (\ref{eq23}) can be solved by the ITM in cartesian coordinates.\cite{Leh07b} Most calculations are carried out in full 3D without taking 
advantage of possible symmetries in the external potential. 
Densities, wave functions, differential operators, etc., are represented on discrete equally spaced cartesian grids. 
The spatial step employed in these calculations is typically \textit{ca.} 0.4~\AA{}. 
The differential operators (first and second derivatives) are represented by $k$-point formulas or evaluated directly in the Fourier space using the  split operator technique.\cite{Leh04} 
In the former case, 13-point formulas have been found accurate enough. 

Since the integral terms in OT-DFT can be expressed as convolutions,\cite{Dal95,Anc05a,Leh04} they can be conveniently computed in the Fourier space. Therefore, a key tool for an efficient numerical implementation of OT-DFT is the Fast Fourier Transformation (FFT) technique.\cite{Fri05} FFT algorithms are well established in the literature and have efficient parallel implementations. Note that many of the transformations required for evaluating the OT-DFT functional need to be carried out only once.

\subsection{Introduction of vorticity}
\label{3.2}

In order to represent a sustained current in liquid helium, the order parameter must be a complex valued function.
This is the case for a vortex line, which involves liquid circulation around its core. 
Vorticity around the symmetry axis ($z$) of an axially symmetric helium droplet can be represented by  
\begin{equation}
\Psi(r,\theta) \equiv  \rho^{1/2}(r) \, e^{\imath m \theta} 
\label{eq27}
\end{equation}
where $r$ is the distance from the symmetry axis, $\theta$ the polar angle, and $m$ the circulation quantum number. 
This is an eigenfunction of the total angular momentum operator,
$\hat{L}_z  \Psi(r,\theta)  = m N_4 \hbar  \Psi(r,\theta) $.
In practice, only configurations with circulation $m=\pm 1$ are relevant. 
This is because the kinetic energy of a vortex line is proportional to $m^2$ and therefore a vortex line with $m=\pm 2$ is energetically less favored than two 
separate vortex lines with $m=\pm 1$. A single vortex line in a pure $^4$He$_{500}$ droplet, in a mixed $^4$He$_{500}$+$^3$He$_{100}$ droplet, 
and the same systems doped with an HCN molecule can be seen in Fig. 30 of Ref. \onlinecite{Bar06}.

The EL equations are as for vortex-free droplets, Eqs. (\ref{eq6}) and (\ref{eq21}), but the effective wave function has to be complex valued. 
Since the ITM can only converge to a solution that has overlap with the initial order parameter, starting the calculation with an initial guess similar to Eq. (\ref{eq27}) will automatically yield the vortex solution. For instance, a vortex line along the $z$ axis can be produced by starting the imaginary-time calculation with the following initial order parameter 
\begin{equation}
\Psi(\mathbf{r}) = \frac{\rho_0^{1/2}(\mathbf{r})}{\sqrt{x^2 + y^2}} \, (x + \imath y)
\label{eq28}
\end{equation}
where $\rho_0(\mathbf{r})$ is the density corresponding to either a pure or  doped droplet without vortex. 
In cylindrical coordinates, this expression reduces to Eq. (\ref{eq27}) with $m=1$ provided that the density is axially symmetric. For a more detailed discussion, see e.g. Ref. \onlinecite{Pi07}. 

The energetics of pure and doped helium hosting vortices are usually characterised by the following quantities \cite{Pi07,Anc15,Dal00,Mat15a}

$\bullet$
Solvation energy of the impurity $X$: $S_X = E(X@^4{\rm He}_N) - E(^4{\rm He}_N)$

$\bullet$
Vortex energy: $E_V= E(V@^4{\rm He}_N) - E(^4{\rm He}_N)$

$\bullet$ 
Binding energy of the impurity $X$  to the vortex: 
\[B_X =  S_X  - \{E[(X+V)@^4{\rm He}_N] - E(V@^4{\rm He}_N)\}\]
The binding energy is the result of a delicate balance between the contributing terms and the resulting values are typically rather small. For example, 
the binding energy of a Xe atom to a vortex line is only 3--5 K.\cite{Anc14,Dal00}

The kinetic energy of the superfluid flow in the volume excluded by the impurity intuitively corresponds to $B_X$ and for this reason it is also called 
`substitution energy'.\cite{Don91} Using a classical sharp wall model for the impurity bubble and vortex line, the binding energy can be approximated as\cite{Don91}
\begin{equation}
B_X = 2 \pi \frac{\hbar^2}{m_4} \rho_0  \,R_X \, \left\{\left(1 +  \frac{a^2}{R_X^2}\right)^{1/2} 
\,ln \left[\frac{R_X}{a} + \sqrt{\left(\frac{R_X}{a}\right)^2+1}\right] - 1 \right\} 
\label{eq29}
\end{equation}
where $a$ is the radius of the vortex core and $R_X$ the radius of the atomic bubble. 
Using the Xe atom as an example, setting the liquid density to the $T = P = 0$ value $\rho_0 = 0.0218$ \AA{}$^{-3}$,  $a = 1$ \AA{},  and the
 bubble radius to the value where the Xe-He pair potential becomes repulsive, 
 $R_X = 3.5$ \AA{},  Eq. (\ref{eq29}) yields a binding energy  $B_X = 6.1$ K. 

The critical angular velocity for nucleating the vortex line represented by Eqs. (\ref{eq27})   or (\ref{eq28}) in a droplet consisting of $N_4$ helium atoms is given by\cite{Dal96}
\begin{equation}
\omega_c  = \frac{1}{\hbar}\,\frac{E_V}{N_4} 
\label{eq30}
\end{equation}
where $E_V$ is the vortex energy as defined above. 
For a $^4$He$_{1000}$ droplet this gives  $\omega_c  = 0.127\, {\rm K}/\hbar  =  0.0166\,{\rm  ps}^{-1}$. 

The above approach can be used to create individual vortex lines. 
A different strategy has to be employed to generate an array of vortex lines. 
A  rotational constraint is imposed in the rotating frame of reference (`co-rotating frame') by solving the following EL equation 
\begin{equation}
[{\cal H}-\omega \hat{L}_z] \,  \Psi  (\mathbf{r})  =  \,\mu \,
\Psi (\mathbf{r}) \;,
\label{eq31}
\end{equation}
where $\cal{H}$ is the DFT Hamiltonian (Eq. (\ref{eq6})),  $\hat{L}_z$ is the $z$-component of the angular momentum operator, and $\omega$ is the angular velocity of the co-rotating frame. 

Note that for a vortex array $\Psi(\mathbf{r})$ is no longer an eigenvector of the angular momentum.

The initial guess for imaginary-time evolution can  be obtained by  the `imprinting'  method;
for $n_v$ vortex lines, the initial guess $\Psi(\mathbf{r})$ is written as
\begin{equation}
\Psi(\mathbf{r})=\rho_0^{1/2}(\mathbf{r})\, \prod _{j=1}^{n_v} \left[ {(x-x_j)+\imath (y-y_j) \over \sqrt{(x-x_j)^2+(y-y_j)^2}}  \right] 
\label{eq32}
\end{equation}
where $\rho_0(\mathbf{r})$ is the density of the vortex-free droplet and $(x_j, y_j)$ is the initial position of the $j$\textsuperscript{th} linear vortex core  
parallel to the $z$-axis. Note that the expression for $\Psi(\mathbf{r})$ was incorrectly written in Refs. \onlinecite{Anc14,Anc15}. During the imaginary-time relaxation, the positions 
of the vortex lines will change until convergence 
to the lowest energy configuration for a given $\omega$ is reached. Complex configurations hosting several vortex lines (vortex arrays) will be described in Sec. \ref{5.13}.

\section{Dynamics}
\label{4}

Given a static initial configuration $\Psi(\textbf{r})$ and a known additional perturbation to drive the system, its dynamic evolution can be followed in real-time. 
The additional perturbation can be, for instance, a sudden photoionisation or photoexcitation of the impurity.
As discussed above, a classical or  quantum description is employed to propagate the impurity degrees of freedom, depending on its mass as compared to a helium atom.

\subsection{Heavy impurities}
\label{4.1}

Heavy impurities with no evolution in their electronic degrees of freedom can be treated using classical mechanics. Examples include photoexcitation of heavy alkali metal atoms (e.g. Rb, Cs) from the $n$s electronic ground state to the $(n+1)$s excited state\cite{Van14} and photoionisation of a Ba atom\cite{Mat14} (see also Ref. \onlinecite{Lea14b}) in helium droplets. 
Typically, these photoexcitation and photoionisation processes are considered to be instantaneous, which means that 
the light pulse is short enough that the nuclei do not have time to move, but is long enough that its energy spread covers only one (excited or ionised) electronic state.

After ionisation or electronic excitation, the total energy of the system is written as
\begin{equation}
E[\Psi, \mathbf{r}_I] = \int d \mathbf{r} \, \frac{\hbar^2}{2m_4}|\nabla \Psi|^2 + \frac{p^2_I}{2 m_I} + \int d \mathbf{r} \, {\cal E}_c(\rho)
+ \int d \mathbf{r} \, \rho(\mathbf{r}) \,  V_{X^*}(|\mathbf{r}- \mathbf{r}_I|) 
 \label{eq33}
\end{equation}
where $I$ denotes the impurity and $V_{X^*}$ is the $X$-He pair potential for the excited or ionised state. 
$V_{X^*}$  (and $V_X$ in Sec. \ref{3.1})  are usually obtained from high-level {\it ab initio} calculations\cite{Kry08} or accurate semi-empirical methods.
Since helium mostly interacts with other species through weak van der Waals forces, accurate treatment of electron correlation is very important.

The time evolution of the helium order parameter $\Psi(\textbf{r},t)$ and the impurity position $r_I(t)$ can be obtained from the TDDFT and Newton equations, respectively
\begin{eqnarray}
i\hbar\frac{\partial}{\partial t} \Psi
&=&
\left[
  -\frac{\hbar^2}{2m_4} \nabla^2 +
  \frac{\delta {\cal E}_c}{\delta \rho}
  +
  V_{X^*}(|\mathbf{r}- \mathbf{r}_I|)
\right]
\Psi
\nonumber
\\
m_I \ddot{\mathbf{r}}_I
&=&
- \nabla_{\mathbf{r}_I}
\left[  \int d \mathbf{r} \,\rho(\mathbf{r}) V_{X^*}(|\mathbf{r}- \mathbf{r}_I|)  \right]  =
-  \int d \mathbf{r} \, V_{X^*}(|\mathbf{r}- \mathbf{r}_I|)  \, \nabla \rho(\mathbf{r})  
 \;
\label{eq34}
\end{eqnarray}
For a light impurity (i.e. quantum mechanical treatment), Eqs. (\ref{eq33}) and (\ref{eq34}) become
\begin{eqnarray}
E[\Psi, \phi] &=& \int d \mathbf{r} \, \frac{\hbar^2}{2m_4}|\nabla \Psi|^2 + \int d \mathbf{r}_I \, \frac{\hbar^2}{2m_I}|\nabla_{\mathbf{r}_I} \phi|^2 +
 \int d \mathbf{r} \, {\cal E}_c(\rho)
 \nonumber
 \\
&+& \int \int  d \mathbf{r}  \, d \mathbf{r}_I \, \rho(\mathbf{r},t) \,  V_{X^*}(|\mathbf{r}- \mathbf{r}_I|) |\phi(\mathbf{r}_I,t)|^2
 \label{eq35}
\end{eqnarray}
and 
\begin{eqnarray}
& &i\hbar\frac{\partial}{\partial t} \Psi(\mathbf{r}, t) =
\left[
  -\frac{\hbar^2}{2m_4} \nabla^2 +
  \frac{\delta {\cal E}_c}{\delta \rho}
  +
 \int \, d \mathbf{r}_I \, V_{X^*}(|\mathbf{r}- \mathbf{r}_I|) |\phi(\mathbf{r}_I,t)|^2
\right]
\Psi(\mathbf{r}, t)
\nonumber
\\
& & i\hbar\frac{\partial}{\partial t} \phi (\mathbf{r}_I, t) =
\left[
  -\frac{\hbar^2}{2m_I} \nabla^2 _I+
 \int \, d \mathbf{r} \, V_{X^*}(|\mathbf{r}- \mathbf{r}_I|) \rho(\mathbf{r},t) \,
\right]
\phi(\mathbf{r}_I, t) 
\label{eq36}
\end{eqnarray}
where $\phi(\mathbf{r}_I, t)$ is the wave function for the impurity. 
Since the dynamics of the impurity and that of the liquid tend to have very different time scales, the overall time step has to be chosen with care. 
A safe choice of the shorter one for both equations can, however, increase the computational time significantly. 
Using sub-steps for the faster component can in part alleviate such issues. 
Another problem can arise from the spatial grids. 
Unless interpolation techniques are employed, both the impurity and the liquid grids must have the same size and step length. 
Since light impurities are usually fast and therefore require fine grids, this also increases the computational time required for the liquid. 
An elegant way out of this problem is to propagate the impurity using the so-called `test particle' method.\cite{Wya05} 
This approach has been used to simulate the Na and Li atom dynamics in helium droplets following the 
$(n+1)$s $\leftarrow$ $n$s excitation.\cite{Her12a}

A more complicated situation is encountered when the impurity electronic degrees of freedom must also be included in the dynamics. 
For example, when an impurity is excited from a spherical $n$s to a $n'$p state, the three degenerate p states are split by the dynamic Jahn-Teller effect. 
The interaction between a He atom and the $L=1$ state impurity can be decomposed into $\Sigma$ ($\Lambda=0$) and a doubly degenerate $\Pi$ ($\Lambda = \pm 1$) 
state, where $\Lambda$ is the projection of the orbital angular momentum on the interatomic axis.  So far, only the case where the impurity can be treated 
classically has been considered.\cite{Mat13b} To account for the dynamic orientation of the p-orbital, a simple diatomics-in-molecules (DIM) model can be
 applied.\cite{Ell63,Her08a,Elo01b} Its basic ingredients are given below.

The electronic structure of a $n'$p-state impurity (i.e. effective one-electron excited $^2$P atomic state) interacting with He atoms can be expressed in an effective one-electron p-orbital basis. 
In the diatomic frame coinciding with the $n^{\rm th}$ helium atom ($^1$S) along the $z_n$-axis, the minimal DIM basis set is $|p_{xn}\rangle$, $|p_{yn}\rangle$, $|p_{zn}\rangle$, and the helium-impurity interaction is given by
\begin{equation}
U(r_n)= 
V_{\Pi}(r_n)\mathbf{I}+\{V_{\Sigma}(r_n)-V_{\Pi}(r_n)\}|p_{zn}\rangle\langle p_{zn} |  
\label{eq37}
\end{equation}
where $r_n$ is the interatomic distance and $V_\Pi(r)$ and $V_\Sigma(r)$ are the $\Pi$ and $\Sigma$ impurity-He pair potentials in the absence of spin-orbit coupling.

For a system consisting of $N_4$ helium atoms and an excited p-state impurity, the total potential energy is constructed using the DIM model\cite{Ell63}
\begin{equation}
U=\sum_{n=1}^{N_4}\left\{V_\Pi(r_n)\mathbf{I}+[V_\Sigma(r_n)-
V_\Pi(r_n)] R_n |p_z\rangle\langle p_z | R^{-1}_n\right\}  
\label{eq38}
\end{equation}
where $R_n$ is a rotation matrix which transforms the common laboratory frame to the diatomic frame corresponding to the $n^{\rm th}$ He atom. 
In cartesian coordinates
\begin{equation}
\langle p_i|  R_n |p_z\rangle\langle p_z| R^{-1}_n |p_j\rangle = \frac{r_{in}~r_{jn}}{r_n^2}
\label{eq39}
\end{equation}
where $r_{1n}\equiv x_n$, $r_{2n} \equiv y_n$, $r_{3n} \equiv z_n$, and $r_n^2=x_n^2+y_n^2+z_n^2$ for the $n^{\rm th}$ He atom. 
The matrix elements of the DIM Hamiltonian are then
\begin{equation}
\langle p_i | U | p_j \rangle\equiv U_{ij}=
\sum_{n=1}^{N_4}\left\{V_\Pi(r_n)\delta_{ij}+
[V_\Sigma(r_n)-V_\Pi(r_n)]\frac{r_{in}~r_{jn}}{r_n^2}\right\}  
\label{eq40}
\end{equation}
Since DFT provides a continuous distribution, the discrete sum over helium atoms is replaced by integration over the density 
($\sum_n\rightarrow\int\mathrm{d}^3\mathbf{r''}\rho(\mathbf{r''})$),
which gives
\begin{equation}
U_{ij}(\mathbf{r})= \int\mathrm{d}^3\mathbf{r'}\rho(\mathbf{r'}+\mathbf{r})
\left\{V_\Pi(r')\delta_{ij}+[V_\Sigma(r')-V_\Pi(r')]\frac{r'_i~r'_j}{r'^2}\right\} 
\label{eq41}
\end{equation}
The eigenvalues $V^{\mathrm{ex}}_m(\mathbf{r})$ of this real symmetric matrix define the potential energy curves (PEC) as a function of the distance between the surrounding helium and the impurity.

The above model assumes that spin-orbit (SO) coupling is negligible. 
However, when it becomes comparable to the helium induced splitting of the p-orbitals, it must be included in the calculation. 
The total Hamiltonian is then given by $U_T = U+U_{SO}$ where $U_{SO}$ is the SO hamiltonian matrix, usually approximated by that of the free atom.\cite{Jak97}
The previously mentioned minimal DIM basis set can be extended to include the electron spin: $s = \uparrow (m_s = $ $^1\!/_2)$, $s= \downarrow (m_s= -^1\!/_2)$, i.e. 
$|i,s \rangle \equiv |p_x, \uparrow \rangle, |p_x, \downarrow \rangle, |p_y, \uparrow \rangle, |p_y, \downarrow \rangle, |p_z, \uparrow \rangle, |p_z, \downarrow \rangle$.

Kramers' theorem states that the two-fold degeneracy of the levels originating from total half-integer spin cannot be broken by electrostatic interactions.\cite{Nak01} 
Thus, all the electronic eigenstates of $U_T$ are doubly degenerate. 
Diagonalization of $U_T$ yields three doubly degenerate PEC between the impurity and surrounding helium. 
This method has also been extended to impurities in D electronic states.\cite{Lea16,Mel14}

The DIM wave function of the impurity, $|\lambda\rangle$, is determined by a six-dimensional state vector
\begin{equation}
|\lambda \rangle = \sum_{\mathop{i=x,y,z}\limits_{s=-1/2,1/2}} \lambda_{is} |i,s \rangle \;\; .
\label{eq42}
\end{equation}
The complete set of variables required to describe the
system consists of the complex valued effective wave function for helium $\Psi(\mathbf{r}, t)$ with
$\rho(\mathbf{r}, t) = |\Psi(\mathbf{r}, t)|^2$, the impurity position $\mathbf{r}_I(t)$, and the 6-dimensional complex vector to determine 
its electronic wave function $|\lambda(t)\rangle$. The total energy of the impurity-$^4$He$_N$ complex after excitation to the $^2$P manifold is
\begin{equation}
E[\Psi, \mathbf{r}_I, \lambda] =
\int d \mathbf{r} \,
\frac{\hbar^2}{2m}|\nabla \Psi|^2
+
\frac{p^2_I}{2 m_I}
+
\int d \mathbf{r} \,
{\cal E}_c[\rho]
+
\langle \lambda | V_{SO} |\lambda\rangle
+ \int d \mathbf{r} \, \rho(\mathbf{r}) \,
 V_\lambda (\mathbf{r}- \mathbf{r}_I ) 
 \label{eq43}
\end{equation}
where $V_{SO}$ is the spin-orbit coupling operator and $V_\lambda$ is defined as
\begin{equation}
V_\lambda(\mathbf{r}) \equiv \langle \lambda | {\mathcal V}(\mathbf{r}) | \lambda\rangle = \sum_{ijss'}\lambda^*_{is}{\mathcal V}^{ijss'}(\mathbf{r})\lambda_{js'} 
\label{eq44}
\end{equation}
with the components of the six-dimensional matrix ${\mathcal V}$ given by
\begin{equation}
{\mathcal V}^{ijss'}(\mathbf{r}) =\left[ V_\Pi(r)\delta_{ij} +
\left(V_\Sigma(r)-V_\Pi(r)\right)\frac{r_i r_j}{r^2} \right]\delta_{ss'}
\label{eq45}
\end{equation}

The time evolution of the system is obtained by minimizing the action
\begin{equation}
{\cal A}[\Psi, \mathbf{r}_I, \lambda] =
\int dt \left\{ E[\Psi, \mathbf{r}_I, \lambda]  -  i \hbar \int d \mathbf{r} \,
\Psi^*(\mathbf{r}) \frac{\partial}{\partial t} \Psi(\mathbf{r})  - i \hbar 
\langle \lambda | \frac{\partial}{\partial t} | \lambda \rangle - \frac{1}{2} m_I \dot{\mathbf{r}}^2_I \right\} \;
\label{eq46}
\end{equation}
Variation of $\cal A$ with respect to $\Psi^*$,  $\langle\lambda|$  and $\mathbf{r}_I$ yields
\begin{eqnarray}
&&i\hbar\frac{\partial}{\partial t} \Psi =
\left[
  -\frac{\hbar^2}{2m}\nabla^2 +
  \frac{\delta {\cal E}_c}{\delta \rho(\mathbf{r})}
  +
  V_\lambda(\mathbf{r}- \mathbf{r}_I)
\right]
\Psi
\nonumber
\\
&&i\hbar\frac{\partial}{\partial t} | \lambda \rangle  =
{\cal H} \; |\lambda\rangle
\nonumber
\\
&&m_I\ddot{\mathbf{r}}_I
=
- \nabla_{\mathbf{r}_I}
\left[
  \int d \mathbf{r} \rho(\mathbf{r})
  V_\lambda(\mathbf{r}- \mathbf{r}_I)
  \right] =  -   \int d \mathbf{r} \,  V_\lambda(\mathbf{r}- \mathbf{r}_I)  \nabla \rho(\mathbf{r})
\label{eq47}
\end{eqnarray}
where the explicit time dependence of the variables is omitted for clarity.
The second line of Eq. (\ref{eq47}) is a $6\times 6$ matrix equation with the matrix elements given by
\begin{equation}
H^{ijss'} = \int d \mathbf{r} \, \rho(\mathbf{r})
 {\mathcal V}^{ijss'} (\mathbf{r}- \mathbf{r}_I)
 + V^{ijss'}_{SO}
 \label{eq48}
\end{equation}

In order to solve Eqs. (\ref{eq34}), (\ref{eq36}) or (\ref{eq47}), initial values for the variables must be specified. 
Their choice is guided by the physics of the process studied. 
The initial helium order parameter and the initial impurity position are usually taken from the static solution of the doped droplet, with the initial impurity velocity set to zero. 
The initial choice for $|\lambda \rangle$ is dictated by the optical excitation process. 
It is often taken as one of the eigenstates of the DIM hamiltonian at the time of the electronic excitation.

All dynamic equations in this Section, as e.g. Eq (\ref{eq47}), have
been solved by using Hamming's predictor-modifier-corrector method,\cite{Ral60} initiated by a fourth-order Runge-Kutta-Gill algorithm.\cite{Ral60,Pre92}
 The integration time step employed in most applications is about 0.5 fs. 

The time-dependent relaxation of liquid helium around excited state impurities leads to the creation of sound waves and even shock waves when steep repulsive interactions are present. 
In helium droplets this can also lead to helium evaporation at the droplet surface. Eventually, evaporated helium and
bulk liquid excitations  will reach the simulation box boundaries and re-enter the box from the 
opposite side [periodic boundary conditions (PBC) are implied by the use of FFT to compute 
the convolution integrals in the OT-DFT equations]. This can  interfere with the system in an unphysical and unpredictable way, and lead to significant errors
in the calculations. 

To avoid such artifacts, absorbing boundaries should be implemented by replacing  $\imath \longrightarrow \imath + \Lambda(\mathbf{r})$ in the time-dependent 
OT-DFT equation.\cite{Mat11a}  The attenuation field $\Lambda(\mathbf{r})$ has the form 
\begin{equation}
\Lambda(\mathbf{r})= \Lambda_0 \left[ 1 + \tanh
\left(\frac{s-s_0}{a}\right)\right], \quad s\equiv|\mathbf{r}| \; .
\label{eq49}
\end{equation}
No attenuation takes place when $s < (s_0 - 2a)$ since  $\Lambda(\textbf{r}) \ll 1$. 
The absorbing region has to be large enough to remove all the unwanted effects
due to the presence of the PBC.
Note that for this method to work for bulk helium, the chemical potential must be included in the external potential during the TDDFT evolution.\cite{Mat11a}

Finally, we mention that exciplex configurations can also be studied by DFT. The method, which was inspired by the molecular model of Ref. \onlinecite{Net05}, is discussed in detail in Ref.
\onlinecite{Lea16}.

\subsection{Test particle method for light impurities}
\label{4.2}

If the impurity-helium interaction is highly repulsive in the impurity excited state, its velocity can quickly become very large. 
Inside the droplet this velocity tends to fall below the Landau critical velocity
 because the kinetic energy is dissipated through efficient coupling to elementary excitations of the liquid. This process is not instantaneous\cite{Bue16,Mat14} 
 and the impurity velocity can remain high during this initial period. Furthermore, in the case of helium droplets velocities may remain high indefinitely if the impurity leaves it. 
The wave packet for a light impurity with high velocity exhibits rapid spatial and temporal oscillations, which require the use of very fine spatial grids and short time steps. 
Since these grids must be compatible with the ones used for helium, the computation especially in 3D becomes quickly unaffordable. 

To avoid this problem, the impurity degrees of freedom can be described by Bohmian dynamics.\cite{Wya05} This approach, 
which is equivalent to solving  the Schr\"odinger equation, has been tested for the dynamics of excited state Li and Na atoms ejected from 
the helium droplet surface.\cite{Her12a} An overview of this method is given below.

The second line in Eq. (\ref{eq36}) can effectively be cast into the format of a time-dependent Schr\"odinger equation
(note that the $I$ index to $\mathbf{r}_I$ is dropped to simplify the  notation)
\begin{equation}
 \imath \hbar\frac{\partial}{\partial t} \phi (\mathbf{r}, t) = \left[  -\frac{\hbar^2}{2m} \nabla^2 + V(\mathbf{r})  \,\right] \phi(\mathbf{r}, t) 
 \; .
\label{eq50}
\end{equation}
Using the hydrodynamic form suggested by Madelung,\cite{Mad27} the complex wave function can be written as
\begin{equation}	
\phi(\mathbf{r}, t) \equiv \chi(\mathbf{r}, t) e^{i {\cal S}(\mathbf{r}, t)}	
\label{eq51}
\end{equation}
where $\chi \ge 0$ 
and ${\cal S}$ are both {\it real} valued functions.\cite{Wya05} 
While the real and imaginary parts of 
$\phi$ may oscillate rapidly, the behavior of $\chi$ and ${\cal S}$ is much smoother than $\phi$  as a function of time.\cite{Her12a} 
The associated velocity field and the current density are defined as $\mathbf{v}(\textbf{r},t) \equiv (\hbar/m) \nabla {\cal S}$ 
and $\mathbf{j}(\mathbf{r}, t) \equiv \hbar/(2 m i)\, [\phi^* \nabla \phi - \phi \nabla \phi^*] = \chi^2 \, \mathbf{v}$. 
Substitution of Eq. (\ref{eq51}) into Eq. (\ref{eq50}) and equating the real and imaginary parts of the left and right hand side terms in Eq. (\ref{eq50})
yields the following -- quantum hydrodynamic -- equations 
for $\chi$ and ${\cal S}$:
\begin{eqnarray}
\frac{\partial \chi^2}{\partial t}&=& - \nabla \cdot \mathbf{j} \;\;\;\;\;\;\; \;\;\;\;\;\;\;\; \;\;\;\;\;\;\;\;\;\;\;\; \;\;\;\;   {\rm [continuity \; equation]}
\nonumber
\\
- \hbar \, \frac{{\partial \cal S}}{\partial t} & = & \frac{1}{2} \, m \mathbf{v}^2  + {\cal Q}(\mathbf{r}, t)  + V(\textbf{r})  \;\;\;\;\;  {\rm [quantum \;  Hamilton-Jacobi \; equation]}
\label{eq52}
\end{eqnarray}
where ${\cal Q}$ is the so-called quantum potential (or quantum pressure)
\begin{equation}
{\cal Q}(\mathbf{r}, t) \equiv - \frac{\hbar^2}{2m} \frac{\nabla^2 \chi}{\chi} 
\label{eq53}
\end{equation}

\begin{figure}[t]
\vspace{42pt}
\begin{center}
\subfigure{
\resizebox*{3.5cm}{!}{\includegraphics{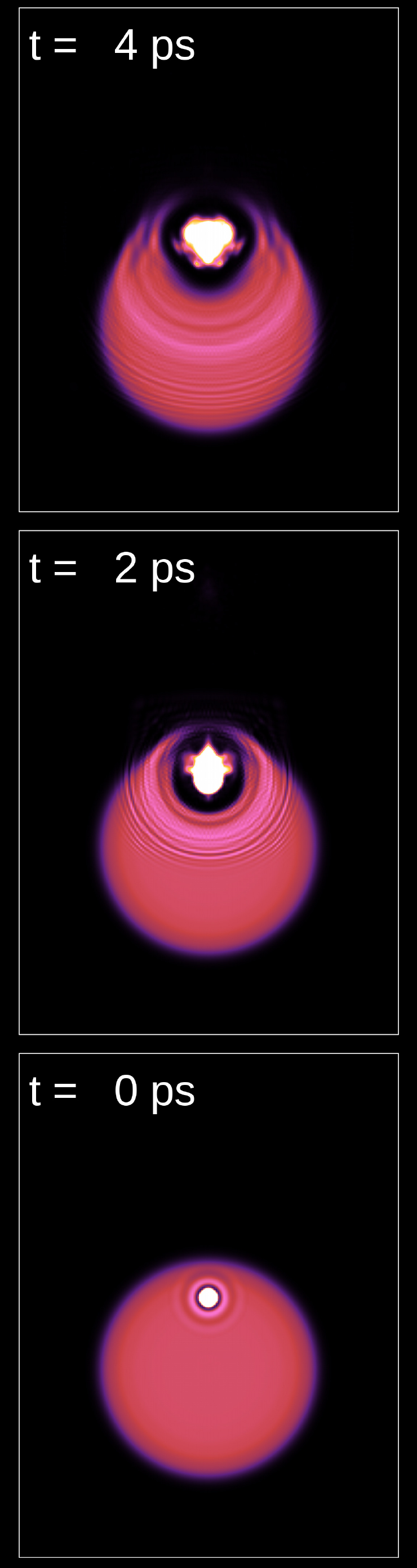}}}%
\subfigure{
\resizebox*{3.5cm}{!}{\includegraphics{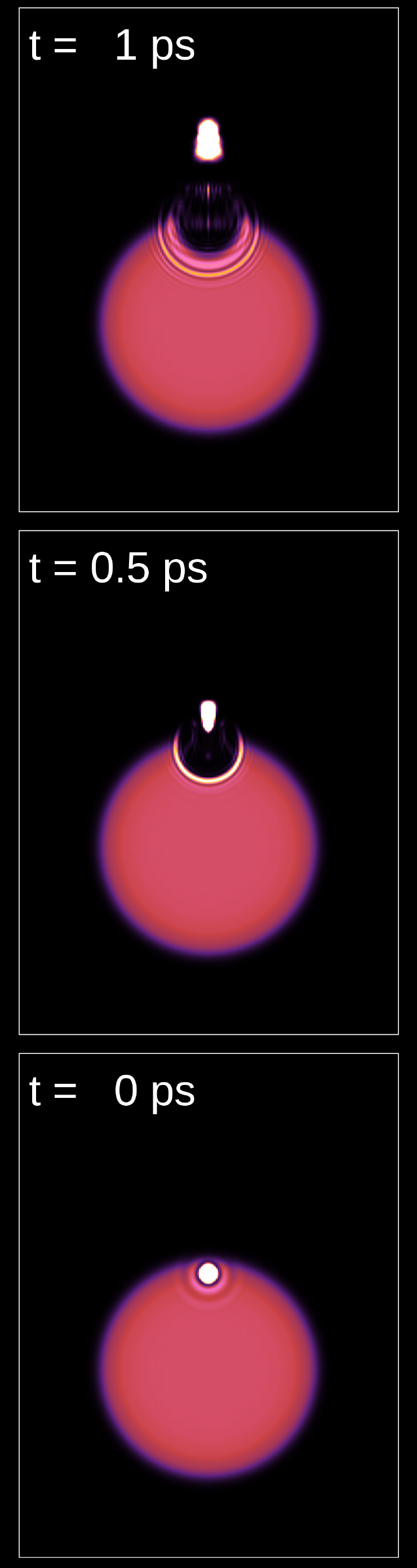}}}%
\caption{\label{fig4}
(a) Snapshots of the He density during the evolution of the He$^*$@$^4$He$_{1000}$ starting from $\mathbf{r}_I$= 15 \AA{}.
The bright yellow spot is the probability distribution of the He$^*$ being ejected. 
(b)  Same as (a) starting from $\mathbf{r}_I$= 20 \AA{}. 
}
\end{center}
\end{figure}

The above Eq. (\ref{eq52}) can be solved by using the test particle method as follows. 
The probability density $\chi^2$  and the current density as a function of time can be constructed from a histogram based on $M$ test particles. 
Given a set of test particle trajectories, $\{\mathbf{R}_i(t)\}_{i=1}^M$, where $\mathbf{R}_i(t)=\mathbf{R}(\mathbf{r}_i,t)$ and
$\mathbf{R}_i(0)=\mathbf{r}_i$, 
$\chi^2$ and $\mathbf{j}$ can be computed as
\begin{eqnarray}
\chi^2(\mathbf{r},t)&=&
\lim_{M\rightarrow\infty}\frac{1}{M}\sum_{i=1}^M\delta[\mathbf{r}-\mathbf{R}_i(t)]
\nonumber
\\
\mathbf{j}(\mathbf{r},t)&=&
\lim_{M\rightarrow\infty}\frac{1}{M}\sum_{i=1}^M\mathbf{v}[\mathbf{R}_i(t)]
\delta[\mathbf{r}-\mathbf{R}_i(t)]
\; .
\label{eq54}
\end{eqnarray}
For example, a value of $M=200\,000$ was used in Ref. \onlinecite{Her12a} to simulate the desorption of Li and Na atoms excited to the 3s and 4s states, respectively. 

The continuity equation is automatically fulfilled provided that $\dot{\mathbf{R}}_i(t)=\mathbf{v}[\mathbf{R}_i(t)]$, i.e.
the test particle velocity must be equal to the value of the velocity field at that point. 
By taking the gradient of both sides of the second line in Eq. (\ref{eq52}) and rewriting it in the Lagrangian reference frame 
($d/dt=\partial/\partial t+\mathbf{v}\cdot\mathbf{\nabla}$), the following equation of motion for the test particles is obtained (`Quantum Newton equation')
\begin{equation}
m\,\ddot{\mathbf{R}}_i(t)=-\left.\mathbf{\nabla}\left[{\cal Q}(\mathbf{r},t)+
V(\mathbf{r},t)\right]\right|_{\mathbf{r}=\mathbf{R}_i(t)}
\label{eq55}
\end{equation}
The quantum potential ${\cal Q}(\mathbf{r},t)$ is computed from the test particle probability density histogram using the  same structure grid 
and $n$-point difference formulas as used for helium DFT calculations. 

The expectation values of $\mathbf{r}(t)$ and $\mathbf{v}(t)$ are often needed for visualization purposes
\begin{equation}
\langle \mathbf{r}(t) \rangle = \int d \mathbf{r} \, \mathbf{r} \, \chi^2(\mathbf{r},t)
\label{eq56}
\end{equation}
\begin{equation}
 \langle \mathbf{v}(t) \rangle 
 =  \int d \mathbf{r} \, \mathbf{v}(\mathbf{r}, t) \, \chi^2(\mathbf{r},t) 
 = \frac{1}{m} \, \int d \mathbf{r} \, \mathbf{j}(\mathbf{r}, t) 
 \label{eq57}
\end{equation}
Furthermore, the energy of the impurity as a function of time is
\begin{equation}
E(t) = \int d \mathbf{r} \, \left[ \frac{1}{2} m \, v^2(\mathbf{r},t)  +{\cal Q}(\mathbf{r},t) + V(\mathbf{r}, t) \right] 
\chi^2(\mathbf{r},t)
 \label{eq58}
\end{equation}
As an example application, 
Fig. \ref{fig4} displays snapshots of the $^4$He$_{1000}$ droplet density on the $x-z$
 plane following  a sudden 1s$^2$ to 1s\,2s  excitation  of a single helium atom  (i.e. formation of He$^*$ as indicated by
 the bright yellow spot in the figure) from  bulk   (15 \AA{} from the center of
the droplet) and  surface (18 \AA{}) locations.\cite{Bar16} 
 The He$^*$ atom ejected from the droplet is  represented by $10^6$ test particle trajectories. 
  Note that, due to the non-spherical liquid distribution at 
  the droplet surface, the normally forbidden s--s transition becomes partially allowed. 
  
\subsection{Simulation of absorption and emission spectra using the density fluctuation method}
\label{4.3}

\begin{figure}[t]
\vspace{42pt}
\begin{center}
\subfigure{
\resizebox*{7cm}{!}{\includegraphics{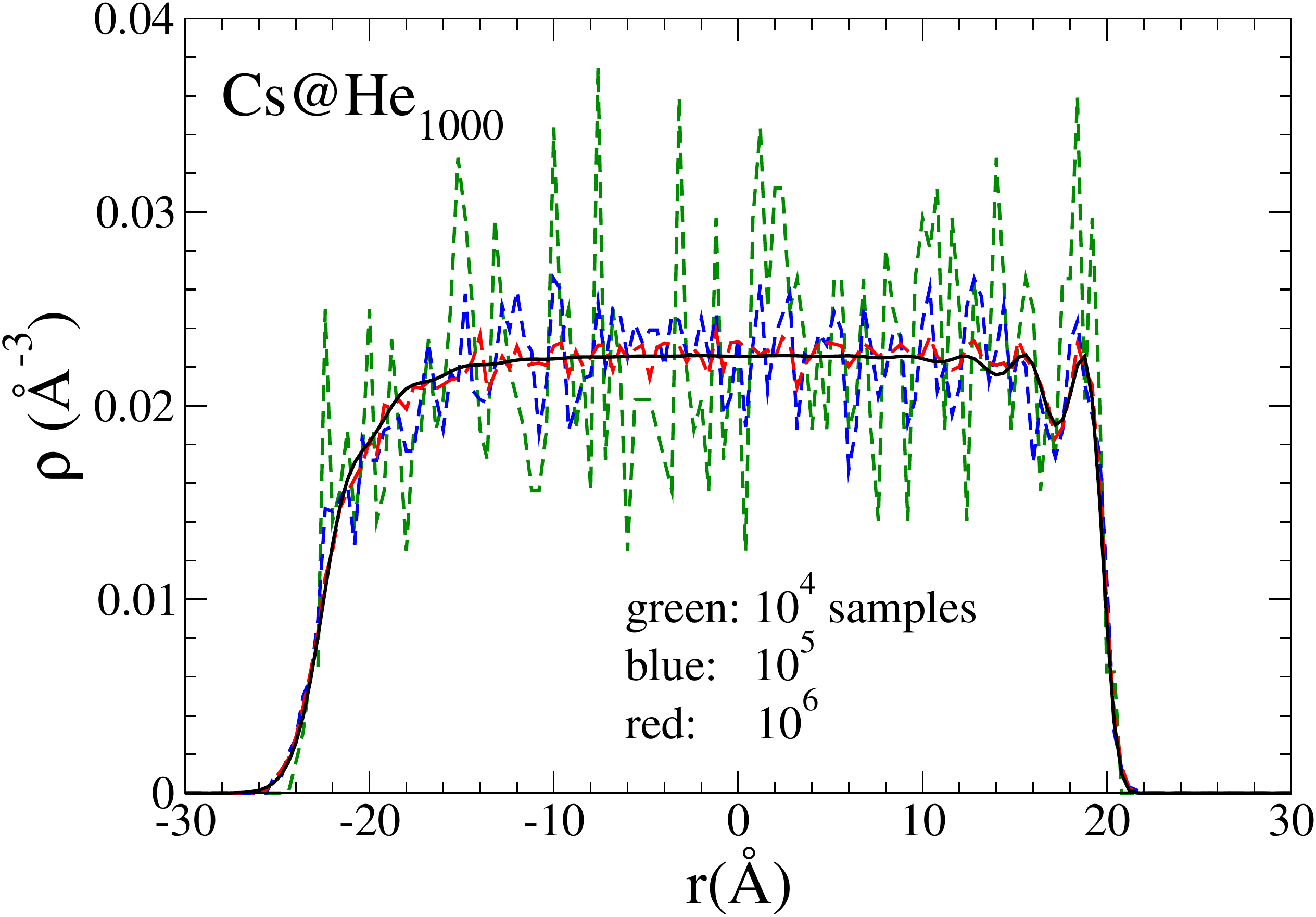}}}%
\subfigure{
\resizebox*{7cm}{!}{\includegraphics{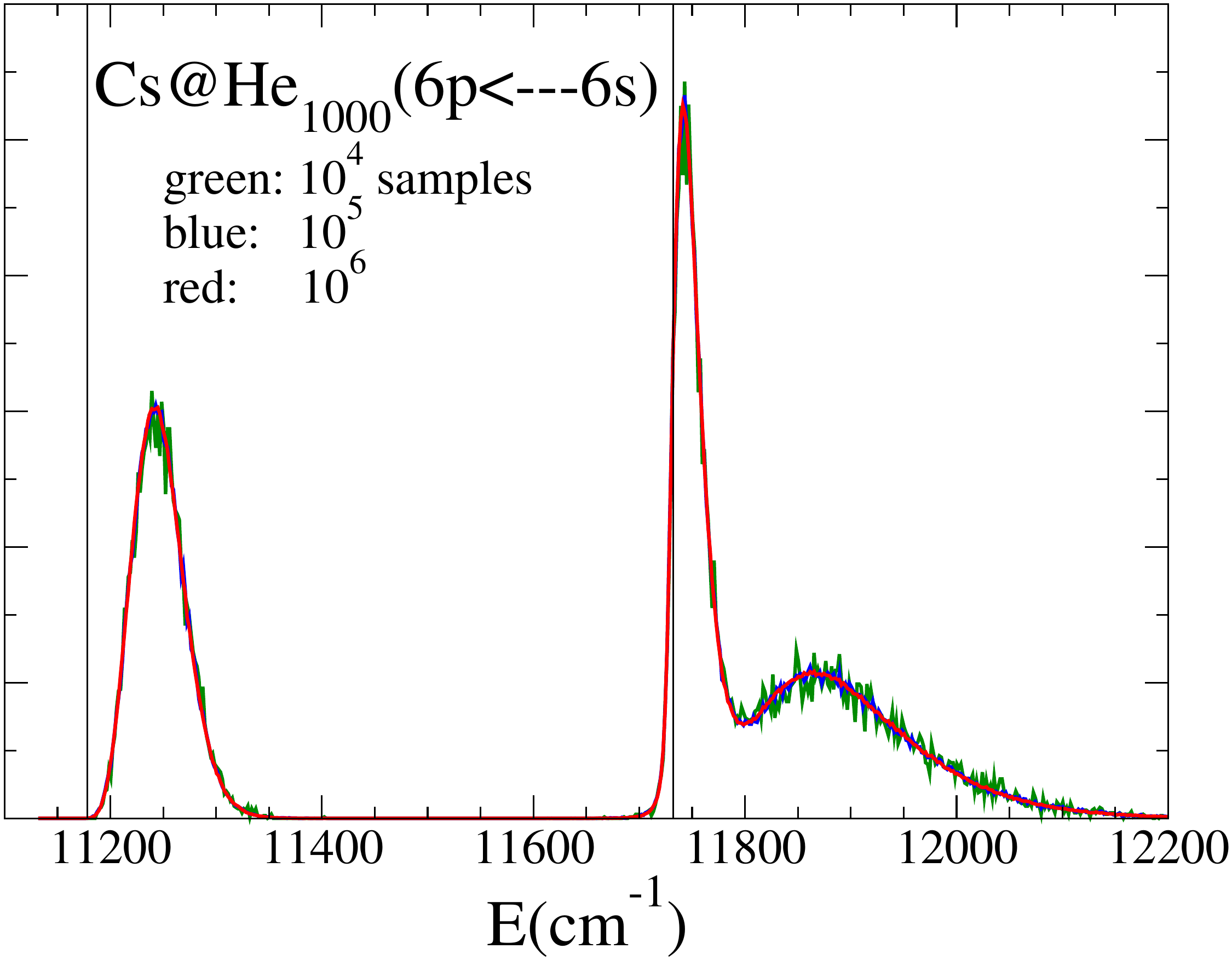}}}%
\caption{\label{fig5}
(a) DFT density profile along the $z$ axis [$\rho(z)$] of the Cs@$^4$He$_{1000}$ droplet (solid line) and the
simulated profile corresponding to $n_c = 10^4$ (green), $10^5$ (blue), and $10^6$ (red) simulations. (b) $6p \leftarrow 6s$ 
absorption spectrum of Cs obtained using the corresponding sampled density distribution. The Patil He-Cs potential\cite{Pat91}
has been used for the ground state and the Pascale He-Cs potentials for the excited states.\cite{Pas83} 
The spectrum is given in arbitrary units; in the cases of $n_c = 10^4$ and $10^5$, it has been multiplied by a factor of 
100 and 10 respectively so that they can be compared to that of $n_c = 10^6$. 
}
\end{center}
\end{figure}

Optical absorption and fluorescence spectroscopy of doped helium droplets establishes an important link between experiments and theory. 
Not only does it provide a test to validate the applied theoretical method, but it can also give a microscopic view into the associated dynamics. 
The latter aspect has, for example, been used to establish the details of impurity solvation in helium droplets (e.g. interior vs. surface solvation).\cite{Sti99}

Provided that the helium dynamics does not contribute to the spectrum significantly, the transition energies can be approximated with the eigenvalues of $U_T$ 
defined after Eq. (\ref{eq41}). Within this model, line broadening originates from fluctuations in the helium density\cite{Mat11b} and/or the zero-point density 
distribution of the impurity $|\phi(\mathbf{r}_I)|^2$.\cite{Her10} An outline of the former case is given below (`DF sampling method').

Within the Born-Oppenheimer approximation, electronic and nuclear degrees of freedom are treated separately. 
The absorption and fluorescence line shapes can then be calculated by Fourier transforming the helium bath time-correlation function.\cite{Tan07} 
Within the semi-classical approximation,\cite{Her08a} the absorption line shape function $I(\omega)$ is
\begin{equation}
I(\omega) \propto\sum_{m}
\int\mathrm{d}^3\mathbf{r}~
|\phi^{\mathrm{gs}}(\mathbf{r})|^2\delta
\left[\omega-(V^{\mathrm{ex}}_m(\mathbf{r})/\hbar-\omega^{\mathrm{gs}})\right]
\label{eq59}
\end{equation}
where `ex' and `gs' refer to the electronic excited and ground states, respectively. The DF sampling method constructs this expression stochastically
 by generating a large number of helium-impurity configurations ($n_c \approx 10^6$). 
Each configuration consists of  $N$ helium atoms positions and, if the impurity is light,  its position as well. For helium, these coordinates are randomly 
generated by importance sampling using the DFT helium one-particle density $\rho(\mathbf{r})/N$ as the probability distribution, where short-range 
correlations from the hard-sphere term are also considered. The impurity positions are sampled using the zero-point distribution $|\phi^{\mathrm{gs}}(\mathbf{r}_I)|^2$. 
Such sampling is obviously  not required for classical impurities. 

Fig. \ref{fig5} shows the one-particle density generated by importance sampling,  compared to the  calculated by  DFT  in the 
case of a (classical) Cs doped $^4$He$_{1000}$ droplet, using $n_c = 10^4, 10^5,$ and $10^6$ configurations.
 Examples  considering the impurity quantum mechanically can be found in  Refs. \onlinecite{Her10,Her11}.
 
\begin{figure}[t]
\vspace{42pt}
\begin{center}
\subfigure{
\resizebox*{7cm}{!}{\includegraphics{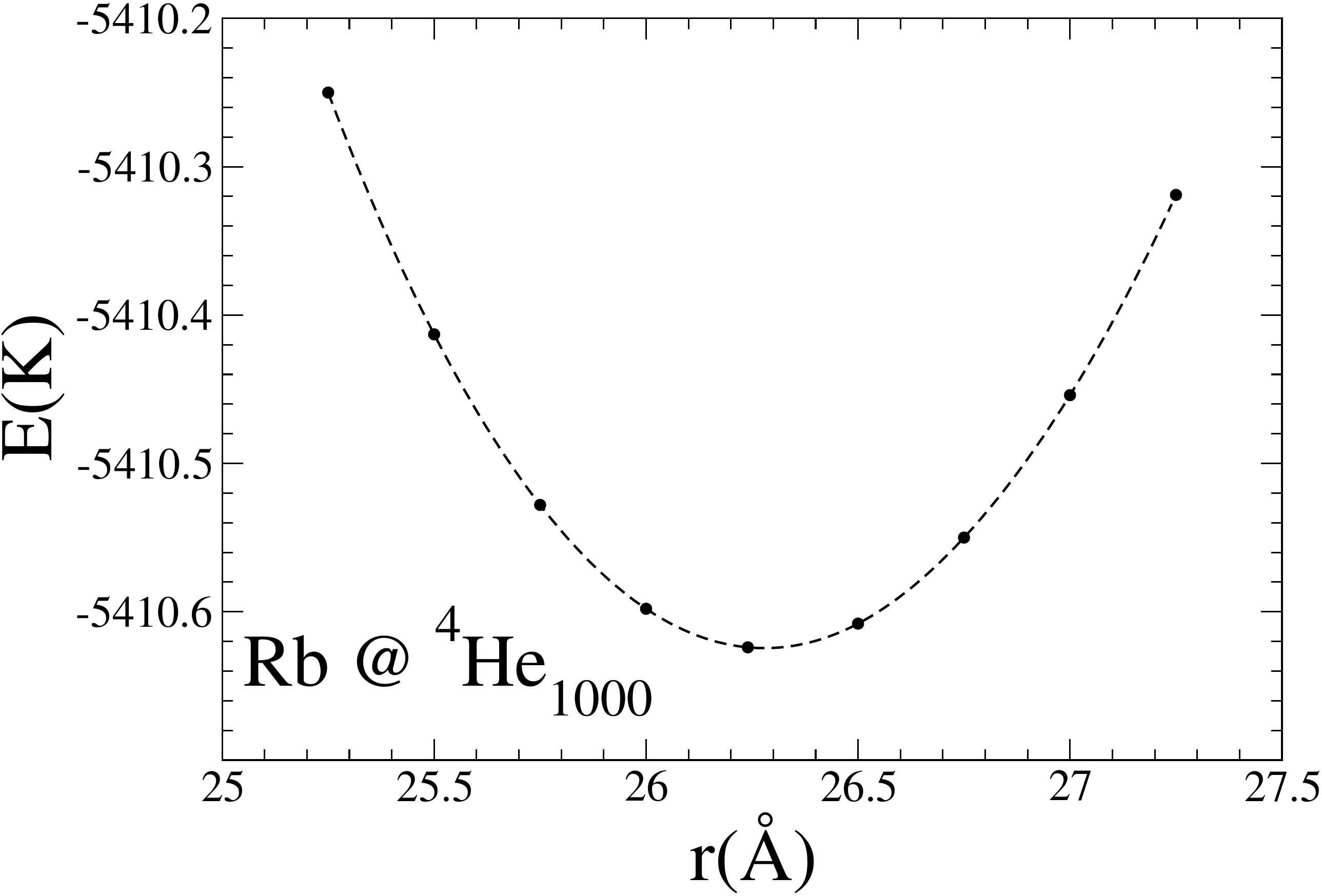}}}%
\subfigure{
\resizebox*{7cm}{!}{\includegraphics{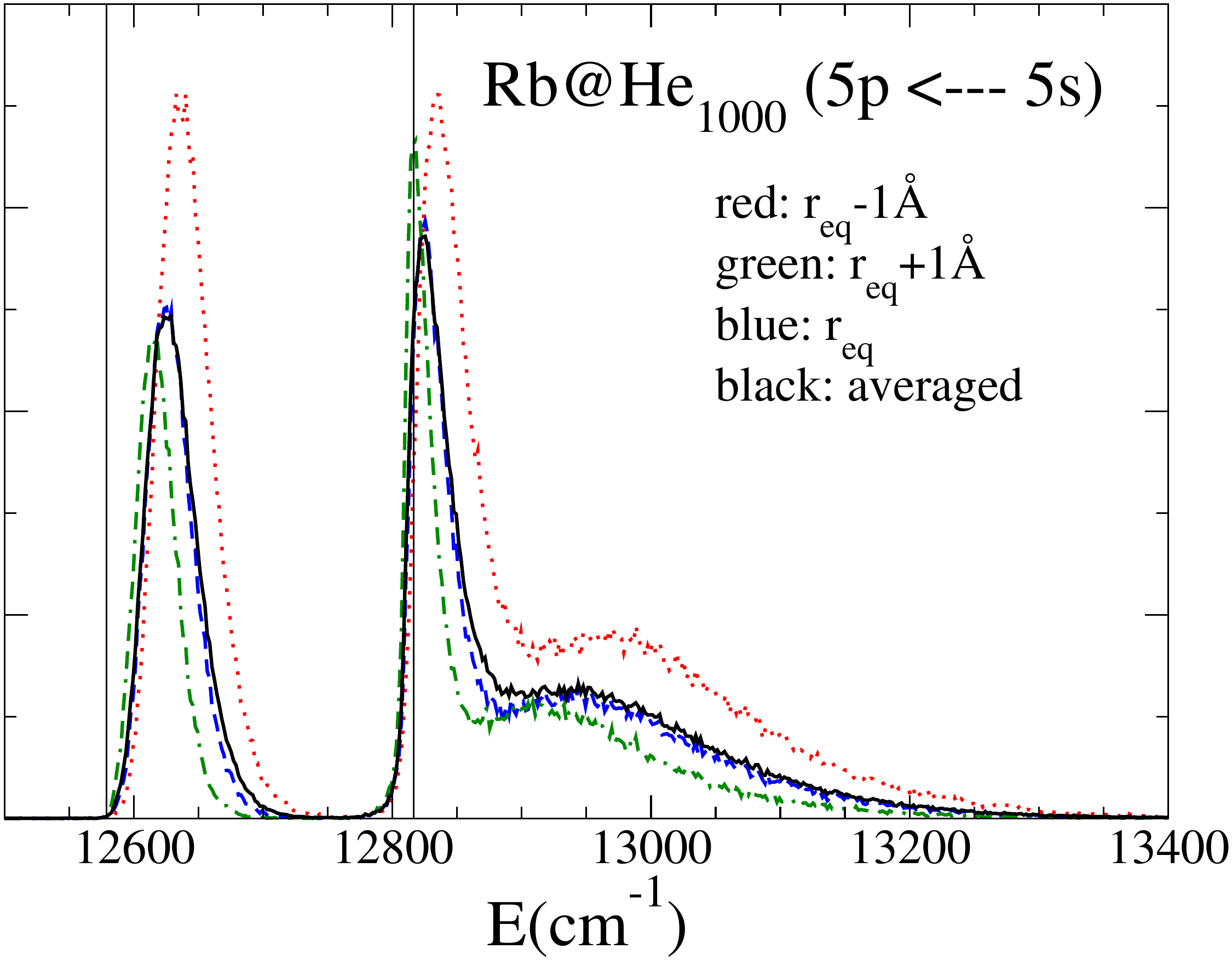}}}%
\caption{\label{fig6}
(a) Energy of  Rb@$^4$He$_{1000}$ as a function of the distance of the Rb atom from the COM of the droplet; the equilibrium position of Rb is $r_{eq}$=26.24 \AA{}.
 (b) Absorption dipole spectrum (arbitrary units) for  several locations of the Rb atom shown in (a). The number of configurations used to simulate the three spectra is $n_c=10^5$. 
}
\end{center}
\end{figure}

To determine the contribution of each configuration $j$ to the overall absorption spectrum, the corresponding line position is computed from the difference between the excited and ground state energies. The latter is simply taken as the sum of pairwise ground state interactions, $V^{\mathrm{gs}}\{j\} = \sum_i V_X(|\mathbf{r}^{\{j\}}_i - \mathbf{r}^{\{j\}}_I|)$, 
whereas the excited state energy is determined by the eigenvalues of $U_T=U+U_{SO}$, where $U_T$ was defined after Eq. (\ref{eq41}) as the sum of the DIM [Eq. (\ref{eq40})] and the SO hamiltonians.
The absorption spectrum is finally constructed as an histogram of  the line positions corresponding to each configuration
\begin{equation}
I(\omega) \propto\sum_{m} \, \sum_{\{j\}}^{n_c}
\delta\left[\omega-
(V^{\mathrm{ex}}_m\{j\} - V^{\mathrm{gs}}\{j\})/\hbar\right]
\; . 
\label{eq60}
\end{equation}
As an example, Fig. \ref{fig5} shows the absorption spectrum for the Cs 6p $\leftarrow$ 6s transition in a $^4$He$_{1000}$ droplet with $n_c = 10^4, 10^5,$ and 
$10^6$. Note that even $n_c = 10^5$ appears to be sufficient to produce a good quality spectrum even though the sampled one-particle density has not fully converged 
yet. It should be stressed that other sources of broadening such as thermal motion,\cite{Her08a} coherent helium bath dynamics,\cite{Elo04} or droplet size distribution 
can also contribute to line broadening and are not included in this model.

The influence of thermal motion on the absorption spectrum can be accounted for
by considering a thermodynamic ensemble of doped droplets at the experimental temperature of 0.37 K. 
This is illustrated in the following for Rb doped $^4$He$_{1000}$ droplets.
By constraining the distance between Rb and the droplet COM, the energy landscape seen by Rb can be computed as shown in Fig. \ref{fig6}. 
The energy  corresponding to the experimental temperature of 0.37~K is obtained for distances of $\sim \pm 1$ \AA{} away from equilibrium.
Fig. \ref{fig6} also shows the Rb 5p $\leftarrow$ 5s absorption spectrum corresponding to selected displacements from equilibrium. 
Indexing them by $i$ and denoting the corresponding spectra by $I_i(\omega)$, the thermally averaged spectrum can be constructed as
\begin{equation}
I(\omega) = \frac{1}{{\cal Z}} \sum_i I_i(\omega) \, e^{-\Delta E_i/(k_B T)}
\label{eq61}
\end{equation}
where $k_B$ is the Boltzmann constant,  $\Delta E_i $  the energy difference from the equilibrium position, $\Delta E_i = E_i - E_{eq}$, and ${\cal Z} =  \sum_i e^{- \Delta E_i/(k_B T)}$ is the partition function. 
At 0.37 K the thermally averaged absorption spectrum of Rb is very close to that obtained at the equilibrium position.

Finally, we note that fluorescence spectra can be calculated in a similar way by exchanging the roles of the ground and excited states.\cite{Lea16} In this case the 
DF sampling employs the helium density around the impurity in its excited electronic state instead of the ground state.

\section{Recent applications of DFT for impurity doped superfluid helium}
\label{5}

This section gives an overview of selected results for impurity doped superfluid helium systems obtained with DFT over the past ten years.
In addition to covering the wealth of activity on helium droplets doped with alkali and alkaline earth metal atoms, which have been thoroughly studied 
from both experimental and theoretical points of view, special attention is paid on reviewing the real-time capture of simple atoms by helium droplets (with or
 without vortex lines) and the dynamics following excitation of impurities attached to helium droplets. 
Furthermore, other aspects that have also drawn much attention recently, such as soft-landing of doped helium droplets on solid surfaces and
the appearance of vortex arrays in helium droplets, are included. Last but not least, impurity dynamics in liquid helium is also considered due to the recent activity
 in this area. The choice of these topics was motivated by the previous experimental work as well as their successful study by DFT or TDDFT.

\subsection{Alkali metal doped helium droplets: solvation and absorption spectra}
\label{5.1}

\begin{figure}[t]
\vspace{42pt}
\begin{center}
\resizebox*{13cm}{!}{\includegraphics{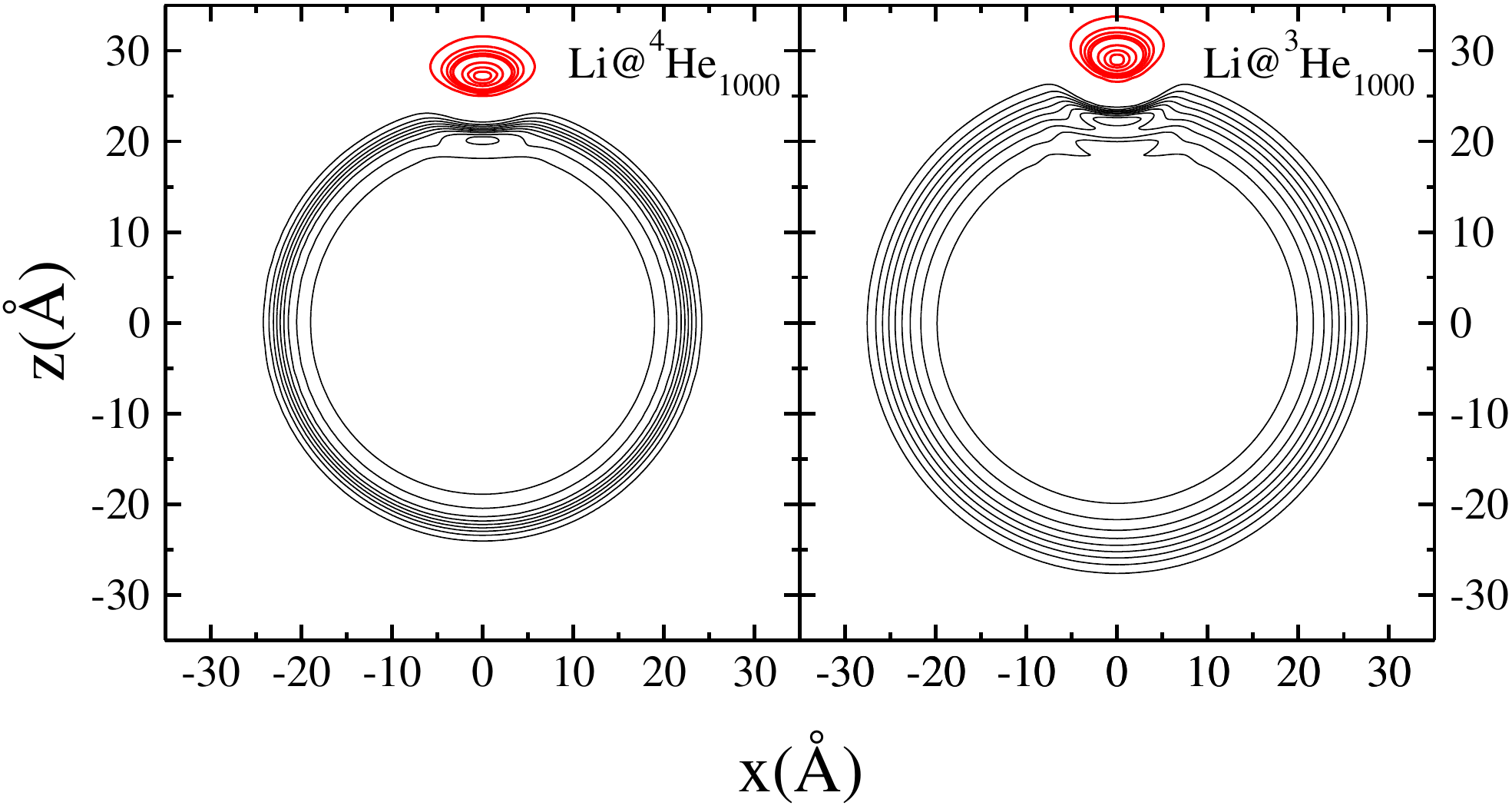}}
\caption{\label{fig7}
Helium equidensity lines in a symmetry plane of a Li@He$_{1000}$ droplet.
There are nine lines with values ranging from $0.1\rho_0$ to $0.9 \rho_0$ in $0.1 \rho_0$ steps with $\rho_0 = 0.0218$ \AA$^{-3}$ for $^4$He, and 0.0163 \AA$^{-3}$ for $^3$He. 
Also shown are equiprobability density lines for Li between
 $0.001\times \max  \{|\phi^{gs}|^2\}$ and $0.99\times \max \{|\phi^{gs}|^2\}$. Left panel: Li@$^4$He$_{1000}$. Right panel: Li@$^3$He$_{1000}$.\cite{Her11}
} 
\end{center}
\end{figure}

Since quantum mechanics dominates the behavior of $^4$He droplets, even the solvation of neutral atomic impurities depends on a subtle interplay between the 
impurity-helium interaction potential and the liquid energetics (e.g. surface tension). A simple procedure to predict whether an impurity solvates in superfluid helium
 (heliophilic) or resides on the surface of helium 
droplets (heliophobic) was introduced in Ref. \onlinecite{Anc95}. If the impurity is treated classically and interacts with helium through a simple Lennard-Jones potential 
(`spherical' impurity), the solvation behavior can be inferred from the value of a dimensionless parameter $\lambda$
\begin{equation}
\lambda = 2^{-1/6}\rho_0 \,\epsilon \,r_{min}/\gamma
\label{eq62}
\end{equation}
where $\gamma$ is the liquid surface tension, $\rho_0$ is the bulk density,
and $\epsilon$ and $r_{min}$ are the well depth and equilibrium distance
of the Lennard-Jones potential, respectively. DFT calculations\cite{Anc95} suggest that  
if $\lambda > 1.9$ the impurity is heliophilic and solvates inside helium droplets, whereas 
if $\lambda < 1.9$ the impurity is heliophobic and resides on the droplet surface instead. 
The validity of treating the impurity classically can be assessed by the de Boer parameter $\lambda_{dB}$
\begin{equation}
\lambda_{dB} = \frac{h^2}{m\,\epsilon\, r_{min}^2}
\label{eq63}
\end{equation}
where $h$ is the Planck constant and $m$ is the impurity mass. 
For light impurities (e.g. H, Li, Na) $\lambda_{dB} > 1$, whereas for heavier impurities that can be treated classically (e.g. Ar, SF$_6$) the value of $\lambda_{dB}$ is typically less than 0.15.

Based on Eq. (\ref{eq62}), all alkali metal atoms have $\lambda $ values much lower than 
the threshold value $\lambda =$1.9, which indicates that they should reside on the droplet surface.\cite{Anc95} 
This prediction was confirmed by subsequent experimental work\cite{Sti96} in which the observed impurity line positions were very close to their gas phase values. 
Surface location is a direct consequence of the very weak binding between alkali metal atoms and helium. 
The prediction based on Eq. (\ref{eq62}) is less conclusive for alkaline earth metals. 
For example, $\lambda$ is very close to 1.9 for Ca, Sr, and Ba, whereas a value of 2.6 is obtained for Mg. 
All available experimental evidence indicates that the former species are located on the droplet surface. 
For Mg the value of $\lambda$ indicates that this species is heliophilic, which is confirmed by both DFT  and QMC calculations. 
This is discussed further in Sec. \ref{5.2}.

Two major achievements of DFT applied to doped helium droplets are the determination of the resulting solvation structures and the associated optical spectra. 
In addition to the work reviewed earlier in Ref. \onlinecite{Bar06}, joint experimental-theoretical studies on alkali metal atoms from Li to Cs in both $^4$He and $^3$He 
droplets have been published since.
Impurities were treated either classically (i.e. as an external field)\cite{Bue07} or quantum mechanically\cite{Her10,Her11} depending on their masses. 
Optical absorption spectra in these studies were computed from the Frank-Condon factors,\cite{Bue07} the DF sampling method,\cite{Her10} 
or Fourier transformation of the time-correlation function.\cite{Her10,Her11} 
In addition, evaluation of time-dependent first-order polarization based on the superfluid helium response has been used for calculating the optical spectrum of intrinsic 
helium impurities,\cite{Elo04} which will be discussed in more detail in Sec. \ref{5.8}.

The above mentioned studies on alkali metal atoms have demonstrated good agreement with existing experimental results. 
The calculations were able to reproduce not only the general features of the absorption spectra for $^4$He vs. $^3$He droplets, but also the fine details observed for Li and Na coupled to either a bosonic or a fermionic helium surface. 
As an example, Fig. \ref{fig7} shows the helium density and Li probability density on the symmetry plane of a Li@He$_{1000}$ droplet (`dimple' surface structure). 
The droplet surface region is contained between the inner and outer equidensity contour lines.
Since both the surface tension and the equilibrium density of $^3$He are smaller than for $^4$He, the surface width of $^3$He droplets is larger. The resulting 
dimple solvation structure for other alkali metal-doped $^3$He$_{1000}$ and $^4$He$_{1000}$ droplets can be found in Fig. 3 of Ref. \onlinecite{Bue07}. 

The dimple solvation structure is deeper on a $^3$He than on a $^4$He surface. This is a direct consequence of the smaller surface tension of 
$^3$He, which also yields a wider surface region. A deeper dimple  increases the interaction between the impurity and the droplet. 
For this reason, the absorption spectra exhibit larger blue shifts in $^3$He vs. $^4$He droplets. This trend has been observed experimentally and confirmed by 
DFT calculations, which are also able to reproduce the fine details in the spectra. For example, in addition to the different absorption line shifts observed for Li/Na 
doped $^4$He vs. $^3$He droplets, the appearance of weak sidebands in $^4$He is  reproduced by DFT.\cite{Her10,Bue07,Her11} 
This methodology has also been extended to molecular species such as Li$_2$.\cite{Lac13}

Both DFT and QMC calculations for doped-helium systems 
require an accurate representation of the helium-impurity interaction as  input. 
Since the excited electronic states are typically much higher in energy than the ground state, DFT calculation of solvation structures only requires the ground state interaction. 
For the spectroscopic applications discussed above, the corresponding excited state interaction with helium must also be known. 
Since a helium atom usually introduces only a small perturbation to the electronic structure of the impurity, the pairwise potential approximation is often very accurate. 
Pair potentials can be obtained with high accuracy from \textit{ab initio} electronic structure calculations such as full configuration interaction or coupled-cluster theory. 

When the pair potential approximation is not sufficient, a perturbative configuration interaction (PCI) method can sometimes be employed.\cite{Cal11b} 
This method was used for excited states of alkali metal atoms where the electronic degrees of freedom couple significantly to the nearby helium atoms. 
PCI solves the electronic Schr\"odinger equation numerically in the valence orbital basis set for a free atom and includes an additional potential due to the valence electron-helium density interaction. 
This method can be applied to highly excited states of alkali metals where the conventional approach would fail. 
In a series of joint theoretical-experimental studies, it has been applied to model one- and two-photon spectroscopy of highly excited states of Rb, K, and Cs atoms in $^4$He droplets,\cite{Pif10}
 the spectroscopy of Rydberg states of Na atoms in $^4$He droplets, \cite{Log11a} and the photoionisation and imaging spectroscopy of Rb atoms attached to $^4$He droplets.\cite{Fec12} 

\begin{figure}[t]
\vspace{42pt}
\begin{center}
\resizebox*{7cm}{!}{\includegraphics{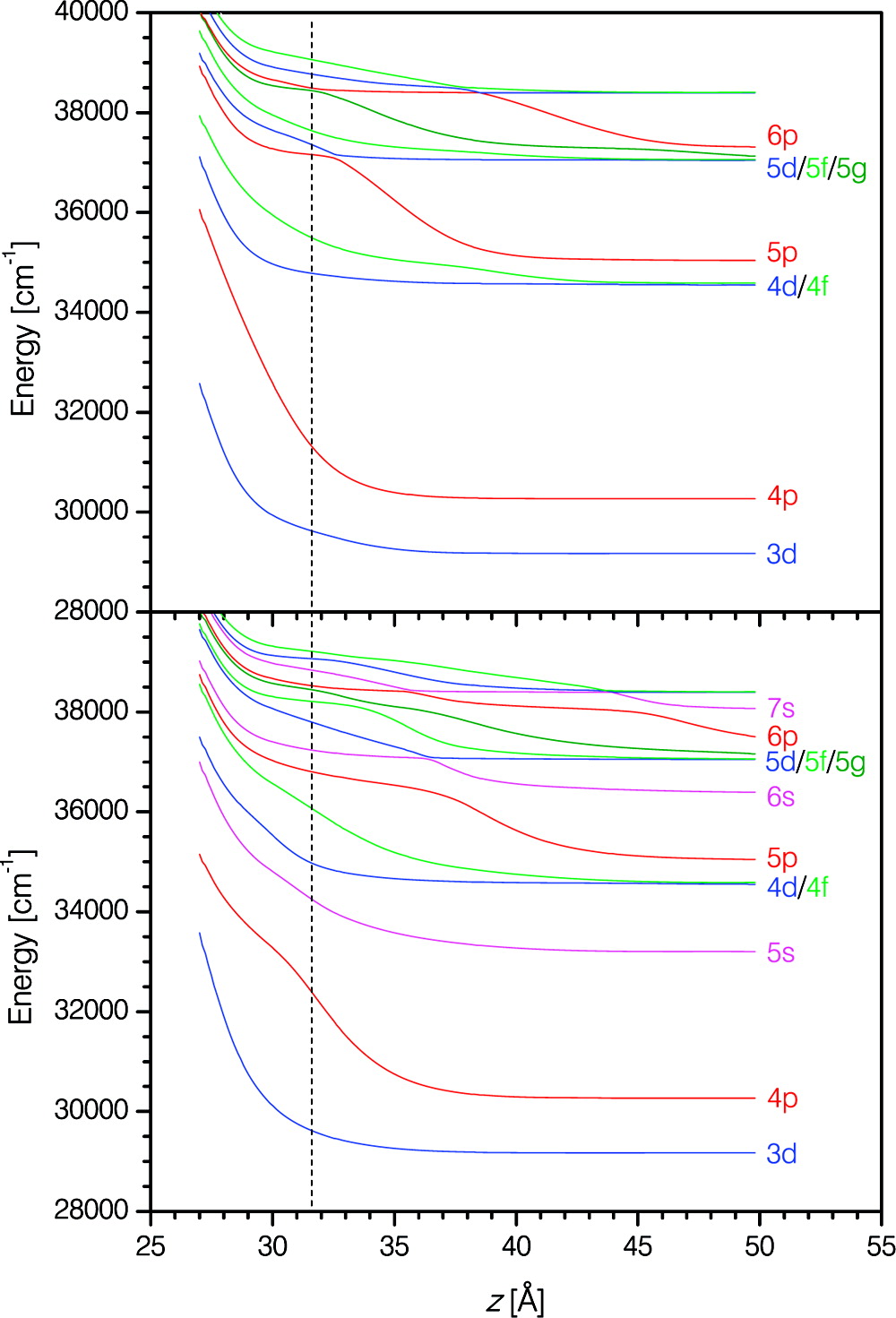}}
\caption{\label{fig8}
Effective interaction potential curves obtained by the PCI method for a Na atom attached to a $^4$He$_{2000}$  droplet. 
Upper panel: $\Pi$ symmetry states. Lower panel: $\Sigma$ symmetry states. The dotted line gives the average distance between the sodium atom
 in the ground state and the center of the helium droplet.\cite{Log11a}
} 
\end{center}
\end{figure}

\begin{figure}[t]
\vspace{42pt}
\begin{center}
\resizebox*{8cm}{!}{\includegraphics{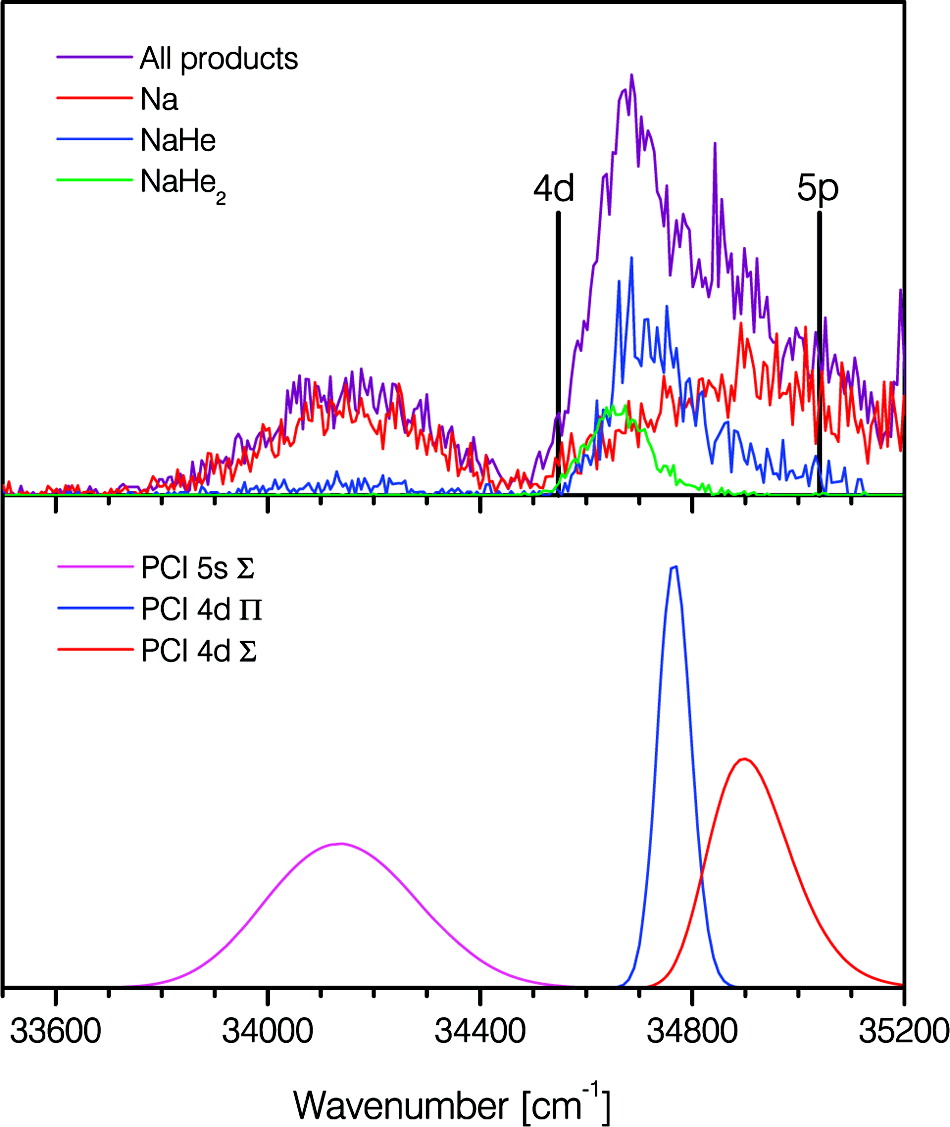}}
\caption{\label{fig9}
Upper panel: Experimental excitation spectra of sodium 
attached to helium droplets recorded by monitoring the 
yield of Na$^+$, NaHe$^+$, and NaHe$_2^+$ ions (arbitrary units). 
The positions of the free atom transitions are indicated by 
vertical lines. Lower panel: Theoretical spectra obtained by DFT employing PCI potentials.\cite{Log11a}
} 
\end{center}
\end{figure}

As an example, Fig. \ref{fig8} shows the PCI potential energy curves for Na@He$_{2000}$. 
Based on these potential energy curves, the electronic excitation spectra of surface bound Na can be calculated and compared directly with experiments. 
This is demonstrated in Fig. \ref{fig9} for one-photon excitation spectra of surface bound Na, which were obtained by monitoring Na$^+$, NaHe$^+$, and NaHe$_2^+$ ion masses. \cite{Log11a} 
The level of agreement obtained is excellent when PCI potentials are employed. 
In contrast, simulations based on pairwise additive potentials (not shown) considerably overestimate the helium induced spectral shift. 
This difference can be attributed to helium-induced mixing of the electron configurations.\cite{Log11a}

Spectroscopy of alkali metal atoms located on the surface of helium droplets has provided a wealth of detailed information on these systems.\cite{Sti06,Cal11a,Mud14} 
In addition to fluorescence excitation and emission spectra, angular distributions of the ejected atoms have been measured .\cite{Log11a,Her12a,Fec12} 
Considering the impurity-droplet system as a pseudo-diatomic molecule, these experiments can clearly distinguish between the $\Sigma$ and $\Pi$ states of the system. 
Alignment of the electronic angular momentum $\mathbf{j}$ for Na$^*$(3p\,${}^2$P$_{3/2}$) obtained by photoejection from $^4$He$_{200}$ droplets was modelled in Ref. \onlinecite{Her13}. 
Together with the angular distribution parameter $\beta$, the coefficient for alignment of $\mathbf{j}$ was obtained from the simulation of the fragment state-resolved photoabsorption spectrum. 
The alignment coefficient exhibits clear oscillations as a function of the excitation energy.
These oscillations were attributed to coherent population of the dissociative $\Sigma$ and $\Pi$ states within the Franck-Condon region. 
They could be observed experimentally through fluorescence polarization, provided that their dependence on the droplet size is not very strong as this could wash them out by averaging. 
They could also be visible in the photoelectron yields following ionisation of the atomic fragment. 
These predictions have not  yet been confirmed by experiments.

The helium degrees of freedom are often involved in the relaxation of photoexcited impurities. 
As discussed before, optical excitation of Na $3$p $\leftarrow 3$s transition populates the pseudo diatomic $\Sigma$ and $\Pi$ states on the droplet surface. 
It was shown that  $\Pi$ state excitation produces both bare Na and NaHe$_n$ exciplexes.\cite{Log15}
Based on their measured velocity distributions, the bare Na atoms appear to be
produced by an impulsive mechanism whereas exciplex production is thermally driven.
The $\Sigma$ state is very repulsive and leads to impulsive desorption of bare Na.\cite{Log15} 
Based on the spin-adiabatic approximation, these bare atoms should only be produced in the $^2$P$_{3/2}$ state. 
However, the experiment measured population in both the $^2$P$_{3/2}$ and the $^2$P$_{1/2}$ states. 
This has been attributed to a curve crossing taking place between the pseudo-diatomic states at long range.\cite{Log15} 
Similar curve crossings have also been reported for other alkali metal-rare gas systems.\cite{Ger16}

\subsection{Alkaline earth metal doped helium droplets:  solvation and absorption spectra}
\label{5.2}

Helium droplets doped with alkaline earth metals have been experimentally  studied and modeled by DFT. Due to their larger binding towards helium 
as compared to heliophobic alkali metal atoms, the resulting solvation structure depends on the species considered and on the isotopic composition of the droplet. 
DFT calculations predict that alkaline earth atoms from Mg to Ba reside inside $^3$He droplets; Ca, Sr, and Ba occupy dimple states on the $^4$He droplet surface, 
and Mg is heliophilic.\cite{Her07}
The DFT results are consistent with the available spectroscopic data for $^4$He droplets, see Ref. \onlinecite{Her07} and references therein. 
No data is available for Be but it is presumably heliophilic. 

\begin{figure}[t]
\vspace{42pt}
\begin{center}
\resizebox*{8cm}{!}{\includegraphics{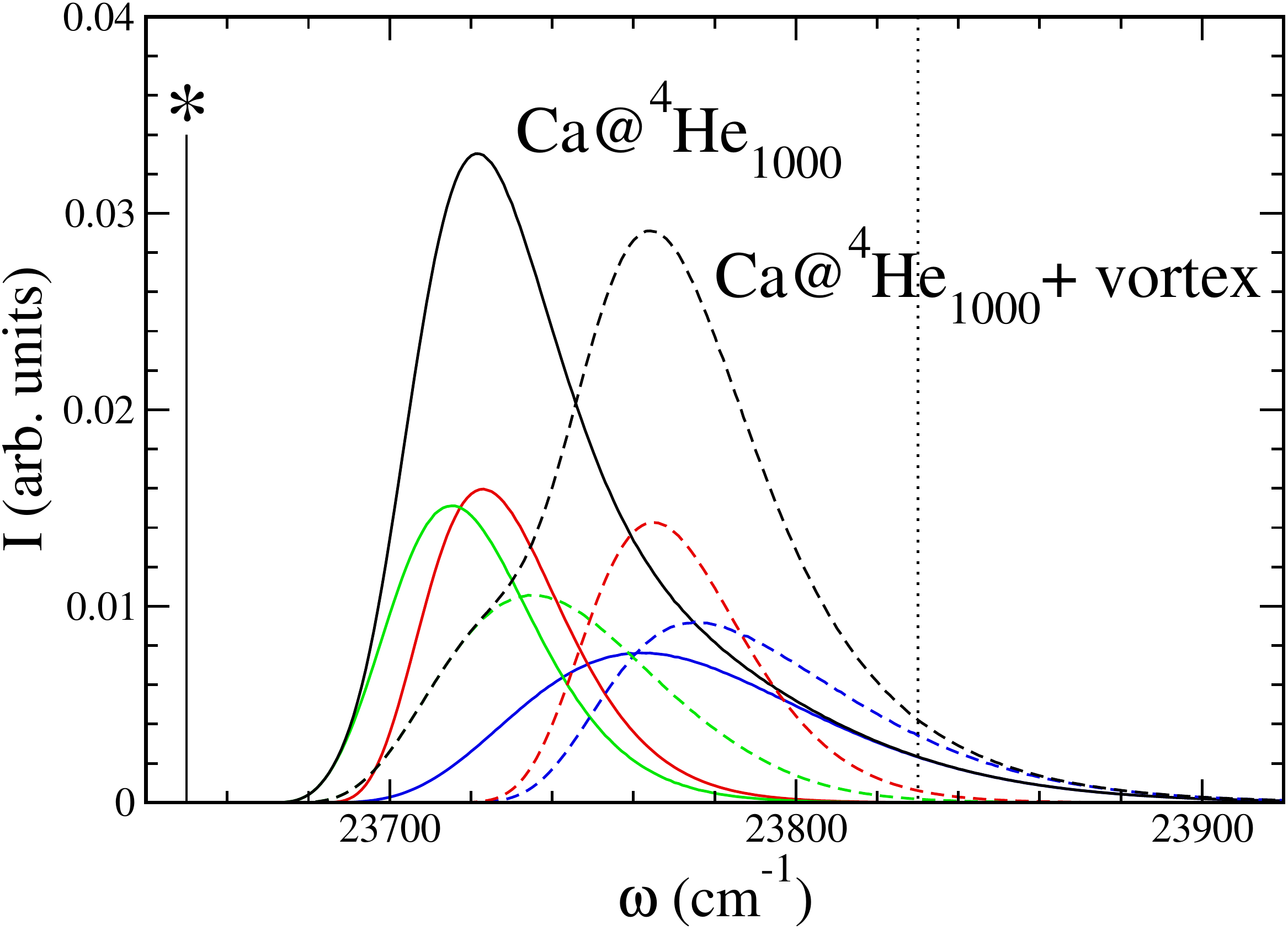}}
\caption{\label{fig10}
Absorption spectrum of Ca in the $^4$He$_{1000}$ droplet with (dashed lines) and without  (solid lines) a vortex line along the symmetry axis.\cite{Her08a} 
The absorption spectrum (black) is split into the $\Sigma$ (blue) and $\Pi$ components (green and red). The starred vertical line represents the gas-phase 
line position and the dotted vertical line represents the experimental value for bulk liquid $^4$He.\cite{Mor05}
} 
\end{center}
\end{figure}

Large helium droplets made up of a few thousand atoms may host vortex lines that are created during the gas condensation phase. Since trapping of impurities 
at vortex lines alters the surrounding liquid density distribution, it has been proposed that absorption spectroscopy of alkali metal atoms, excited state helium 
atoms, or electrons could be used to detect vorticity.\cite{Clo98,Mat10b,Mat15a} 
Unfortunately, the spectral changes are predicted to be very small.
For this reason,  the Ca atom may be a better candidate because it is just barely localised on the droplet surface.\cite{Sti97,Her08a} In the presence of a vortex line, 
Ca atoms could be drawn into the vortex core and sink inside the droplet.\cite{Anc03} Such a change in the solvation environment should produce a more pronounced effect 
in the absorption spectrum. Indeed, DFT calculations confirm this idea and the predicted changes in the absorption spectra are shown in Fig. \ref{fig10}.\cite{Her08a} 
However, the experimental absorption spectrum\cite{Sti97} does not exhibit any structure that could be attributed to the presence of vortices to which  Ca atoms are attached. 
It was concluded that the proportion of droplets with vortex lines in the beam is probably too small to produce a noticeable effect in the spectrum. 
Note that vortex arrays in large helium droplets have been observed experimentally\cite{Gom14,Jon16,Ber17} and modelled by DFT.\cite{Anc15} This will be discussed in more detail in Sec. \ref{5.13}.

\begin{figure}[t]
\vspace{42pt}
\begin{center}
\resizebox*{15cm}{!}{\includegraphics{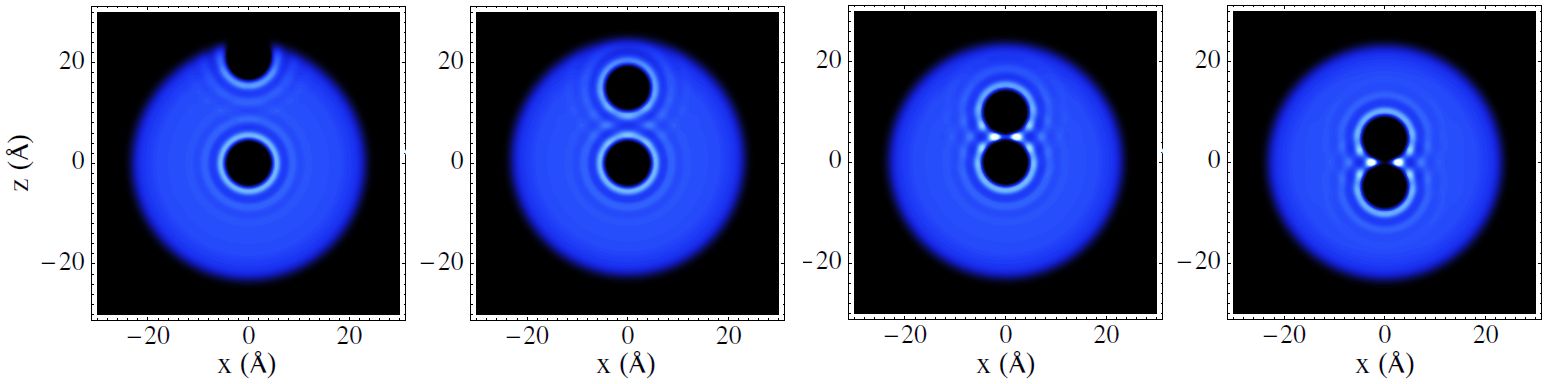}}
\caption{\label{fig11}
From left to right, (Mg+Mg)@$^4$He$_{1000}$ metastable configurations for Mg--Mg interatomic distances 
18.5 \AA{}, 12.9 \AA{}, 9.3 \AA{}, and 9.5 \AA{}. The corresponding total energies are $-$5567.8 K, $-$5573.9 K, $-$5580.3 K and $-$5581.4 K, respectively.
The bright regions correspond to high density helium.\cite{Her08b} 
} 
\end{center}
\end{figure}

Calculations of Mg atom-doped $^4$He droplets have revealed an interesting solvation behavior as a function of the droplet size, indicating that Mg atoms are highly delocalised in the $^4$He 
droplets. Indeed, DMC calculations for small droplets up to $N_4=50$ predict that Mg  is not fully solvated below $N_4\sim 30$.\cite{Mel05,Nav12} 
Recent PIMC calculations have found  that Mg atoms are solvated in  $^4$He$_{100}$  droplets.\cite{Kro16} thus confirming that the droplets must have at least several tens of helium atoms 
to fully solvate the Mg atom.
DFT calculations for small and large  helium droplets\cite{Her07,Her08b} are in agreement  with the QMC findings.  
 These results are also consistent with
  the analysis of Laser Induced Fluorescence (LIF)\cite{Reh00} and Resonant Two-Photon-Ionisation (R2PI) experiments.\cite{Prz08} 
  For very large droplets, 
  $N_4 \sim 10^4$, electron-impact ionisation measurements suggest that Mg atoms are located on the surface,\cite{Ren07} which is in clear disagreement with the previously 
  mentioned LIF and R2PI experiments and with the calculations.   
The origin of this discrepancy has not been identified yet.

When $^4$He droplets are doped with more than one impurity, their free motion and strong mutual attraction  are expected to lead to efficient clustering inside the droplet. 
The formation and properties of metal clusters isolated in helium droplets has been reviewed in~ Ref. \onlinecite{Tig07}. 
If the long-range part of the impurity-impurity interaction becomes comparable to that of the impurity-helium interaction, a dilute loosely bound `bubble foam' structure
 (also called `quantum gel') may form Ref. \onlinecite{Prz08}. Such a foam consists of separated impurities trapped in their own solvation bubbles within the droplet. 
 A similar scenario was put forward to explain experimental 
findings related to the successive capture of impurities in helium droplets\cite{Lew95} and in the bulk liquid.\cite{Gor04} 

The first DFT calculation to model the formation of bubble foam in bulk superfluid helium was carried out for Ne atoms.\cite{Elo08} 
The Ne-He interaction is strong enough to produce a localised solvent shell structure around Ne. 
The calculated interaction energy as a function of the distance between two Ne atoms, including the liquid contribution, exhibits local maxima when the solvent shells centered around each atom overlap. 
This creates a liquid induced energy barrier to recombination, which may localise the atoms far away from their gas phase equilibrium positions, provided the barrier is higher than the thermal energy. 
Similar calculations have been published for Ag-Ag, Cu-Cu, Au-Au, and F-F interactions in superfluid helium.\cite{Hau15,Elo11} 
One of the  goals in these studies was to address the timescale for metal-cluster formation in $^4$He droplets by using a mixed DFT-classical molecular dynamics approach.\cite{Hau15}

Motivated by the experimental work on multiply doped Mg droplets,\cite{Prz08} the above mentioned DFT approach was also used to study Mg-Mg recombination 
in $^4$He$_{1000}$ droplets.\cite{Her08b} By carrying out the same calculation for $^3$He droplets where the solvation shells are less pronounced, it was conclusively 
shown that the solvent shell structure around the impurity plays a key role in  the foam formation. As an example, Fig. \ref{fig11} shows several configurations for 
(Mg+Mg)@$^4$He$_{1000}$. Note that a ring of high density helium forms around the diatomic axis (see also Ref. \onlinecite{Elo08,Prz08}).
The extreme right configuration, where the Mg-Mg distance is 9.5 \AA{}, corresponds to the metastable foam configuration. At shorter Mg-Mg distances the energy increases 
and prevents the recombination into the Mg$_2$ dimer.\cite{Her08b} Based on experimental data, this metastable complex collapses into 
a tightly bound cluster in \textit{ca.} 20 ps.\cite{Prz08} 
The response of Mg atoms embedded in $^4$He nanodroplets 
was later studied by femtosecond dual-pulse spectroscopy, which yielded results consistent with the hypothesis of isolated atoms arranged in a foam-like structure.\cite{God13}

\begin{figure}[t]
\vspace{42pt}
\begin{center}
\resizebox*{8cm}{!}{\includegraphics{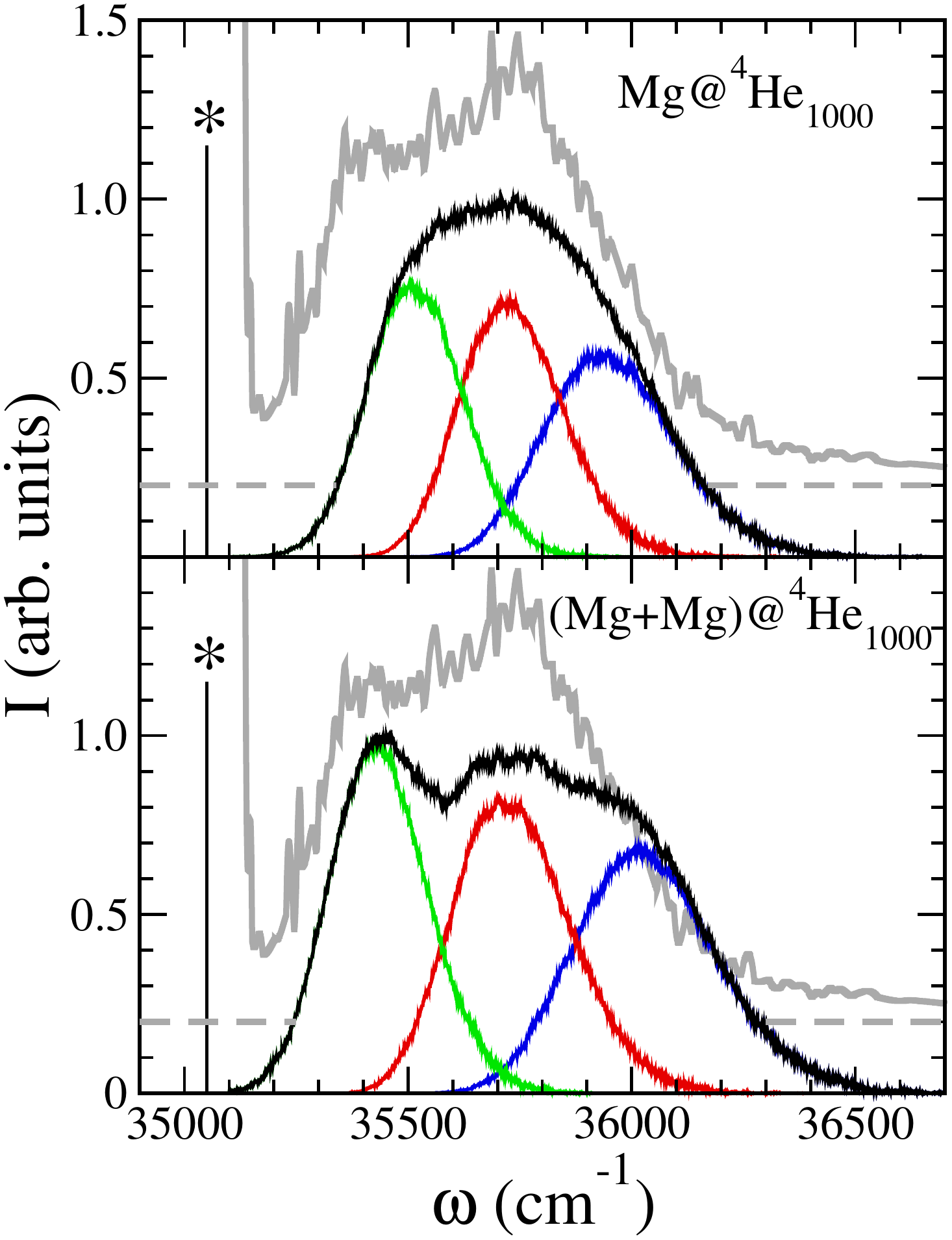}}
\caption{\label{fig12}
Top panel: 3s3p $^1$P$_1$ $\leftarrow$ 3s$^2$\,$^1$S$_0$ absorption spectrum of a single Mg atom in a $^4$He$_{1000}$ droplet.
The starred vertical lines indicate the position of the corresponding gas-phase transition. The experimental curve is shown in grey.\cite{Reh00}
Bottom panel: Same as top panel but the Mg atom resides in a distorted environment created by the presence of another nearby Mg atom.\cite{Her08b,Her09a}
} 
\end{center}
\end{figure}

The effect of the above bubble foam configurations on LIF and R2PI spectra in $^4$He droplets was found to be in   good agreement with the experimental data, as shown in  Fig. \ref{fig12}.\cite{Her08b}
 The experiments show that doping helium nanodroplets with more than one Mg atom leads to a shift of the atomic absorption line from 279 nm to 282 nm due to the additional 
 perturbation produced by the neighboring Mg solvation bubbles.\cite{Her08b} It is worth mentioning that recent QMC calculations on the Mg pair in $^4$He droplets 
 did not yield any barrier for dimer formation;\cite{Kro16}  no alternative interpretation for the R2PI experiments was presented. 

The foam structures correspond to loosely bound clusters. Clusters  may grow  inside helium droplets 
with different structures, depending on the size of both the droplet and the cluster itself. The formation of Ag clusters up to a few thousand atoms in He droplets was studied via optical laser 
spectroscopy.\cite{Log11c} It was found that small Ag clusters ($N_{\rm Ag} \sim 100$) exhibited a plasmon resonance at about 3.7 eV, similar to that previously obtained 
for dense spherical clusters. However, larger Ag 
 clusters ($N_{\rm Ag} > 1000$) formed in $^4$He$_{N_4}$  droplets, $N_4 \sim 10^7$, exhibited an unusually broad spectrum extending into the infrared spectral range. The dramatic change in the spectrum has been associated with a transition from single-cluster to multi-centre growth regime when the droplet size increases. The structure of the cluster aggregates formed inside He droplets remains unknown; it is conceivable that they are loosely packed and may even exhibit a fractal-like structure.

\subsection{Droplets doped with more than one species}
\label{5.3}

Helium droplets doped with two different impurity species with opposite solvation behavior have been investigated by  DFT.
Such studies have been inspired by  experimental work showing that  an otherwise heliophobic Ba atom could be solvated in helium droplets  which already contain
a heliophilic xenon cluster in their center.\cite{Lug00}

In a joint experimental and DFT work,\cite{Dou10}  droplets doped with HCN-M  (M$=$Na, Ca, and Sr) have been studied.  The calculations for these systems show 
a strong surface-bound state for Na, a purely solvated state for Ca, and both surface and solvated 
states separated by a barrier for Sr. The results for Ca and Sr were consistent with the appearance of the infrared spectrum for these complexes.

\begin{figure}[t]
\vspace{42pt}
\begin{center}
\resizebox*{8cm}{!}{\includegraphics{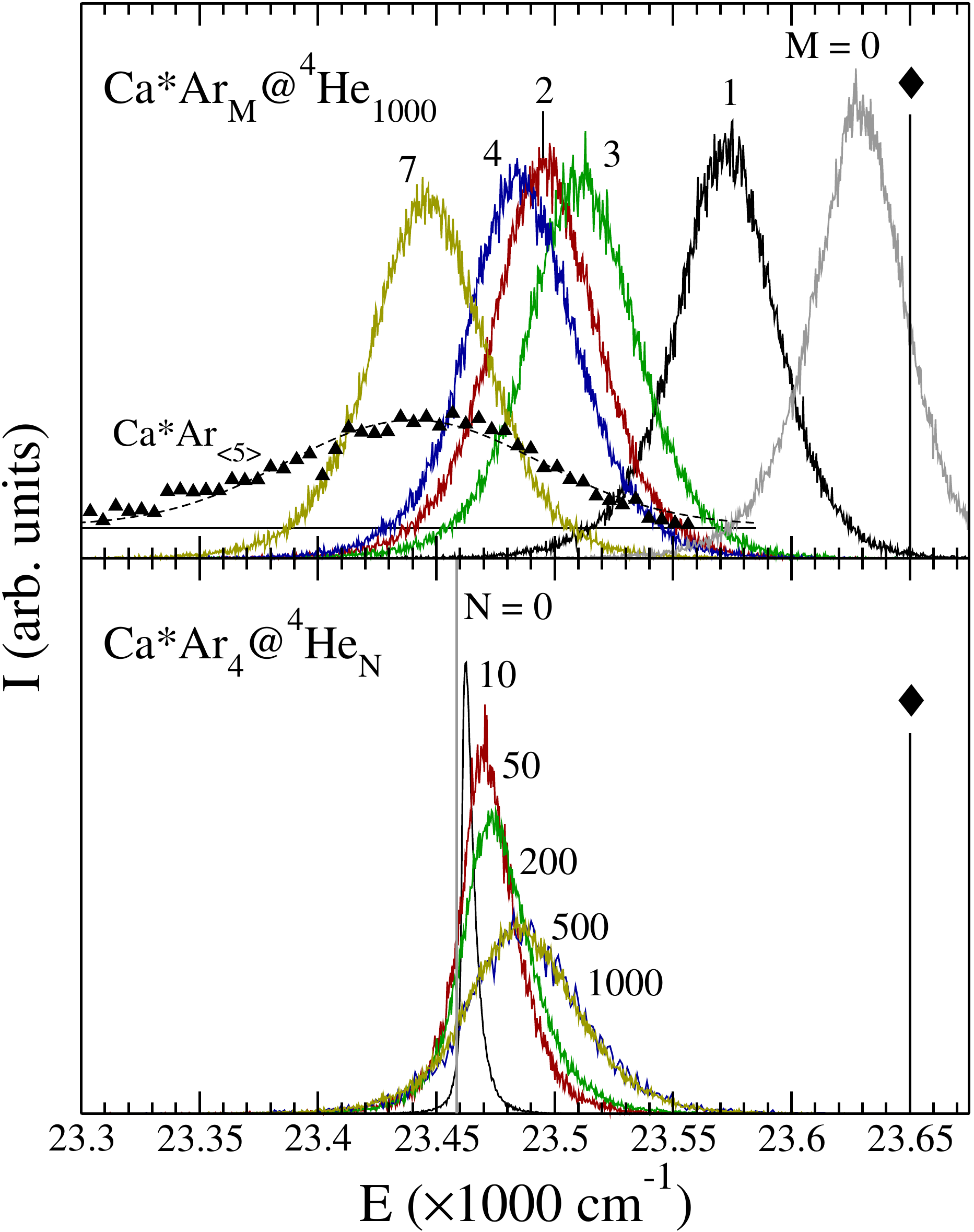}}
\caption{\label{fig13}
Overview of Ca$^*$-Ar$_M$He$_N$ $^1$P$\rightarrow ^1$S fluorescence emission spectra obtained by DFT. The upper panel shows the emission
 spectra for varying number of Ar atoms $M$=1, 2, 3, 4, and 7 in  a $^4$He$_{1000}$ droplet. The bottom panel shows the dependence of the Ca$^*$-Ar$_4$ 
 spectrum on the number of helium atoms in the droplet ($N$). The grey, vertical line labelled $N=0$ indicates the position of the isolated Ca$^*$Ar$_4$ vertical emission. 
 The line labelled with a diamond is the atomic Ca$^*$ emission line. The triangles in the upper plot are experimental data for $\langle M\rangle$=5 
 (hence a Poisson distribution of sizes with a weight of e.g. 0.03 for $M=1$, 0.08 for $M=2$, 0.14 for $M=3$, 0.18 for $M=4$ and $5$, 0.15 for $M=6$, 0.10 for $M=7$). 
 The lower plot shows that, if the emitting species is inside a smaller droplet, the calculated spectra would shift by about 20 cm$^{-1}$ to the red, giving a better agreement with experiment.
 \cite{Mas12,Her12b}
} 
\end{center}
\end{figure}

In another joint experimental and theoretical project,\cite{Mas12,Her12b} the influence of heliophilic argon doping on the solvation of heliophobic calcium atoms in helium droplets has been studied.
The experiment considered the photodissociation of  Ca$_2$ to Ca(4s4p $^1$P) + Ca(4s$^2$ ${}^1$S) in the presence of a varying number of Ar atoms in the droplet. 
The absorption and emission spectra of Ca-Ar$_M$ ($M = 0-7$) complexes were calculated by using the DF  sampling method described in Sec. \ref{4.3} with the Ca and Ar atoms treated classically. 
It was found that even a single Ar atom is enough to trigger Ca atom sinking into the He droplet, where they form a  Ca$_2$ dimer. 
Furthermore, by studying the emission spectrum as a function of the droplet size (Fig.~\ref{fig13}), it was concluded that the emitting species was 
Ca$^*$Ar$_M$ attached to the  droplet that has shrunk down to a size less than 200 helium atoms by either evaporation or detachment of helium atoms from the complex.

In another DFT study,\cite{Pom12} a heliophilic Xe atom was placed in the bulk  of a $^4$He$_{500}$ droplet with a heliophobic Rb atom located on the surface. 
The Rb-Xe van der Waals attraction was not sufficiently high to overcome the 23.4 K barrier induced by the presence of helium between the dopants and therefore Rb remained 
on the surface. Clearly, this is a droplet-size dependent effect. Furthermore, it was concluded that the order in which the dopants are introduced to the droplet plays 
an important role in the formation of such dimers, as they can only form on the droplet surface. In a recent study,\cite{Ren16} evidence has emerged that sodium and cesium 
clusters, and even single Na atoms (but not Cs), can enter $^4$He droplets (average size $N_4 \sim 5 \times 10^5$) in the presence of a fully solvated C$_{60}$ fullerene.

\subsection{Cluster-doped helium droplets}
\label{5.4}

Despite their practical and conceptual importance,\cite{Tig07}  theoretical studies simulating atomic clusters embedded in helium droplets are scarce. 
A major difficulty in these studies is to obtain reliable cluster-droplet interaction potentials. Besides the work on Ca-Ar$_M$ clusters mentioned before,\cite{Mas12,Her12b} 
the interaction of two Ne clusters in liquid $^4$He has been studied in Ref. \onlinecite{Elo08}. Furthermore, Path-Integral MC (PIMC) calculations of Mg and Na clusters 
in both helium droplets and the bulk liquid have been carried out.\cite{Hol15} These calculations show that Mg clusters are heliophilic whereas small Na$_M$ 
clusters with $M=7,9$ remain on the surface. Recall that a single Mg atom is also heliophilic when $N_4 \lesssim 30$.

Alkali atom clusters are especially interesting because the individual atoms reside on the droplet surface whereas larger clusters may become heliophilic and sink inside the droplet. 
The critical cluster size $N_c$ for switching from heliophobic to heliophilic behavior  has been  determined  from the  energy balance 
between the metal-helium van der Waals attraction, the short-range repulsion, and the liquid surface tension.\cite{Sta10} The following values have been 
predicted for $N_c$: Li,Na/$^4$He $\sim$20; Rb/$^4$He $\sim$100; Li,Na/$^3$He $\sim$5; and Rb/$^3$He $\sim$20 The values of $N_c$ in $^3$He
 are smaller than in $^4$He because of the lower value of the surface tension and saturation density. 
 The prediction for Na in $^4$He droplets was later confirmed by the experiments;\cite{Lan11}
 a recent study  on the submersion of Na clusters 
 in $^4$He and para-H$_2$ clusters employing path-integral molecular dynamics has also found the submersion of Na$_N$ clusters in $^4$He droplets around $N\sim 20$.\cite{Cal17}

Superfluidity of the helium surrounding Mg$_{11}$ clusters in $^4$He droplets consisting of up to a few hundred helium atoms has been studied by 
QMC.\cite{Hol14} Furthermore, the commensurate-incommensurate transition of the $^4$He atoms adsorbed on the surface of C$_{20}$ and C$_{60}$ 
was characterised by PIMC.\cite{Kwo10,Shi12}

In a more approximate way, the dissociation dynamics of neon clusters upon ionisation has been studied in a $^4$He$_{100}$ droplet using 
molecular dynamics corrected for delocalisation of the helium atoms and DIM based interaction potentials.\cite{Bon07,Bon08}
The results showed two interesting processes, one in which the ionic core of the cluster, usually Ne$_2^+$, is expelled from
 the rest of the droplet, and another showing a very efficient cooling effect 
by helium atom ejection rather than evaporation, with a wide kinetic energy distribution.

Sequential doping of helium droplets allows for the synthesis of core-shell clusters (`nanomatryoshkas'). Bimetallic clusters have been formed via sequential 
pickup of gold and silver atoms by helium droplets.\cite{Moz07} The resulting structure persists upon `soft-landing' of the clusters on a solid surface. Another 
nanomatryoshka, made of an Ag core coated by a shell of ethane molecules, has been studied.\cite{Log13} 
These systems are currently beyond the reach of a DFT-based description.

The DFT approach has also been used to simulate  the solvation of single-walled carbon nanotubes consisting of up to 360 carbon atoms in a $^4$He$_{2000}$ 
droplet using  an {\it ab initio} He-nanotube interaction potential,\cite{Hau16} see also Ref. \onlinecite{Hau17}. Depending on the nanotube diameter, the outer and inner walls are covered 
by one or more dense 
layers of helium. This structure, which was also found earlier on He-wetted graphite,\cite{Cle93} forms 
as a consequence of the strong surface-He interaction and geometric effects.\cite{Hern11} 

\subsection{Doped mixed $^3$He-$^4$He and $^3$He droplets}
\label{5.5}

Experiments employing mixed helium droplets have been integral to the discovery of $^4$He 
droplet superfluidity by rotational spectroscopy.\cite{Gre98} At low temperatures the two isotopes separate such that the inner part of the droplet consists of $^4$He 
whereas $^3$He resides on the outside.\cite{Bar06} Depending on the strength of the impurity-He interaction, the impurity may reside on the droplet surface, at
the $^3$He-$^4$He interface, or fully solvated inside the $^4$He core.\cite{Mat11b,Bue09}

\begin{figure}[t]
\vspace{42pt}
\begin{center}
\resizebox*{13cm}{!}{\includegraphics{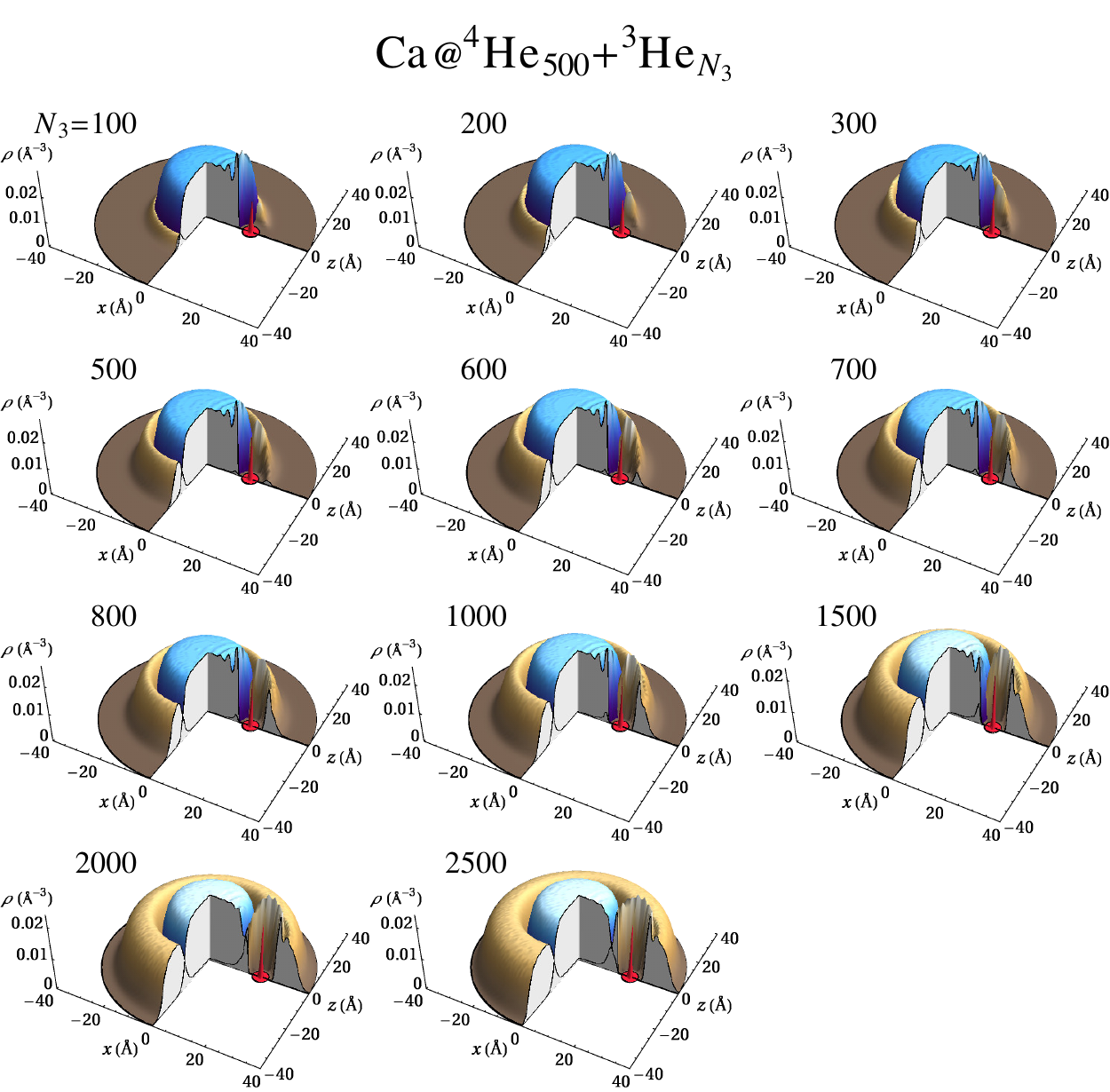}}
\caption{\label{fig14}
A three-dimensional view of Ca@$^4$He$_{500}$+$^3$He$_{N_3}$ helium droplets with $N_3$ varying from 100 to 2500. The $^4$He core and $^3$He shell are pictured in blue and in brown, respectively. The probability density of the Ca atom is also displayed along the density cut (red spot; specified in arbitrary units).}
\end{center}
\end{figure}

The structure and energetics of small mixed He droplets doped with Mg and Ca impurities has been studied by the quantum Diffusion Monte Carlo (DMC) 
method with the aim of determining their solvation behavior in pure $^4$He and $^3$He droplets.\cite{Nav12,Gua09} Since Ca is heliophilic in $^3$He droplets 
but heliophobic in $^4$He, it was expected to reach the $^4$He core surface while remaining inside the $^3$He shell. This is indeed confirmed by DFT calculations
 as illustrated in Fig. \ref{fig14} for Ca@$^4$He$_{500}$+$^3$He$_{N_3}$ droplets.

\begin{figure}[t]
\vspace{42pt}
\begin{center}
\resizebox*{10cm}{!}{\includegraphics{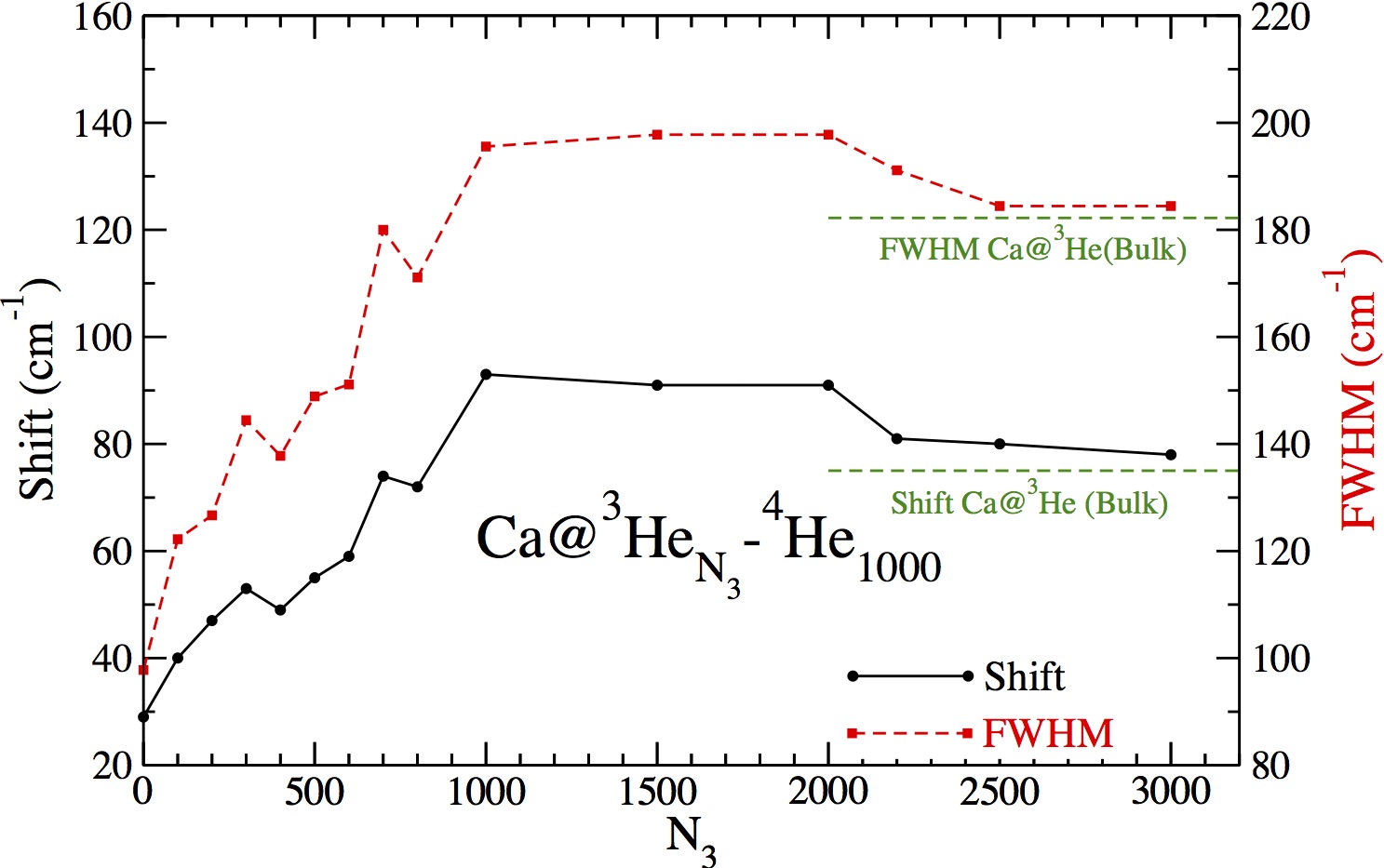}}
\caption{\label{fig15}
Calculated shift and full width at half maximum (FWHM) of the Ca absorption spectrum around the 4s4p$\leftarrow$4s$^2$ transition as a function of the number of $^3$He 
atoms $N_3$  for $N_4$ = 1000.\cite{Bue09}
 }
\end{center}
\end{figure}

The interfacial location of Ca has also been verified by independent QMC calculations and absorption spectroscopy experiments.\cite{Gua09,Bue09} 
Figure \ref{fig15} shows the calculated spectral shift and full width at half maximum (FWHM) as a function of the number of $^3$He atoms 
$N_3$ for $N_4$ = 1000. Direct comparison with experimental data is difficult because the $^3$He-$^4$He composition of the gas used may not directly carry over to the droplets.\cite{Bue09}

\begin{figure}[t]
\vspace{42pt}
\begin{center}
\resizebox*{10cm}{!}{\includegraphics{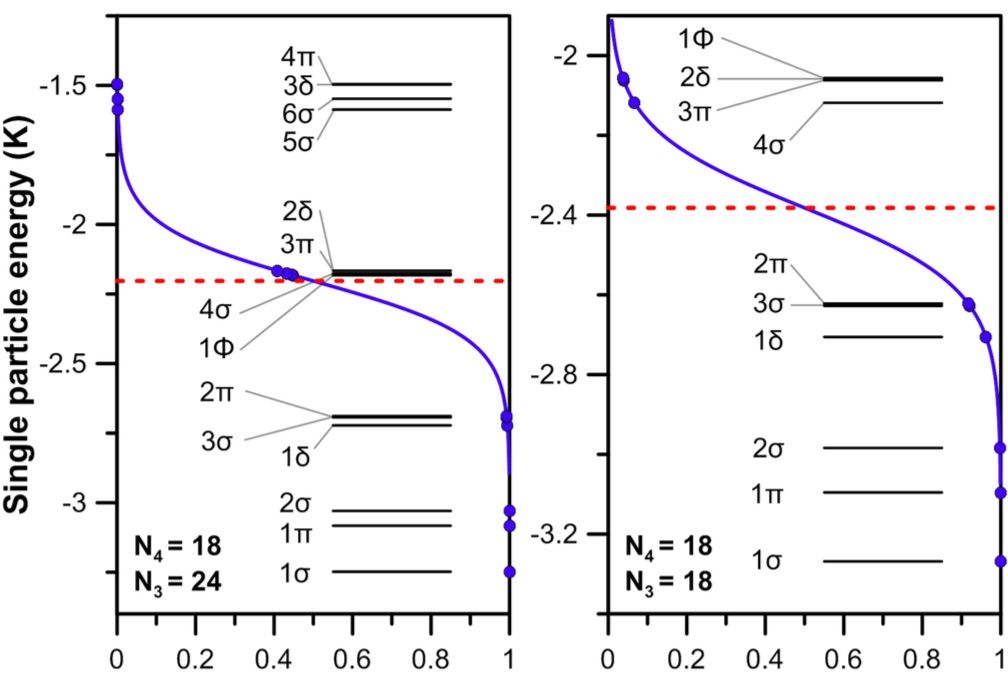}}
\caption{\label{fig16}
Single-particle energies $\epsilon_i$ (vertical scale, K) of the $^3$He component of an OCS@$^3$He$_{N_3}$+$^4$He$_{18}$ droplet for 
 two selected values of $N_3$.\cite{Lea13} The dashed line represents the $^3$He chemical potential $\mu_3$. The occupation numbers are represented by 
 dots (horizontal scale) and the line connecting them represents the function $n_i =1/\{1+exp[(\epsilon_i-\mu_3)/k_BT]\}$.
}
\end{center}
\end{figure}

Interatomic Coulombic decay (ICD)\cite{Ced97} has been proposed as a tool 
for studying the interface of isotopically mixed helium 
droplets doped with Ca atoms,\cite{Kry13} since ICD is highly sensitive to the solvation environment. 
In a previous ICD study, isotopically pure $^3$He and $^4$He droplets doped with Ne and Ca were studied.\cite{Kry07} The aim was to provide 
observables that would be sensitive to helium density around the impurity atom and compare them with DFT results.  
The first experimental study of ICD in $^4$He nanodroplets, induced by photoexcitation of the $n=2$ excited state of $^4$He$^+$, 
has been carried out  recently.\cite{Shc17} It was found that the $^4$He$^+$ kinetic energy distribution was strongly affected by the droplet environment,
 depending on whether ICD occurred inside the droplet or within the droplet surface region.

In DFT calculations of large mixed helium droplets,  the kinetic energy of the $^3$He component  is treated  using the Thomas-Fermi-Weizs\"acker approximation, 
where the kinetic energy density  is written as a sum of two terms, one  proportional to $\rho_3^{5/3}(\mathbf{r})$ and  the other proportional to  $(\nabla \rho_3)^2/\rho_3$.\cite{Gui93} 
However, this is only justified for large $^3$He droplets. For small droplets, the Kohn-Sham (KS) orbitals must be employed, which introduces an additional 
complication as the systems of interest are not spherical. 
DFT-KS studies have  been published on small mixed helium droplets doped with Ca.\cite{Mat09b} The single $^3$He atom excitation spectrum 
in $^4$He$_{N_4}$ droplets 
with $N_4$= 8, 20, 40, and 50 has been obtained and compared with DMC results,\cite{Mat09a} and  the effect on the $^3$He excitation 
spectrum of doping the $^4$He$_{50}$ droplet with Ca  was discussed.

\begin{figure}[t]
\vspace{42pt}
\begin{center}
\resizebox*{14cm}{!}{\includegraphics{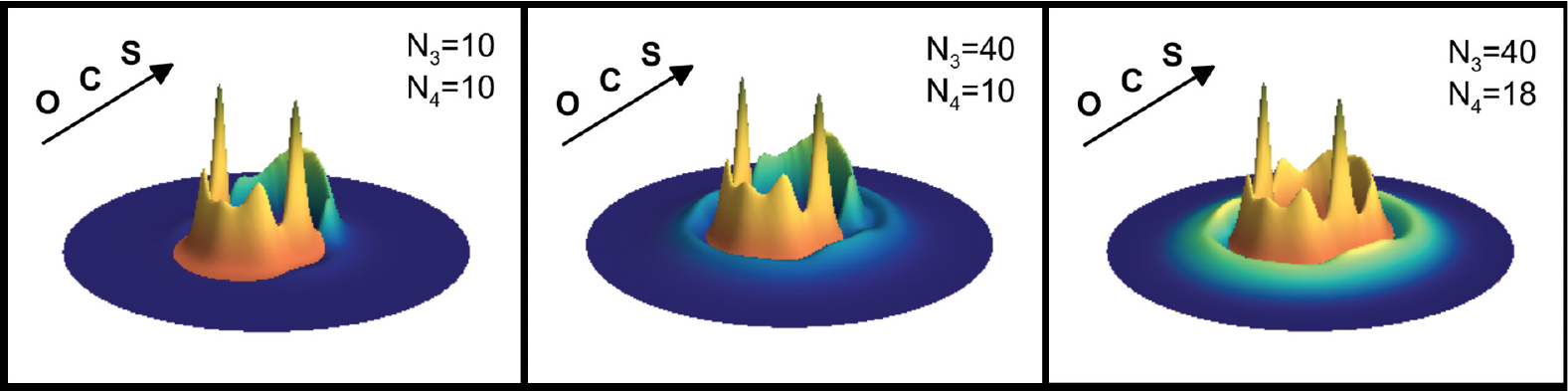}}
\caption{\label{fig17}
Three-dimensional visualization of the atomic densities around OCS for mixed $^3$He-$^4$He droplets.\cite{Lea13}
The $N_3$ and $N_4$ values are indicated.
 Blue/green gradient represents the density of $^3$He and
orange/yellow gradient the density of $^4$He. The OCS molecule sits at the center and its orientation is indicated schematically.}
\end{center}
\end{figure}

Despite the conceptual relevance of  addressing an OCS molecule embedded inside mixed helium droplets, DFT-KS calculations 
 for the structure of small OCS@$^3$He$_{N_3}$+$^4$He$_{N_4}$
  systems, where the OCS molecule  was treated as an external field, have only appeared 
recently.\cite{Lea13}  
One interesting aspect of this work is that $^4$He has been described at $T=0$ whereas a finite temperature DFT-KS approach was used for $^3$He. 
This can be justified by considering that the elementary excitations 
  of $^4$He droplets are collective and their energies are of the order of several Kelvin, whereas the elementary particle-hole excitations of $^3$He have 
  energies comparable to the temperature of the experiment  ($\sim$ 0.1 K for OCS doped $^3$He droplets).\cite{Sar12} Since mixed droplets cool down 
  by evaporation from the $^3$He free surface, a similar temperature to the previously mentioned particle-hole excitation energy is expected for mixed droplets. 
  Such a small temperature has a negligible effect on the bosonic component of the droplet, but it may influence the fermionic component provided that the level 
  spacing of the single-particle (s.p.) energy levels is of the order of $k_B T$. In this case, a large density of states with fractional occupation $n_i$ 
  is expected around the Fermi level.

Given an ensemble $\{n_i\}$ that fulfills $N_3 = \sum_i n_i$, the standard deviation of $N_3$ is given by
\[\Delta N_3 = \sqrt {\sum_i n_i \, (1- n_i)}  \]
This quantity exhibits pronounced local minima at $^3$He shell closures (`magic numbers') and local maxima at $N_3$ values that correspond to half-filled shells. Notice that $\Delta N_3=0$ when all the occupation numbers are either 0 or 1.

Figure \ref{fig16} shows the s.p. structure of the $^3$He component of an OCS@$^3$He$_{N_3}$+$^4$He$_{18}$ droplet with $N_3 = 18$ and 24. 
Notice that $N_3 = 18$ corresponds to a closed-shell configuration whereas $N_3 = 24$ is a half-filled shell. Accordingly, the level scheme in Fig. \ref{fig16} 
displays a fairly large energy gap around the Fermi level for $N_3 = 18$ and shows both large and small occupations. In contrast, the spectrum for 
$N_3 = 24$ is dense around the Fermi level with several s.p. states partially occupied. Since the energy gap is small in open shell droplets, their proper 
description must include thermal effects. Other examples are discussed in Ref. \onlinecite{Lea13}. 
Helium density distributions around the OCS molecule for three selected configurations are shown in Fig. \ref{fig17}.

Helium-3 droplets  doped with OCS have been investigated within DFT-KS.\cite{Mat13a} This work was motivated by the analysis of experimental infrared spectroscopy data for
 the OCS molecule embedded in $^3$He$_{N_3}$ droplets with $N_3 \sim 1.2 \times 10^4$.\cite{Gre98,Sar12} Before this experiment, only the glyoxal molecule had been studied
  in $^3$He droplets through the excitation of electronic and vibronic transitions. It was observed that the zero phonon lines (ZPL) were accompanied by 
  additional broad bands on their red side due to particle-hole excitations of the droplet.\cite{Poe09} Furthermore, a small sharp peak superimposed on the 
  additional band was assigned to vibrations of the snowball structure that surrounds the molecule. These particle-hole excitations accompanying the ZPLs were further analysed in Ref. \onlinecite{Ben12}.

\begin{figure}[t]
\vspace{42pt}
\begin{center}
\resizebox*{9cm}{!}{\includegraphics{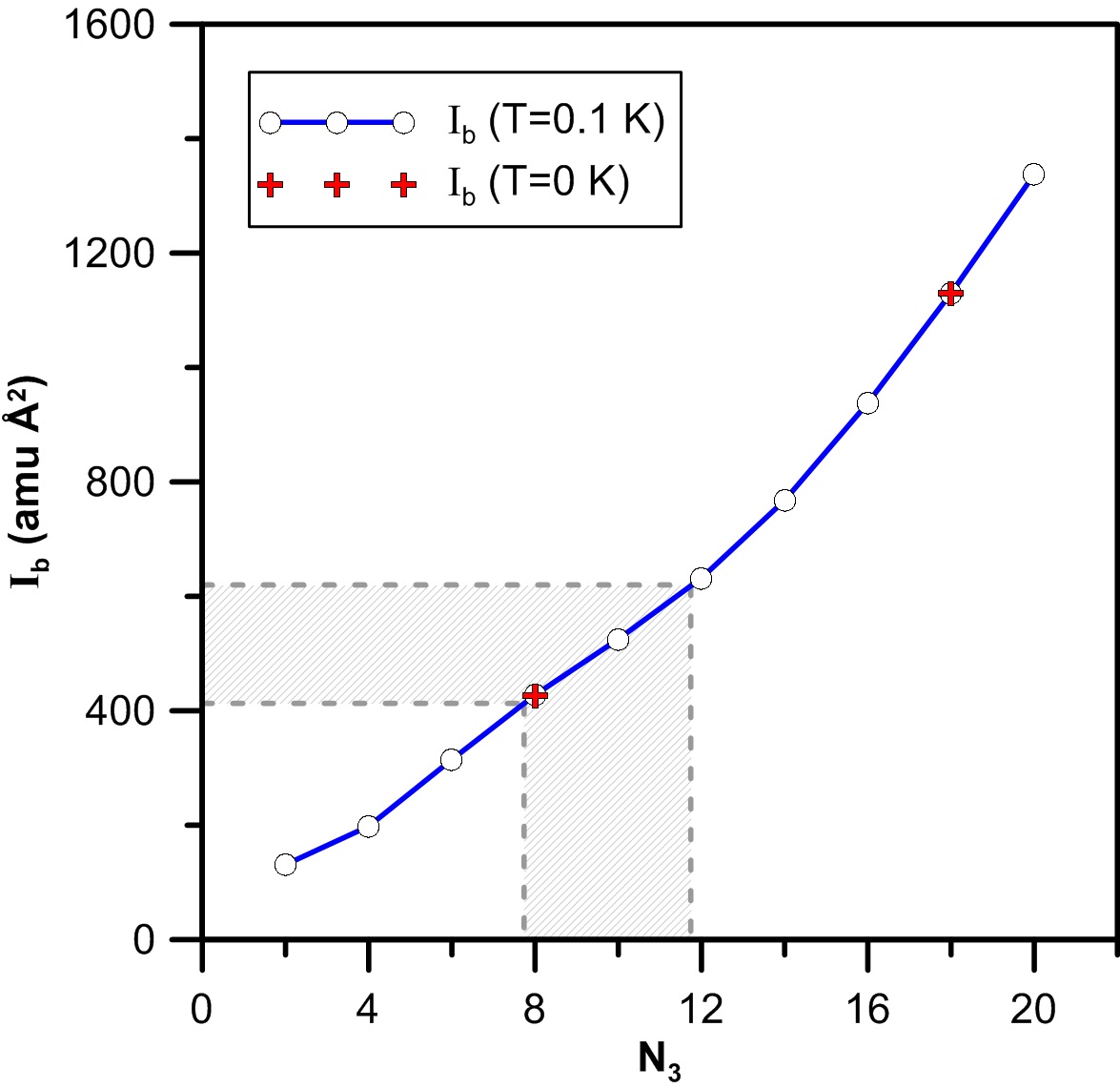}}
\caption{\label{fig18}
Rigid moment of inertia of OCS@$^3$He$_{N_3}$ (in amu \AA$^2$) perpendicular to the symmetry axis as a function of $N_3$.\cite{Lea13} 
The shaded area indicates the experimental values;\cite{Sar12}  the $T=0$ values are from Ref. \onlinecite{Mat13a}.}
\end{center}
\end{figure}

While the Q-branch is missing in the infrared spectrum of OCS in both $^3$He and $^4$He droplets, the effective moment of inertia (MOI) in $^3$He is approximately twice as large as compared to $^4$He. Note that this value is 5.5 times larger than observed in the gas phase. The increase in MOI was attributed to the presence of an evenly distributed shell of 11 $^3$He atoms around the molecule.

The structure and energetics of small OCS@$^3$He$_{N_3}$ droplets was studied in Ref. \onlinecite{Mat13a} by using the functional described in Ref. \onlinecite{Bar97}. 
Similarly to the previously discussed calculations, the OCS molecule was incorporated into the model as an external potential. Since the calculation was carried out at $T=0$, it was restricted to small $N_3$ values that correspond to closed-shell OCS@$^3$He$_{N_3}$ droplets, $N_3 = 8, 18,$ and 40. The $^3$He atoms are expected to fill the waist around OCS between the O and C atoms. The calculated number of atoms present in this ring was about 4 for $^3$He droplets and 5 for $^4$He droplets.

The distribution of He atoms around the OCS molecule affects its rotational properties. The MOI for OCS attached to  $^3$He$_{N_3}$
 droplets with $N_3=8$, 18, and 40 was calculated by the rigid body expression and the results were compared with $^4$He data. The calculated MOI of OCS@$^3$He$_{N_3}$ 
 is shown in Fig. \ref{fig18}.\cite{Lea13} Comparison with $T=0$ results shows that the small non-zero temperature in the calculations does not influence the morphology 
 of the fermionic droplet, but allows to carry out the calculations for any $N_3$ value and not only for the magic numbers.

\subsection{Electrons in liquid helium}
\label{5.6}
 
\begin{figure}[t]
\vspace{42pt}
\begin{center}
\resizebox*{15cm}{!}{\includegraphics{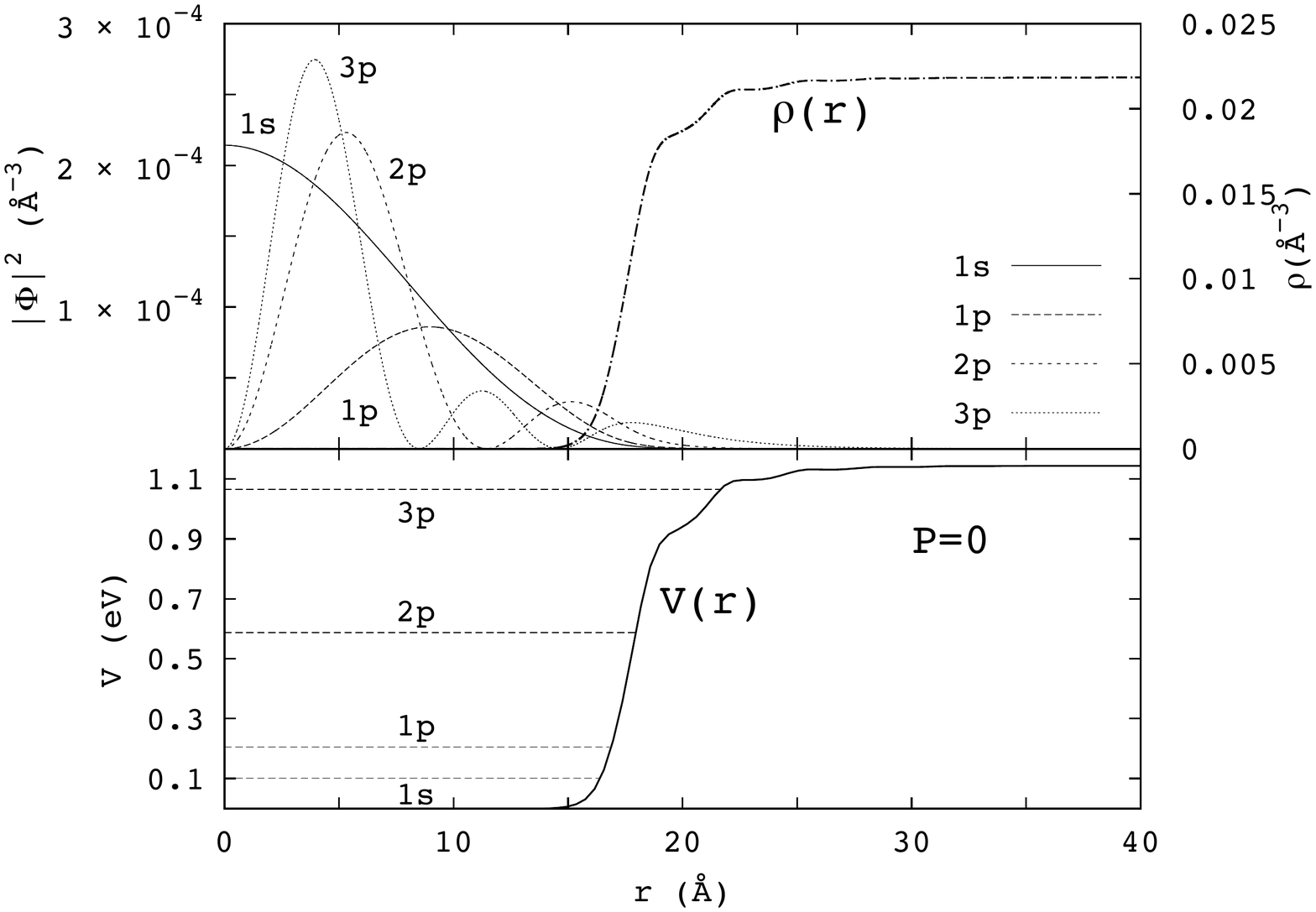}}
\caption{\label{fig19}
Top panel: Electron bubble density profile (right scale) and electron probability density for the 1s, 1p, 2p and 3p states (left scale) at $P=T=0$ in superfluid $^4$He. 
Bottom panel: The corresponding confining potential well and single-electron energies.\cite{Bar13}
}
\end{center}
\end{figure}

DFT has been succesfully used to study excess electrons and homogeneous cavitation in liquid $^4$He and $^3$He. Analysis of phenomena such 
as the crossover from thermal to quantum cavitation in liquid $^4$He, heterogeneous cavitation by excess electrons, and the effect of vorticity on these 
processes have been presented in the literature.\cite{Bar02,Bal02,Mar08} 

Any impurity embedded in liquid helium, including an electron, must loose its excess kinetic energy by ionisation of helium atoms and/or excitation of the liquid excitation modes. 
After the electron has lost most of its kinetic energy and moves at a velocity below the speed of sound in the liquid, it  produces a cavity (`bubble') and becomes localised in it. 
In liquid $^4$He, this cavity -- which is void of He atoms at $T=0$ -- has a radius of $\sim 19$ \AA {} and is the result of  the competition between
the zero-point energy of the confined electron, the surface energy of the bubble, and the work done against the liquid pressure for creating the cavity.

Figure \ref{fig19} shows the helium density profile around the electron at $P=T=0$ along with the electron probability densities for 1s, 1p, 2p, and 3p states, and the confining potential. 
The 3p state is barely bound under SVP as its energy is just below the free electron limit, and it becomes delocalised above 1.7 bar.\cite{Bar13} The energy differences between the np and 1s electron levels correspond to the peak maxima in the electron absorption spectrum.

The properties of electron bubbles (e-bubble) in liquid helium have been reviewed  in Ref. \onlinecite{Mar08,Tab97,Cou09,Mar00,Mar15} with the more recent articles concentrating specifically on the identification of the experimentally observed unusual negative species (`exotic ions'). Furthermore, multi-electron bubbles in liquid helium have also been studied and their properties reviewed
 in Ref. \onlinecite{Tem07}. It was shown  that highly charged multi-electron bubbles are unstable against fission  at positive pressures.\cite{Guo08}
The production of long-lived multi-electron bubbles at negative pressures has been reported recently.\cite{Fan17}
A major difficulty in these studies is that the radius of the bubble must be on the micron-scale so that the electron-electron Coulomb repulsion can be reduced. So far, only two-electron bubbles have been modelled with DFT,\cite{Leh07} and the conclusion was that they are unstable and break into two separate electron bubbles. No real-time simulation of this process has been carried out yet.

While there is no direct way to measure the properties of electron bubbles, absorption measurements can provide relative energetics for the electron level structure. The DFT calculations performed 
in Refs. \onlinecite{Elo02,Gra06} are in excellent agreement with the experimental absorption data of Grimes and Adams.\cite{Gri92} Furthermore, DFT calculations are able to reproduce the experimentally observed negative pressure required to explode electron bubbles in the liquid.\cite{Cla98,Pi05} These comparisons confirm that the combined electron and DFT model can capture the essential physics of the e-bubble state.

\begin{figure}[t]
\vspace{42pt}
\begin{center}
\subfigure{
\resizebox*{7cm}{!}{\includegraphics{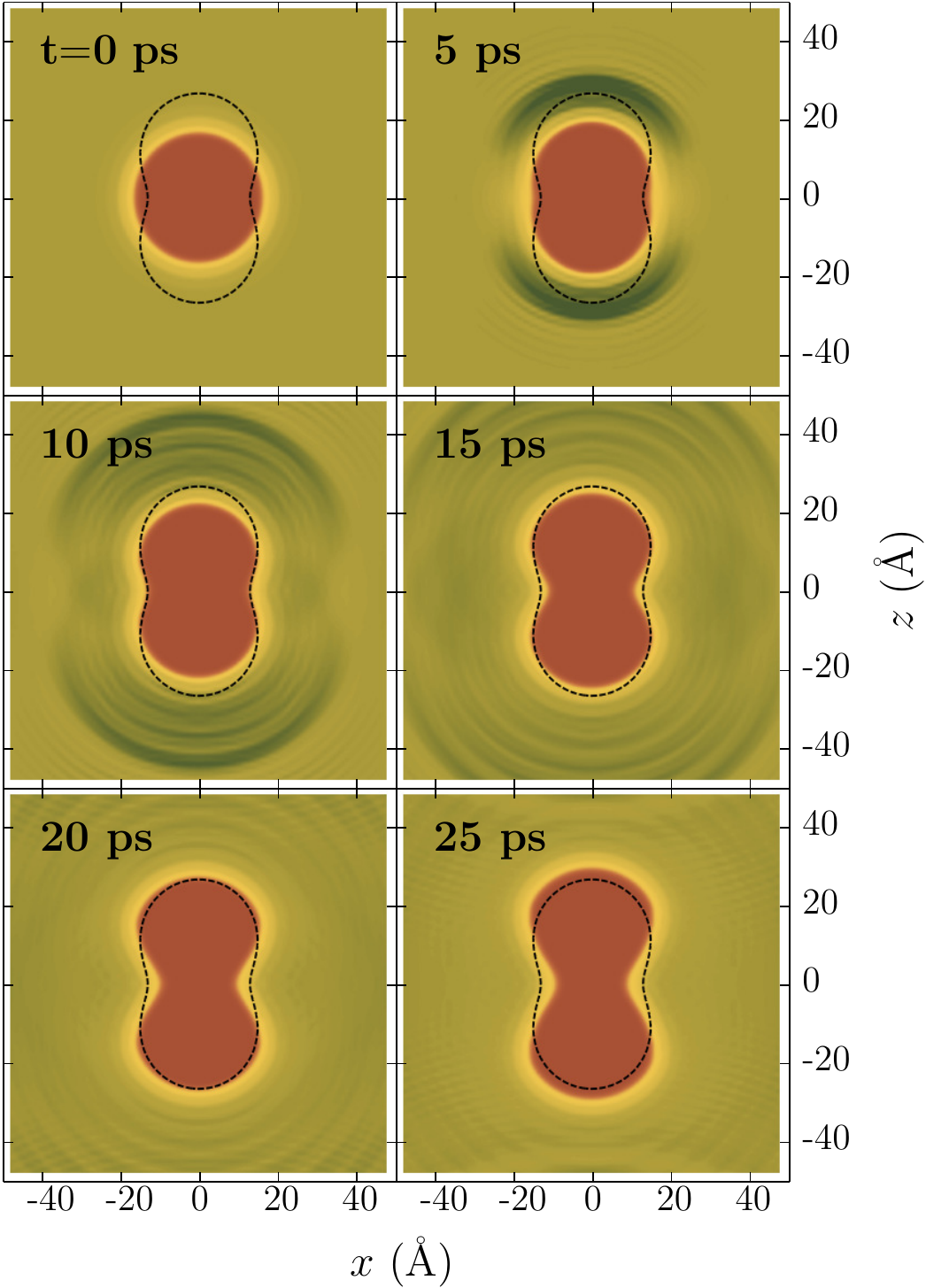}}}%
\subfigure{
\resizebox*{7cm}{!}{\includegraphics{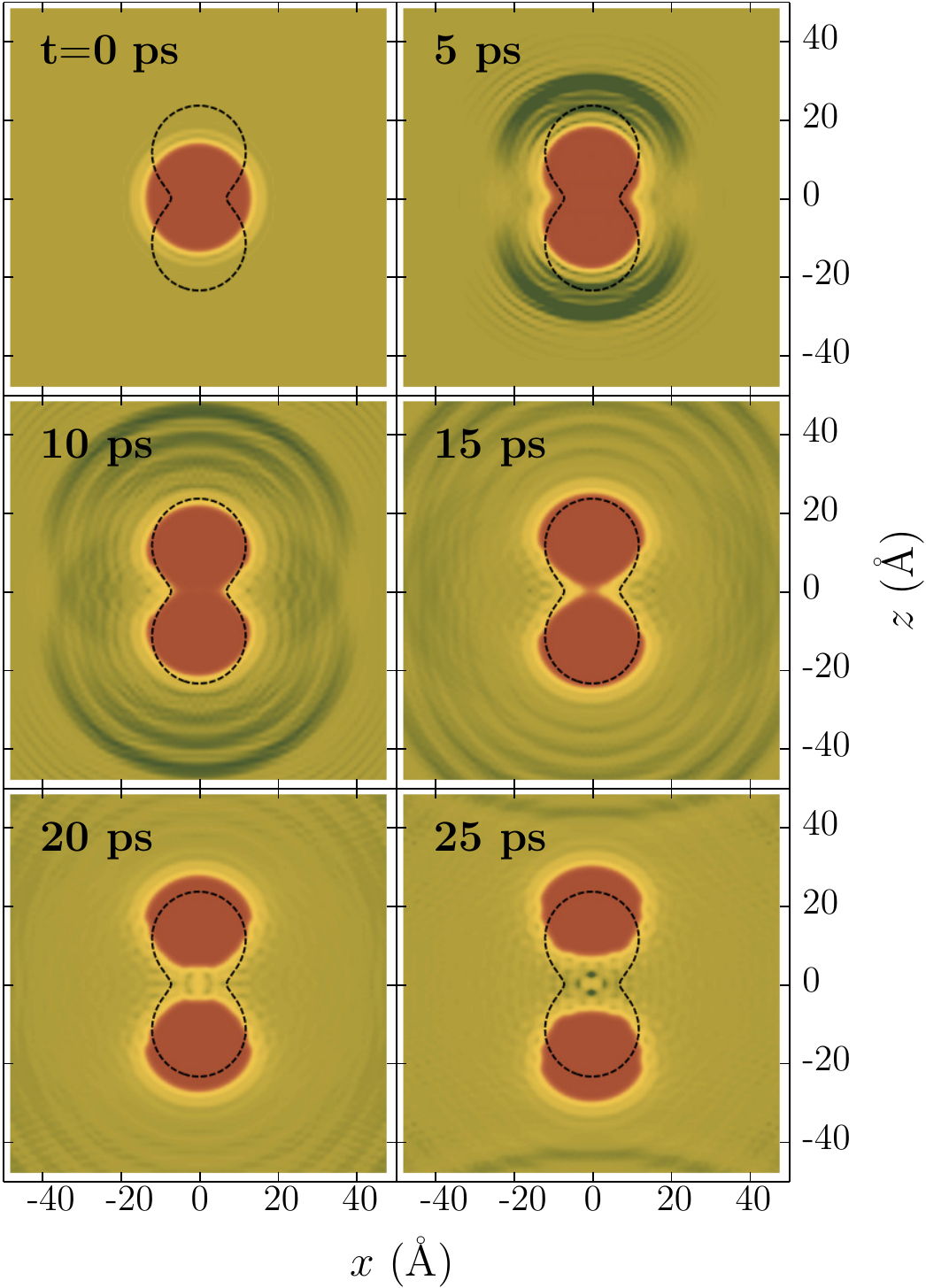}}}%
\caption{\label{fig20}
(a) Adiabatic evolution of the 1p e-bubble at $P=0$. The panels display the helium configurations at the indicated times. 
The dashed line represents the dividing surface, which corresponds to half of the helium equilibrium density for the quasi-equilibrium configuration 
at $P = 0$. (b) Same as (a) but for $P=5$ bar.\cite{Mat10c}
}
\end{center}
\end{figure} 

\begin{figure}[t]
\vspace{42pt}
\begin{center}
\subfigure{
\resizebox*{7cm}{!}{\includegraphics{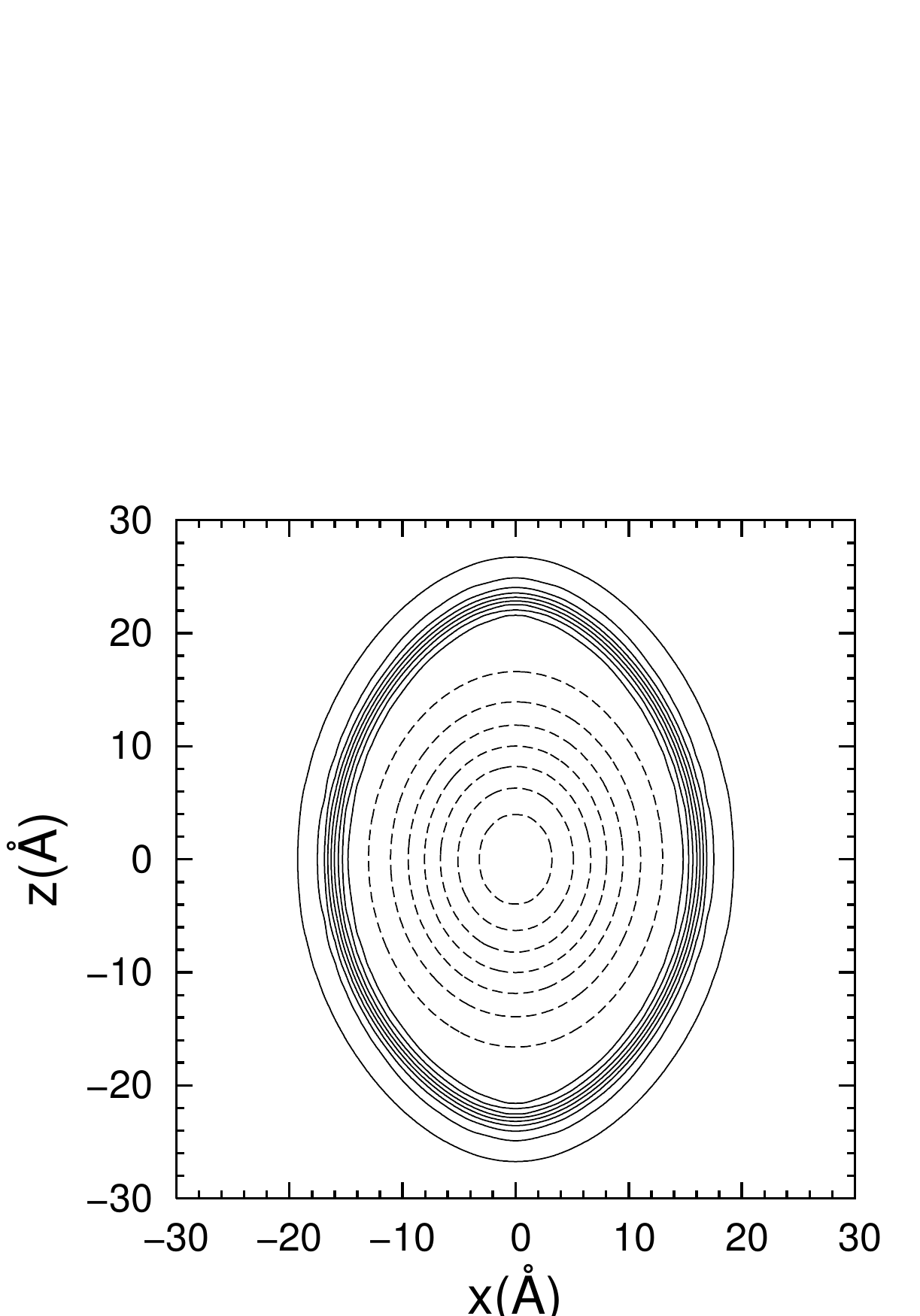}}}%
\subfigure{
\resizebox*{7cm}{!}{\includegraphics{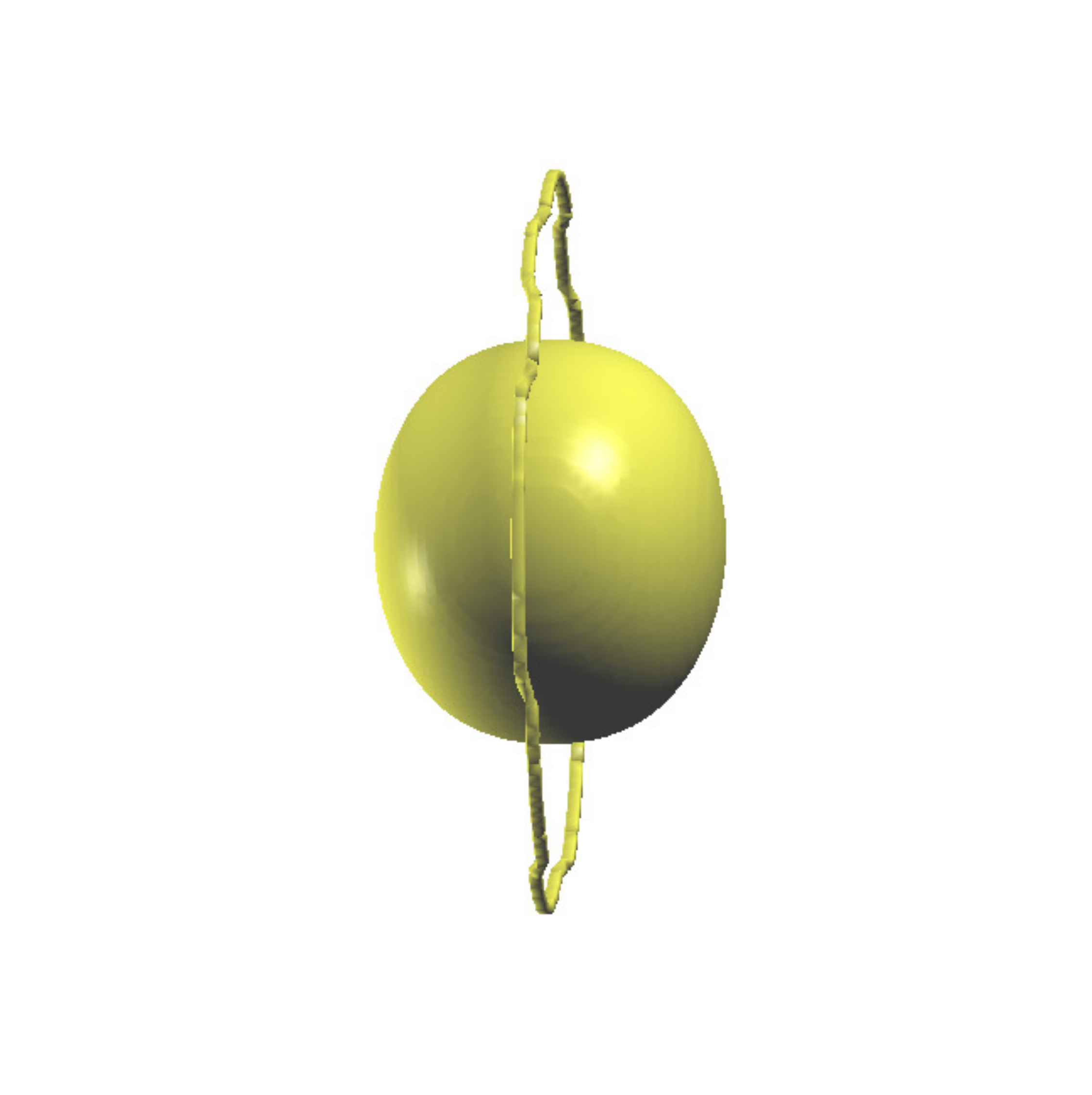}}}%
\caption{\label{fig21}
(a) Contour plot showing the stationary state of the e-bubble at $P=0$ corresponding to $v=50.5$ m/s, which is just below the 
critical value of $v_c=50.7$ m/s. The equidensity lines for the $^4$He density (solid lines) are plotted for values between $0.1\rho _0$ 
and $0.9\rho _0$ in steps of $0.1\rho _0$ with $\rho _0=0.0218$ \AA$^{-3}$. The equidensity lines for the electron probability density (dashed lines) 
are plotted using nine lines between zero and its maximum value. (b) Surface isodensity plot showing a quantised  vortex ring emitted at $P=0$ just above $v_c$.\cite{Anc10}
}
\end{center}
\end{figure}

The structure of e-bubbles hosting an excited electron has been a subject of interest since the experimental work in Refs. \onlinecite{Kon03,Mar08} was interpreted to involve such states. 
Furthermore, these bubble structures determine the emission spectrum of the electron bubble. The relevant DFT work on excited e-bubbles was carried out in Refs.  \onlinecite{Leh07,Mat10c,Jin10c}. 
The most interesting excited states are the 1p and 2p states, which can be accessed by experiments. The evolution of the bubble around these states has been computed by 
DFT.\cite{Mat10c} In this calculation, the liquid degrees of freedom evolve in real time while a time-independent Schr\"odinger equation is solved for the $n$p electron
 at each time step. This adiabatic approximation can be justified by the large helium-electron mass ratio, $m_4/m_e \sim7300$. Note that this approximation would fail if level 
 crossings are encountered during the time evolution. 

The adiabatic evolution of a 1p e-bubble at $P = 0$ and 5 bar is shown in Fig. \ref{fig20} along with the quasi-static configurations obtained by fully relaxing the liquid. At $P=0$ 
the adiabatic evolution leads to a quasi-equilibrium configuration whereas at $P=5$ bar  the e-bubble splits at the waist. The threshold pressure for the bubble 
splitting process is $\sim$ 1 bar. 
This in agreement with the experiments, which indicate that the relaxed 1p bubble is only stable when pressure is smaller than \textit{ca.} 1 bar.\cite{Mar04}

The evolution of a 2p state e-bubble at $P=0$ has been studied within the adiabatic approximation.\cite{Mat10c} The calculations revealed that after 7 ps, the $m = 0$
 levels of the 2p and 1f states become very close and the adiabatic approximation fails. This indicates that quasi-static electron bubble configurations above 1p cannot be reached.  
 A detailed discussion on the validity of the adiabatic approximation for the electron bubble can be found in Ref. \onlinecite{Mat10c} and references therein. 
 Other excited e-bubbles have also been studied by the quasi-static approximation.\cite{Mar08,Leh07} 
 
Real-time propagation of e-bubbles using  finite-range functionals such as OT-DFT is computationally unfeasible due their complexity. 
The use of  a zero-range functional as  e.g. that of  Refs. \onlinecite{Str87a,Str87b}  (ST functional)
 simplifies the calculation of the mean field potential and reduces the computational demand of the calculation.\cite{Mat11a,Jin10b,Jin10a,Jin10c}
  A major shortcoming of zero-range functionals is that they cannot reproduce the maxon-roton portion of the liquid dispersion relation, but only the phonon part 
  up to $q \sim 0.6$ \AA$^{-1}$.\cite{Mat11a} Despite this limitation, the evolution of 1p electron bubbles at various pressures with both the zero and finite-range 
  functionals appear fairly similar. This suggests that zero-range functionals 
can also be applied to study 2p electron bubbles.

Real-time dynamics calculations employing the ST functional have confirmed the two key findings obtained using the adiabatic OT-DFT approximation: 
1) the splitting of the 1p e-bubble above $P \gtrsim$1 bar and 2) the failure of the adiabatic approximation for states higher than 1p.
 The latter finding confirms that quasi-static configurations corresponding to 2p, 2d, $\dots$ do not exist.

Motion of electrons in liquid $^4$He has been studied by Maris \textit{et al.} using zero-range functionals.\cite{Jin10a,Jin10b} They showed that a 1s electron bubble moving at a sufficiently 
high velocity begins to expand, deforms from the spherical symmetry, and nucleates vortex rings. Energy dissipation did not only arise from the creation of vorticity but also
 from the excitation of bubble surface modes. These results were confirmed by imaginary-time OT-DFT calculations in the co-moving frame, see Eq. (\ref{eq67}).\cite{Anc05b,Anc10}
  Based on these calculations, the critical velocity for vortex ring nucleation at $P=0$ is $v_c = 50.7$ m/s. This value is in agreement with the critical electron drift velocity measured at
   low pressures.\cite{Nan85}

The complete mechanism of ring vortex nucleation is not yet fully understood. Based on DFT calculations, the e-bubble becomes compressed along the axis of propagation
 and elongated in the perpendicular direction upon increasing its velocity. Once the bubble exceeds the critical velocity $v_c$, a quantised vortex ring emerges from the equator of the bubble where the local liquid velocity is higher. Fig. \ref{fig21} shows  electron bubbles moving at velocities of 50.5 m/s ($<v_c$) and slightly above $v_c$ (vortex ring emission). Note the significant distortion of the electron bubble geometry that appears below $v_c$ (cf. Fig. \ref{fig19}). The interaction of electrons with vortex lines is discussed in Sec. \ref{5.12.1}.

\subsection{Cations in liquid helium and droplets}
\label{5.7}

\begin{figure}[t]
\vspace{42pt}
\begin{center}
\subfigure{
\resizebox*{7cm}{!}{\includegraphics{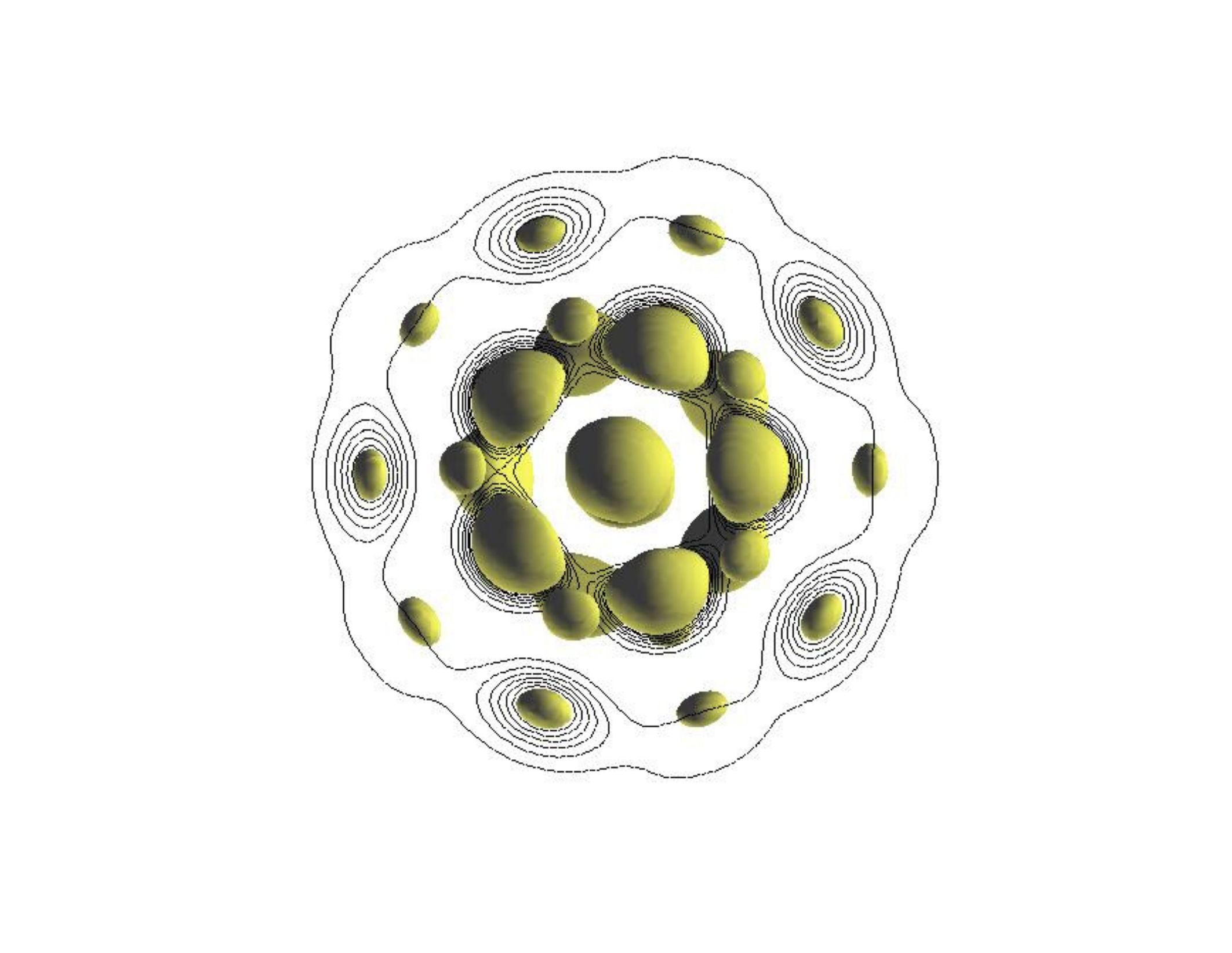}}}%
\subfigure{
\resizebox*{7cm}{!}{\includegraphics{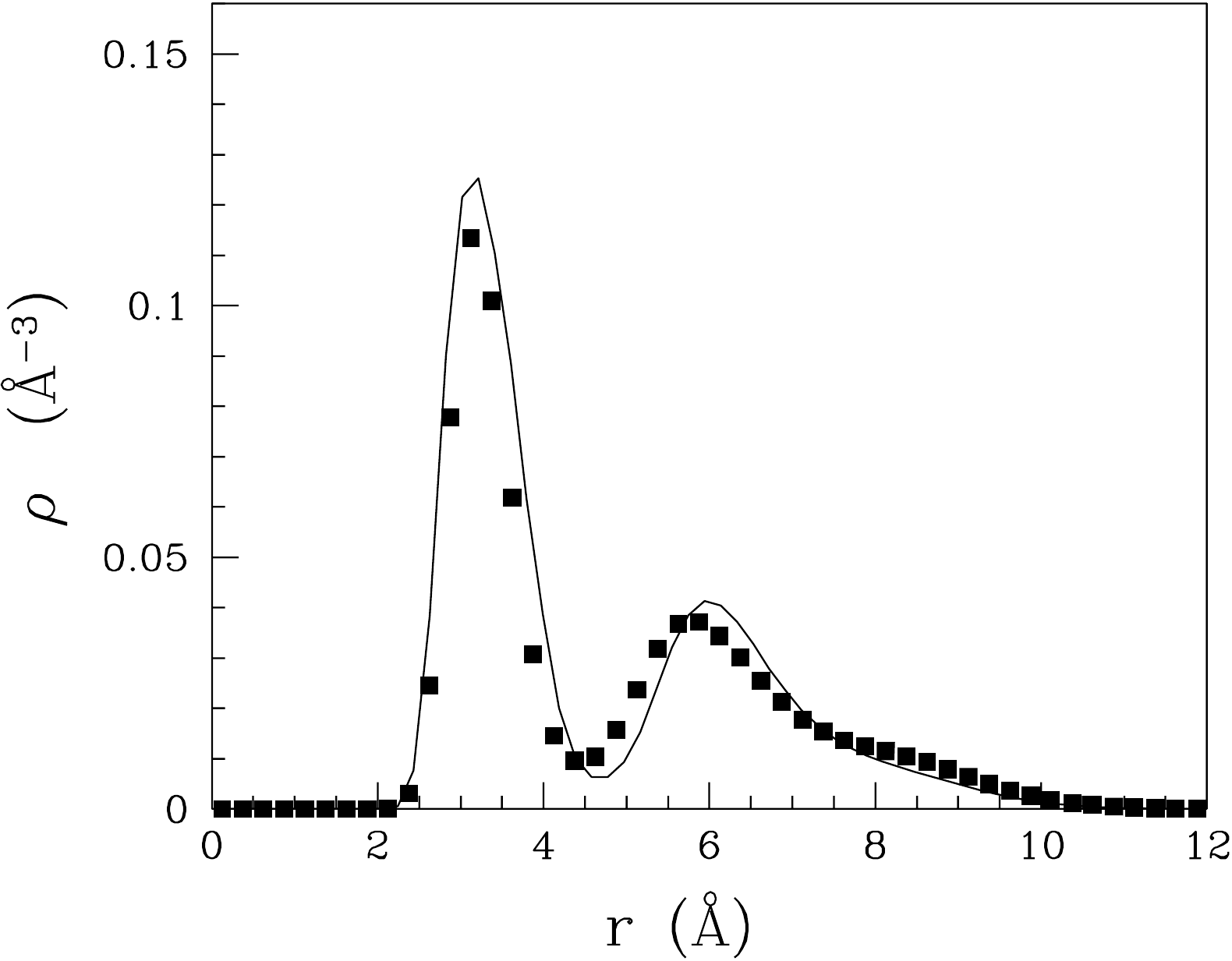}}}%
\caption{\label{fig22}
(a) Helium density distribution around a Be$^+$ ion in a $^4$He$_{70}$ droplet shown by constant density surfaces $\rho= 0.04$ \AA$^{-3}$.\cite{Anc07} 
The lines show equidensity contours along a plane passing through the center of the droplet. (b) Average radial helium density of the Be$^+$@$^4$He$_{70}$ droplet.
 Solid line: DFT result; squares: PIMC result.\cite{Pao07}
}
\end{center}
\end{figure}

Development of new techniques for doping helium droplets with charged impurities allow the controlled study of positive ions in superfluid helium.\cite{Sti03} 
While neutral species could be excited in helium droplets by a resonant laser and detected essentially against a zero background, similar experiments have had limited success for ions. It is only 
recently that experimentalists have found a way to study the dynamics of photoexcited ions in helium droplets. This technique relies on 
the ejection of photoexcited ions from the droplets, 
which increases the yield of unsolvated ions at resonant wavelengths.\cite{Bra04}

\begin{figure}[t]
\vspace{42pt}
\begin{center}
\resizebox*{7cm}{!}{\includegraphics{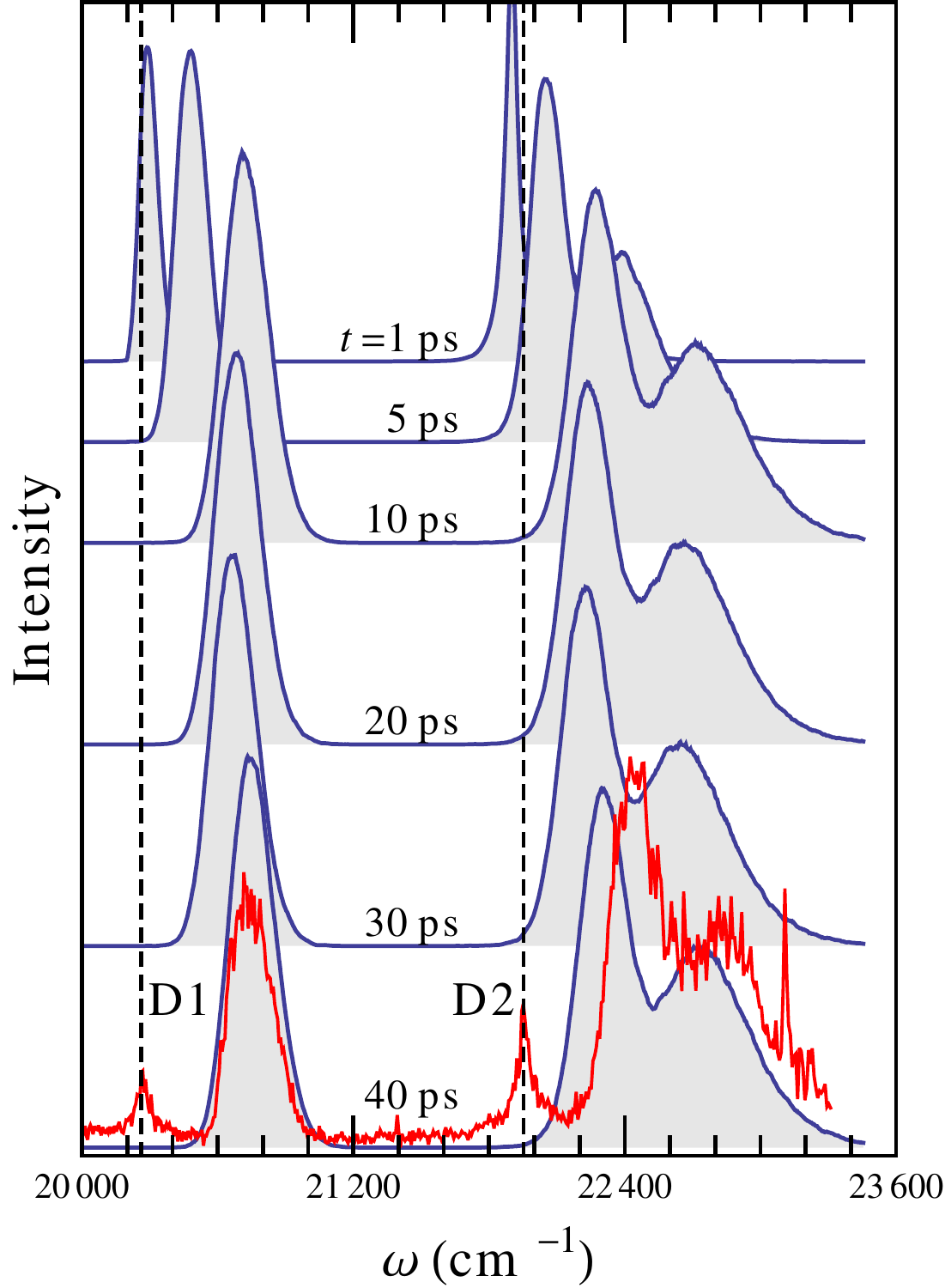}}
\caption{\label{fig23}
TDDFT time-resolved absorption spectrum of Ba$^+$ in a $^4$He$_{1000}$ droplet.\cite{Mat14}  
The experimental spectrum corresponding to helium droplets with an average size of 2700 atoms  is shown in red.\cite{Zha12} 
The vertical lines indicate the D1 and D2 transitions of free Ba$^+$.
}
\end{center}
\end{figure}

\begin{figure}[t]
\vspace{42pt}
\begin{center}
\resizebox*{8cm}{!}{\includegraphics{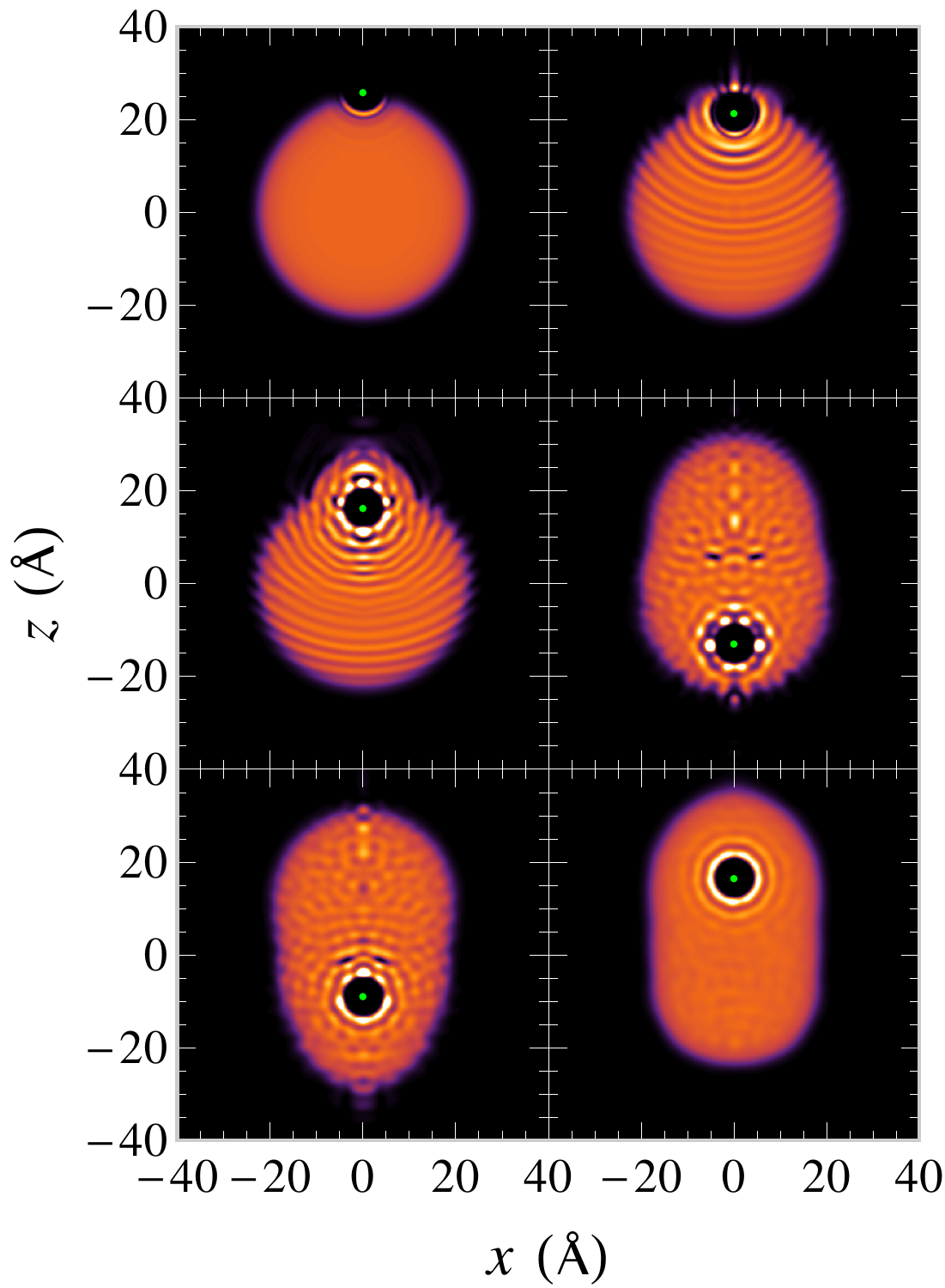}}
\caption{\label{fig24}
Snapshots of the temporal evolution of Ba$^+$@$^4$He$_{1000}$ after photoionisation of the neutral Ba atom located on the droplet
surface.\cite{Mat14} From top to bottom and left to right, the panels correspond to helium densities at times $t= $ 0, 8, 14, 47, 60 and 220 ps.}
\end{center}
\end{figure}
 
Strongly attractive ions tend to form a solid-like helium layer around them (snowball). 
Alkali metal ions are believed to belong to this category.\cite{Pao07,Gal11}
In contrast, singly charged alkaline earth cations 
are expected to produce a cavity due to the outer electron-helium 
repulsion; thus, they are surrounded by a compressed but less 
inhomogeneous liquid.\cite{Ros04,Pao07,Lea14b} 

Alkali metal, alkaline earth metal, and rare gas cations have been recently studied by both experiments and theory.\cite{Lan12,Bar14}
A DMC investigation of Pb$^+$ in small He droplets, motivated by the experimental findings of Refs. \onlinecite{Tig07,Dop07}, is reported in Ref. \onlinecite{Sla10}. 
A similar study was carried out on Na$^+$.\cite{Iss14} 
These studies have highlighted the importance of the many-body interactions that arise from charge-induced dipole interaction between the ion and the surrounding helium atoms. 
Note that this interaction cannot be accounted for by including just 
the $-\alpha e^2/(2r^4)$ polarization term in the cation-helium pair interaction.

DFT calculations have been used to study the solvation structure of Ba$^+$ cation in liquid helium and the stability of the so-called `scolium'. 
Scolium, which was named after Giacinto Scoles,\cite{Sti06} consists of an electron orbiting around a small helium droplet that hosts a positively charged ion. 
The electron cannot penetrate inside the droplet because of the large solvation energy barrier (\textit{ca.} 1 eV). It was suggested\cite{Anc07} that for small droplets, 
the pressure exerted by the orbiting electron further increases the local helium density around the ion due to electrostriction, which consequently turns the whole droplet into a  solid. 
It was also shown that the lowest scolium  state is unstable, the cation being pulled off from the droplet center 
towards the surface where it  undergoes fast charge neutralization. The neutralization time was estimated to be on the order of a few picoseconds for a 50 \AA{} radius droplet. 
As an example of cation solvation in a helium droplet, Fig. \ref{fig22} 
shows the helium density for a Be$^+$ doped $^4$He$_{70}$ droplet where the ion is located at the center.\cite{Anc07} 
Comparison with the QMC data provided in that figure demonstrates a good agreement between the two methods. Notice that such a good agreement is only possible when the `solid' OT-DFT 
described in Sec. \ref{2.3.1} 
is used. The conventional OT-DFT produces  unphysically large
pile-up of helium density around the cation.
A similar level of agreement has been found for other cations.\cite{Fie12,Lea14b} 
The experimental realization of scolium was later achieved by using Na$^+$ rather than Be$^+$.\cite{Log11b} 

QMC and DFT calculations on Rb$^+$ and Cs$^+$ cations in helium droplets show that they are fully solvated and develop snowball structures.\cite{Ros04,Lea14b} 
Based on DFT results, the first solvation shell around Rb$^+$ and Cs$^+$ hosts 19.2 and 21.4 atoms, respectively.
These values are somewhat larger than those found experimentally\cite{Mue09,The11} as well as by QMC calculations.\cite{Gal11} On the contrary, the solvation structure 
around a Ba$^+$ ion was found to be smooth without pronounced structure. This is in agreement with QMC calculations.\cite{Pao07} The difference in the solvation structures
 is a direct consequence of the much weaker interaction between Ba$^+$ and helium compared to that of Rb$^+$ or Cs$^+$ with helium. 
 
In general, cations are solvated in helium droplets because they exhibit very attractive interaction with helium. 
Based on this observation, experimentalists have studied the sinking  of positive ions in $^4$He droplets resulting from  the ionisation of heliophobic alkali and alkaline 
earth metal atoms.\cite{The10,Log11b,Zha12} This activity has motivated the corresponding TDDFT simulations of the ion sinking process.\cite{Mat14,Lea14b,Lea16}

\begin{figure}[t]
\vspace{42pt}
\begin{center}
\resizebox*{10cm}{!}{\includegraphics{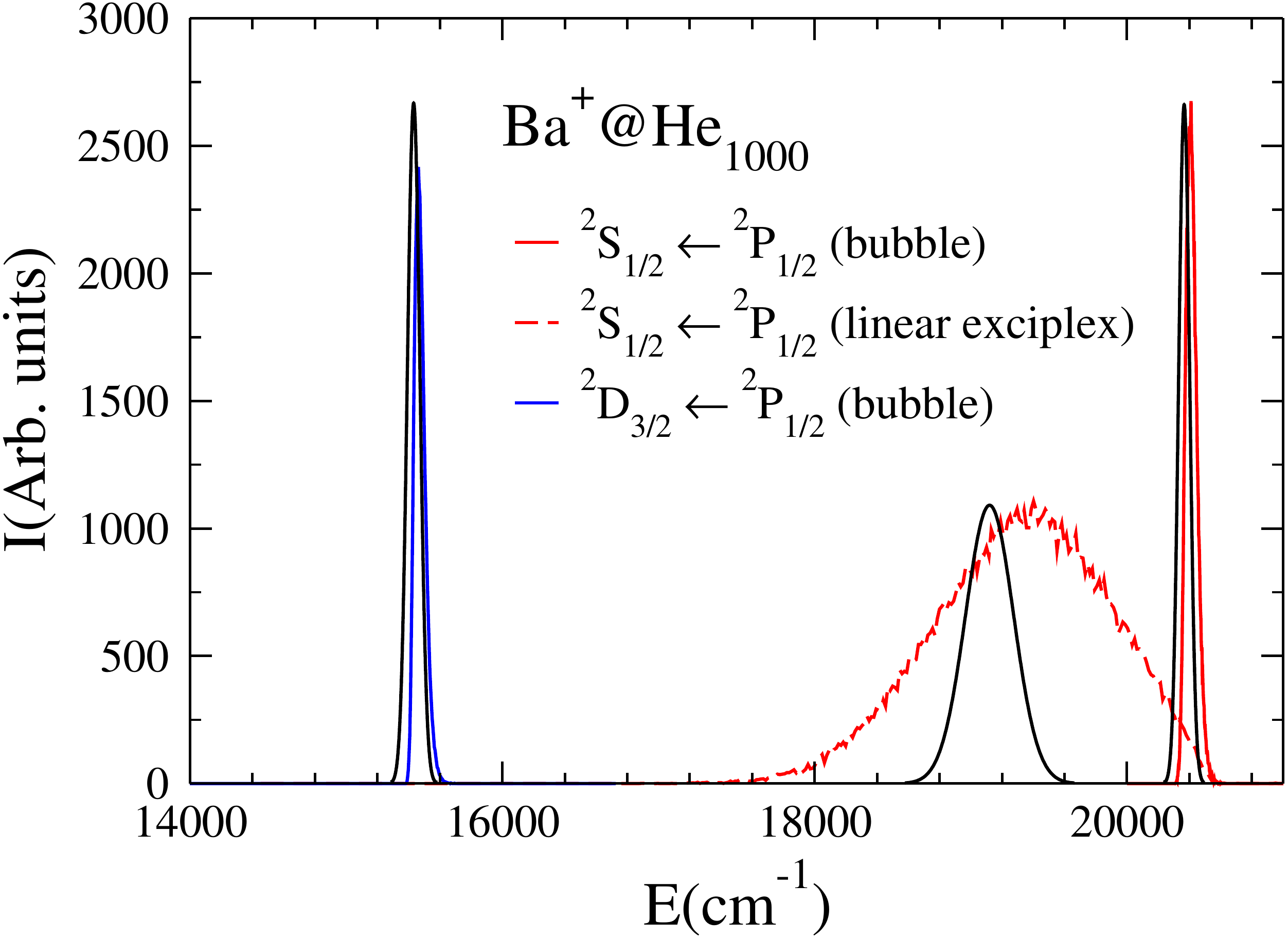}}
\caption{\label{fig25}
Emission spectrum obtained from de-excitation of the relaxed $^2$P$_{1/2}$ state of a Ba$^+$ cation in a  $^4$He$_{1000}$ droplet.\cite{Lea16}  
The gaussian lines shown in black are the experimental results.\cite{Rey86}
}
\end{center}
\end{figure}

Figure \ref{fig23} shows the DFT  time-resolved absorption spectrum of a Ba$^+$@$^4$He$_{1000}$ 6p$ \leftarrow$ 6s transition after photoionisation of the surface-bound atom.
The experimental data\cite{Zha12} show a clear signature of the cation sinking process as the observed spectrum coincides with the corresponding absorption spectrum 
in the bulk liquid.\cite{Rey86} DFT simulations of the ion sinking into the droplet have revealed the nucleation of vortex rings as illustrated in Fig. \ref{fig24}. 
The snowball structure appears dynamically as bright spots around the cation during the initial solvation process, and  wears out when the ion velocity decreases due
 to  kinetic energy dissipation. The latter observation is in accordance with static DFT calculations.\cite{Lea14b}

In addition to the dynamic formation of the Ba$^+$ snowball and the time-resolved absorption spectra of this cation, the most interesting outcome of the calculations is the 
formation of a vortex ring at the equator of the Ba$^+$ solvation structure after about 13 ps. This vortex ring slips around the ion and eventually detaches at 24 ps. 
The cross-section of the vortex ring can be readily identified from the two dark spots behind the ion bubble at 47 ps as shown in Fig. \ref{fig24}. Calculation of the circulation 
around the core yields a value of unity in units of $h/m_4$. Eventually, the vortex ring is destroyed by colliding with the ion bubble.
 
Alkali metal atoms reside on the droplet surface like the heavy alkaline earth atoms, but the interaction of the corresponding ions with helium is strongly attractive. 
Hence their dynamics upon ionisation is expected to be similar to that of barium. 
This motivated the TDDFT study of Rb$^+$ and Cs$^+$ cations produced by the photoionisation of the neutral atom on the droplet surface.\cite{Lea14b} 
Surprisingly, in neither case did the sinking process any vortex rings. 
This was attributed to a subtle effect overlooked in the previous studies. Comparison of the initial surface solvation structures of Ba vs. Rb or Cs shows 
that the latter species are located in a shallower dimple such that the ion is farther away from the droplet surface. 
Upon ionisation, Rb$^+$ and Cs$^+$  pull the lighter helium atoms on the surface 
towards them and the resulting structure `floats' on the droplet. This screens the interaction of the cation with the rest of the nearby He atoms in 
the droplet and, as a consequence, their sinking velocity is lower than for Ba$^+$. Moreover, the sinking process requires a much larger droplet to take place; 
for example, Cs$^+$ did sink inside a $^4$He$_{2000}$ droplet but not in   $^4$He$_{1000}$. 
In the latter case, it was actually expelled from the droplet as a charged minicluster. Both cations were found to sink  when the neutral parent atom was located on the liquid free surface
that might  locally represent  a very large droplet. Interestingly, vortex loops (i.e. vortex segments that start and end on the droplet surface) were nucleated 
 by the appearance of local  distortions in the droplet surface during the sinking process.\cite{Lea14b}

The desolvation dynamics following 6p~$\leftarrow$~6s excitation of Ba$^+$ in helium droplets was further investigated in a joint experimental and theoretical 
work.\cite{Lea16} The experiment showed that the desolvation process yielded mainly bare Ba$^+$ and Ba$^+$He$_n$ exciplexes with $n=1$ and 2. 
In terms of TDDFT simulations, this process is similar to the evolution of photoexcited Ag atoms in helium droplets.\cite{Mat13b} As shown in Fig. \ref{fig25}, 
the calculations reproduced the main features of the experimental Ba$^+$ emission spectrum\cite{Rey86} and, furthermore, they demonstrated the dynamical 
formation of exciplexes. These linear and ring geometry Ba$^+$-He$_n$ exciplexes were previously found by QMC,\cite{Mel14} where the experimentally observed 
line at 19120 cm$^{-1}$ was assigned to the de-excitation of the Ba$^+(^2\Pi_{1/2})$He$_2$ linear exciplex.

Despite the above achievements, the DFT approach did not yield the detachment of  excited Ba$^+$ ions as found in the experiments. 
The origin of this discrepancy was extensively discussed in  Ref. \onlinecite{Lea16} and the ejection mechanism of photoexcited Ba$^+$ from helium 
droplets is still an open question. Elucidating this issue would not only require additional experimental data (e.g. state distribution of the desolvated Ba$^+$ ions), 
but also an improved theoretical model which includes non-adiabatic transitions between electronic states.\cite{Log15} Indeed, non-radiative relaxation 
of the excited states to the $^2$D state or to the $^2$S ground state could deposit a sufficient amount of energy into the system to eject Ba$^+$ and/or blow up the droplet. 
So far, this has not been explicitly shown and this proposition should just be viewed as an `educated guess'.

\subsection{Intrinsic helium impurities}
\label{5.8}

Ionisation of superfluid helium and subsequent charge recombination leads to the generation of intrinsic singlet and triplet state He$^*$ atoms and He$_2^*$ excimers in
 the liquid.\cite{Den69,Hil71} While the singlet states rapidly decay to the electronic ground state through radiative processes, the triplet He$^*$(1s2s) and He$^*_2(^3$a) 
 states are metastable due to the lack of spin-orbit coupling. For this reason, most experimental work has concentrated on employing the triplet species to study the
  response of the surrounding bulk liquid.\cite{Zme13,Ben02} 

To study the solvation of triplet He$^*$ species in superfluid helium by OT-DFT, \textit{ab initio} electronic structure calculations have been conducted  to map out
 the He$^*$-He(1s) interaction in its various electronic states.\cite{Elo01,Bon12,Fie14,Bon16} 
The interaction of these and higher excited triplet species with ground state He atoms was found mostly repulsive, i.e. they are heliophobic and form bubble states in superfluid helium. 
However, nodal planes in excited Rydberg state orbitals can create close range attractive pockets in which helium atoms can accumulate.\cite{Elo02b}
The static solvation structure around He$^*$ was obtained using OT-DFT with He$^*$ treated quantum mechanically 
due to its light mass, see e.g. Eq. (\ref{eq36}).

Since the obtained bubble interfaces have appreciable width, the exact meaning of  the bubble radius must be 
unambiguously defined. For a spherical solvation bubble in the bulk liquid, the interface average radius $R_b$ can be obtained from
\begin{equation}
R_b = \left[\frac{3}{4\pi}\int d \mathbf{r}\left(1 - \frac{\rho(\mathbf{r})}{\rho_0}\right)\right]^{1/3}
\label{eq64}
\end{equation}
Note that this form is only applicable to bubble structures whereas no clear definition for snowball-type solvation cavities can be given.
 Using Eq. (\ref{eq64}), bubble radii in the range of 6-12 \AA{} have been obtained for He$^*$(1s2s) and He$^*$(1s3s) depending on the pressure.\cite{Bon12,Fie14} Since the external potentials have essentially no binding, the interfaces appear smooth with the exception of weak oscillatory structures arising from the correlated nature of the liquid. 

To establish a comparison with existing experimental data, absorption (i.e. He$^*$(1s2p) $\leftarrow$ He$^*$(1s2s)) and fluorescence (i.e. He$^*$(1s3s) $\rightarrow$ 
He$^*$(1s2p)) line shapes were calculated by TDDFT.\cite{Bon12,Fie14} Since the technique for evaluating spectral lineshape is 
different from that presented in Sec. \ref{4.3}, a short description is provided here. 

\begin{figure}[t]
\vspace{42pt}
\begin{center}
\resizebox*{9cm}{!}{\includegraphics{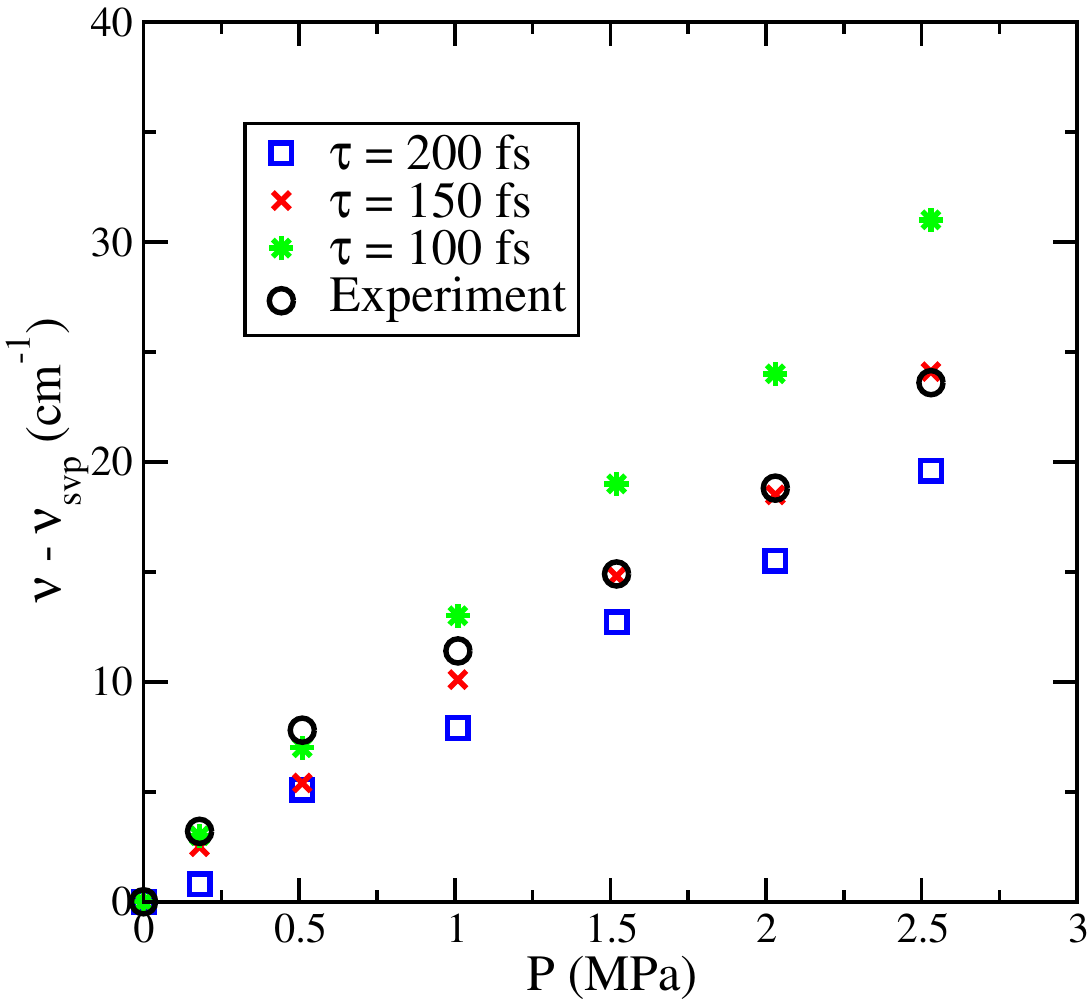}}
\caption{\label{fig26}
Pressure induced shift for the He$^*$ (1s2p) $\leftarrow $(1s2s) absorption line in superfluid helium.\cite{Fie14} 
Comparison between the calculated line shifts (OT-DFT and Eq. (\ref{eq65})) for selected values
 of  the dephasing time constant $\tau$  and the experimental data obtained at 1.6 K temperature.\cite{Hic75,Sol74}
}
\end{center}
\end{figure}

Given the static liquid density profile around the initial state, time evolution of 
the liquid in the final state can be used to determine the first order polarization at excitation angular frequency $\omega$ as\cite{Elo04}
\begin{equation}
P^{(1)}(t) \propto \int_0^t dt'\exp\left(-\frac{i}{\hbar}\int_{t'}^t dt'' \Delta E(t'') - i\omega t'\right) + \textnormal{C.C.}
\label{eq65}
\end{equation}
where C.C. stands for complex conjugate of the preceding term and $\Delta E$ is the energy difference between the two electronic states. For example, for an absorption process 
\begin{eqnarray}
\Delta E(t) & = & \int\int d\mathbf{r} \, d\mathbf{r}' \rho_u'(\mathbf{r}',t)V_u\left(|\mathbf{r} - \mathbf{r}'|\right)\rho_u(\mathbf{r},t)
\nonumber
\\
 & & - \int\int d\mathbf{r} \, d\mathbf{r}' \rho_l'(\mathbf{r}',0)V_l\left(|\mathbf{r} - \mathbf{r}'|\right) \rho_l(\mathbf{r},0)
 \label{eq66}
\end{eqnarray}
where $\rho_l(\textbf{r},0)$ and $\rho_u(\textbf{r},t)$ represent the densities for the initial and final states, respectively. The primed quantities refer to the 
probability density of  the impurity that is 
treated quantum mechanically. Note that for fluorescence the roles of the upper and lower levels are reversed,  the initial upper level contribution is time-independent 
and the dynamics takes place on the lower level potential. To include dephasing in Eq. (\ref{eq65}), the polarization can be multiplied by a phenomenological 
exponential decay, $\bar{P}^{(1)}(t) = e^{-t/\tau}P^{(1)}(t)$, where $\tau$ is the dephasing time constant. During this time, the spectrum is sensitive to the impurity-helium bath interaction. 
The linear absorption or fluorescence spectrum is finally obtained by Fourier transforming the polarization  provided by Eq. (\ref{eq65}). A comparison between 
experimental 2p $\leftarrow$ 2s absorption line shift as a function of external pressure and the OT-DFT results employing this method is shown in 
Fig. \ref{fig26}.\cite{Fie14} The best match with experiments is obtained with $\tau = 150$ fs. Since the response time associated with bubble breathing 
 and interface curvature dynamics is longer than this dephasing time, the resulting absorption spectra appear broad and exhibit no additional structure.

The above line shape model has also been employed to calculate He$^*$ (1s3s) $\rightarrow$ (1s2p) and He$^*_2$ $^3$d $\rightarrow$ $^3$a fluorescence 
line shifts as a function of pressure.\cite{Bon12,Bon16} In general, the calculations show slightly larger blue shifts than the experimental 
data, which may be related to the higher temperature in the experiments vs. calculations or  to the accuracy of the used He$^*$-He and He$^*_2$-He pair potentials. 
Note that the experimental line shifts are sensitive to energy differences of just a couple of cm$^{-1}$ far away from the impurity.

Finally, we mention a time-dependent OT-DFT calculation modelling the superfluid dynamics following a two-photon excitation of the He$^*_2$ excimer from the
$^3$a to the  $^3$d state.\cite{Elo07} These calculations, which were carried out in 1D spherical coordinates, were motivated by earlier optical pump-probe measurements
 that determined the bubble breathing period around the $^3$d state as a function of $P$ and $T$.\cite{Ben02} The period was observed to track the bulk 
 liquid viscosity and reached \textit{ca.} 150 ps at the lowest measured temperature of 1.4 K. The normal fluid fraction at this $T$ is 
only 0.08, which implies that the viscous contribution to the breathing period should be very small. While the OT-DFT calculations carried out in Ref. \onlinecite{Elo07} did not
 include this viscous response, bubble breathing periods in the range of 50 to 120 ps were obtained depending on the He$^*_2$-He potential employed. Two open 
 questions still remain regarding this system: 1) inclusion of the viscous response [see Eq. (\ref{eq69})]; and 2) accurate calculation of the long-range 
 He$^*_2$($^3$d)-He interaction potential. The latter may also require inclusion of many-body corrections beyond the pair potential approximation as the $^3$d 
 Rydberg orbital is somewhat compressible.\cite{Elo01}

\subsection{Translational motion of ions below the Landau critical velocity}
\label{5.9}

In addition to the above mentioned solvation dynamics of ions in superfluid helium, the hydrodynamic response of the surrounding liquid due to translational motion
 of various ions has been studied by OT-DFT.\cite{Fie12,Mat15b} At $T=0$, provided that the ion velocity remains well below the Landau critical value, dissipation 
 of energy can only take place through the emission of sound when the ion accelerates or decelerates in the liquid. 
In the presence of thermal excitations (i.e. thermal phonons and rotons), the viscous drag force also opposes the ion motion. The most important experimentally 
accessible parameters that are sensitive to this dissipative liquid response are the ion hydrodynamic mass $(m_{add})$ and the ion mobility ($\mu$). The former quantity  
corresponds to the difference between the bare ion mass in vacuum and its effective mass in the liquid whereas the ion mobility is determined by the ion steady-state 
velocity in the liquid.\cite{Bor07}

Hydrodynamic added masses for several halogen anions\cite{Fie12} as well as bare positive (i.e. He$_3^+$) and negative charges\cite{Mat15b} in superfluid $^4$He 
have been calculated by OT-DFT. In the latter work, the ion mass was computed by imaginary-time OT-DFT in the co-moving reference frame (see also Sec. \ref{5.6})
\begin{equation}
\left[\hat{H} - v_{0,z}\,\hat{P}_z\right]\Psi(\mathbf{r}) = \mu\, \Psi(\mathbf{r})
\label{eq67}
\end{equation}
where $v_{0,z}$ is the constrained liquid velocity along the $z$-axis and $\hat{P}_z$ is the $z$-component of the momentum operator. 
An estimate for the added mass during imaginary-time iterations can be computed from
\begin{equation}
\frac{m_{add}}{m_4} = \frac{1}{v_{0,z}}\int d \mathbf{r} \, \rho(\mathbf{r}) \, v_z(\mathbf{r})
\label{eq68}
\end{equation}
where $v_z$ is the $z$-component of the liquid velocity $\mathbf{v}(\mathbf{r}) = \mathbf{j}(\mathbf{r})/\rho(\mathbf{r})$. 
This model yields results consistent with the available experimental information (i.e. positive and negative charges)\cite{Mat15b} as well as the
 independently obtained QMC data (i.e. K$^+$).\cite{Fie12,Buz01} 

In ion mobility experiments, an external electric field accelerates the ion until the electrostatic and hydrodynamic drag forces cancel. The resulting steady-state velocity $v_z$
 is directly related to the ion mobility, $\mu = ev_z/F_z$, where $F_z$ is the electrostatic force acting on the ion. In a  liquid at $T \neq 0$,  the drag force arises from collisions
  with thermal excitations,  whereas in the limit of 0 K the ion could in principle accelerate until the critical value for the creation of vorticity or turbulence is reached 
  (see  Sec. \ref{5.6} for the case of an electron). 

To help comparison with the available experimental mobility data,\cite{Bor07} the standard OT-DFT functional must be extended to include 
the liquid viscous response. At temperatures higher than 1.4 K, the roton density is sufficiently high for a continuum-based model 
to be applicable. The viscous response term from the Navier-Stokes equation can be adapted to DFT by employing the Madelung transformation,\cite{Ait16} see also Sec. \ref{4.2}. This gives the following equation for the associated non-linear potential, $V_{NS} = V_{NS}\left[\rho,\textbf{v}\right]$,
\begin{equation}
\nabla^2 V_{NS} = -\nabla\cdot\left\{\frac{1}{\rho}\left[\eta\left(\nabla \mathbf{v} + \left(\nabla\mathbf{v}\right)^{\rm T} - \frac{2}{3}\left(\nabla\cdot \mathbf{v}\right)\mathbf{1}\right)\right]\right\}
\label{eq69}
\end{equation}
where $\eta = \eta(\rho,T)$ is the liquid shear viscosity, $\mathbf{1}$ denotes the unit tensor, and superscript T denotes matrix transpose. Note that this form allows 
for both liquid compression ($\nabla\cdot \mathbf{v} \ne 0$) and rotation ($\nabla\times \mathbf{v} \ne 0$) as well as spatial variation of the viscosity. The discrete form of this 
equation reduces to the Poisson problem, 
which can be efficiently solved in Fourier space using standard techniques.\cite{Pre92} 
 
An additional complication arises from the presence of a wide gas-liquid interface surrounding most ions in superfluid helium. 
To obtain agreement with the experimental electron mobility data, the shear viscosity in this region must be modified from the bulk value,
 $\eta\left(\rho(\textbf{r}),T\right) = \left[\rho(\textbf{r})/\rho_0(T)\right]^{\alpha(T)}\eta_0(T)$.\cite{Ait16} where $\alpha(T)$ determines the spatial variation of the viscosity 
 across the interface and $\eta_0$ is the bulk shear viscosity.
 
\begin{figure}[t]
\vspace{42pt}
\begin{center}
\resizebox*{10cm}{!}{\includegraphics{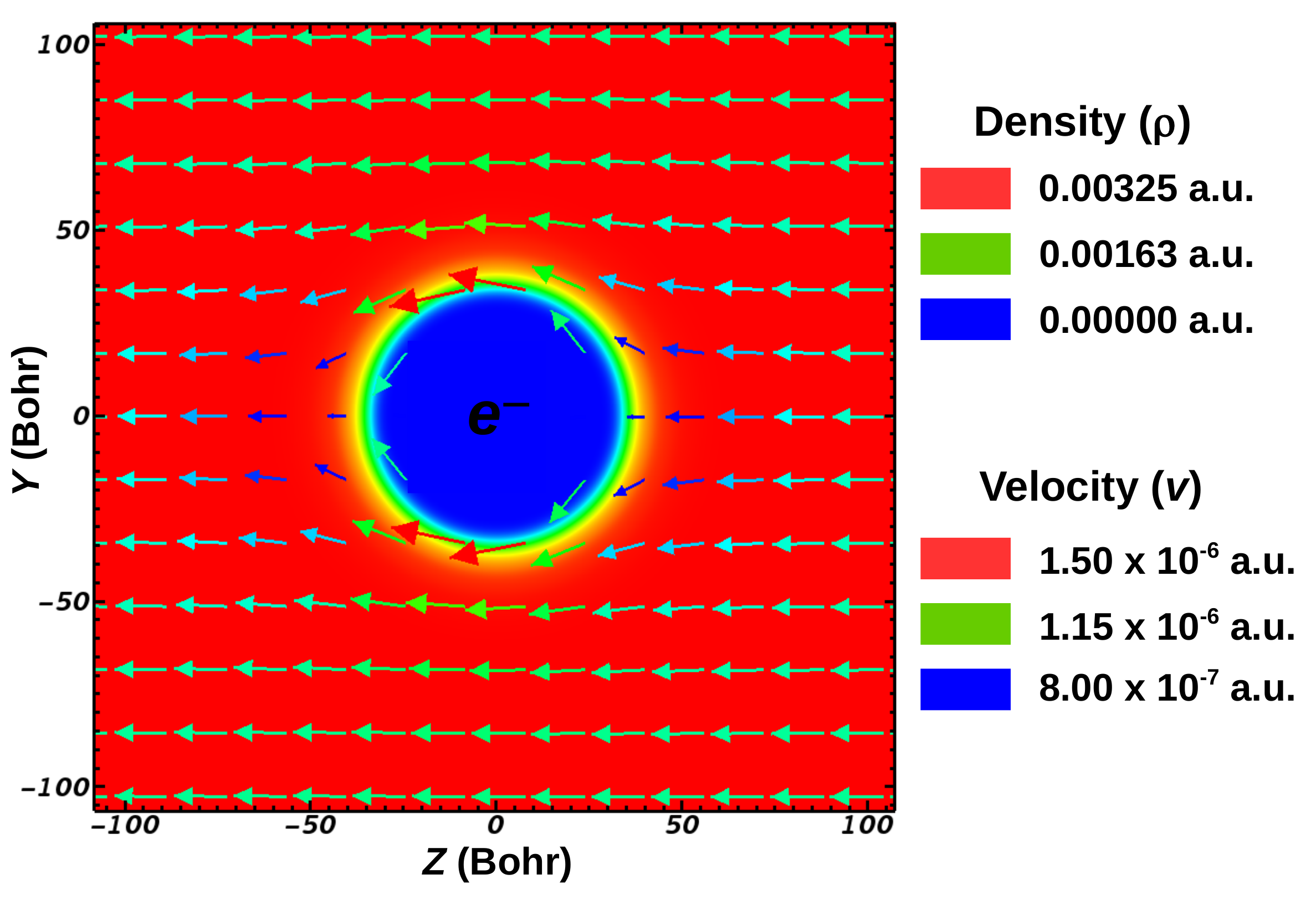}}
\caption{\label{fig27}
Steady-state helium density contours ($\rho$) and velocity field ($v$) around an electron moving in superfluid helium at $T=2.1$ K. 
The velocity component shown along the $z$-axis was shifted by $v_0 = -2.3$ m/s.\cite{Ait16}
}
\end{center}
\end{figure}

The steady-state liquid flow solution around the ion can be obtained by including Eq. (\ref{eq69}) in the OT functional and propagating the system in imaginary-time 
according to the velocity constraint of Eq. (\ref{eq67}).
 Under this condition, the electrostatic and drag forces cancel out and the ion mobility can be evaluated. The hydrodynamic drag on the ion can simply be obtained
  by calculating the force due to the surrounding liquid for classical impurities (see Eq. (\ref{eq47})). If the impurity (e.g. electron) is treated quantum mechanically, 
  integration over the impurity coordinate must also be included. This model was shown to reproduce the known electron mobility data between 1.4 K and the lambda 
  point along the saturated vapor pressure line.\cite{Ait16} 
As an illustration, Fig. \ref{fig27} shows the results for an electron moving in superfluid helium at 2.1~K. 

\subsection{Critical Landau velocity in small $^4$He droplets}
\label{5.10}

\begin{figure}[t]
\vspace{42pt}
\begin{center}
\resizebox*{7cm}{!}{\includegraphics{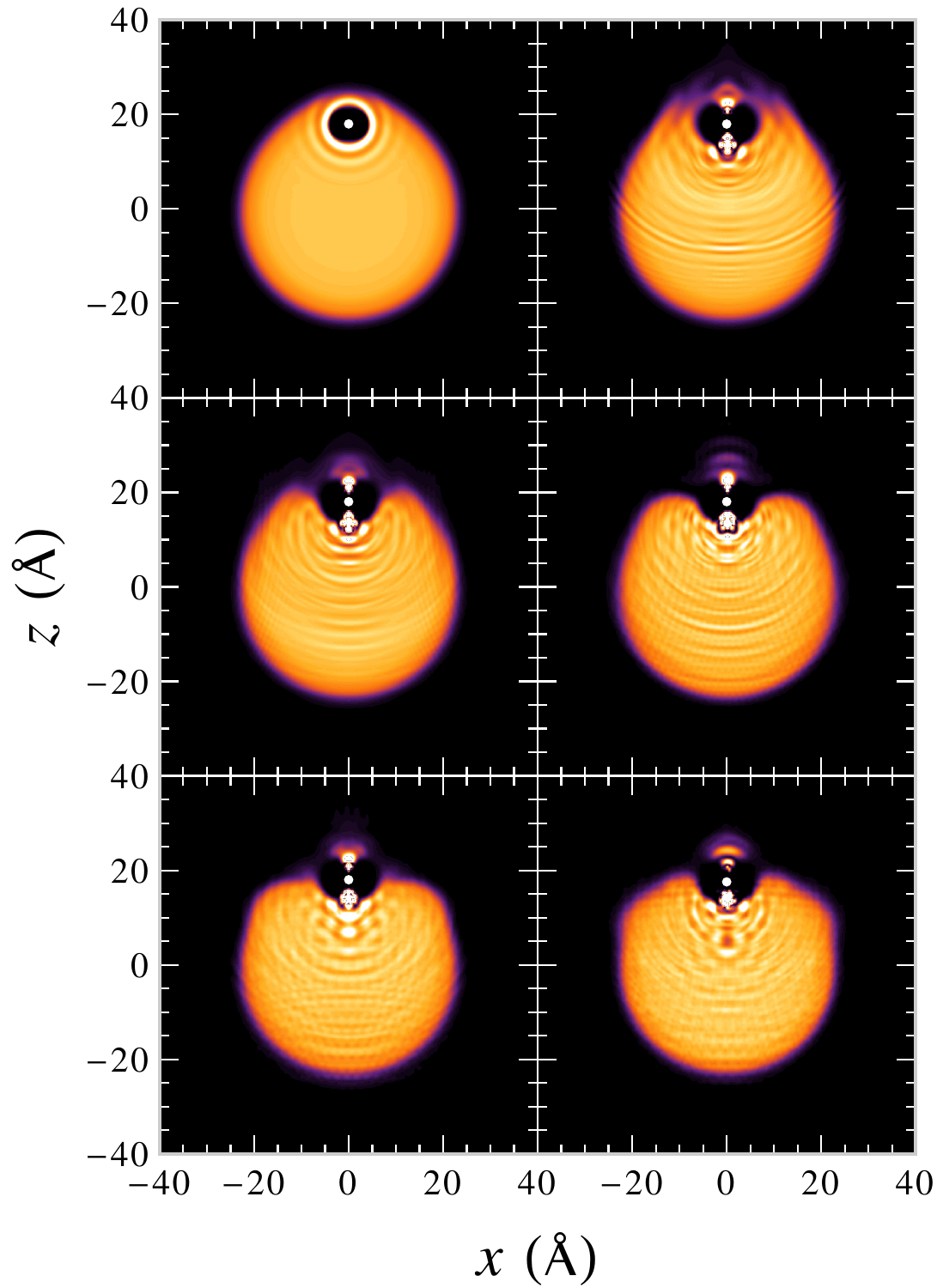}}
\caption{\label{fig28}
Dynamic evolution of an Ag@$^4$He$_{1000}$ complex upon sudden excitation of Ag to the $^2$P$_{3/2}$ state 
displaying the appearance of a linear AgHe$_2$ exciplex on the droplet surface.
The Ag atom is  initially at rest 18 \AA{} off the center of the droplet.
Snapshots are shown every 4 ps starting from the top left frame.\cite{Mat13b}}
\end{center}
\end{figure}

Many properties of helium nanodroplets have been characterised during the last two decades by using solvated molecules as spectroscopic probes. In particular, 
vibrational and rotational spectroscopy of solvated carbonyl sulfide (OCS) provided evidence for microscopic superfluidity in these finite size systems.\cite{Toe04,Gre98} 
However, this raises the question to what extent can microscopic superfluidity be related to the frictionless flow of superfluid helium.

The first point considered was the existence of Landau critical velocity in helium nanodroplets, which are microscopic objects of only 10$^3$-10$^6$ atoms, 
and the possibility of using  atoms or molecules as probes for it. 
This was the starting point of a joint experimental and theoretical search\cite{Bra13} for the existence of a limiting velocity for species ejected from helium droplets.

The following scheme was designed for that purpose.
A probe atom or molecule, initially located in the bulk portion of the droplet, was suddenly optically excited to an electronic state with repulsive interaction with helium. 
As a consequence, the probe is accelerated and ejected from the droplet.
If there is a critical velocity, the probe cannot accelerate beyond it and this will be reflected in its final velocity.
Experimental measurements of the velocity distributions of atoms/molecules ejected from various size helium droplets have revealed the 
existence of a critical velocity threshold even for droplets consisting of a thousand helium atoms. 
In particular, $^2$P$_{1/2}$ excitation of Ag leads to its ejection with a velocity distribution peaking around 55 m/s. A similar, although not identical, velocity 
distribution was obtained upon excitation to the 
$^2$P$_{3/2}$ state, where the ejected species was either Ag or AgHe. 
DFT simulations on the dynamic evolution of this system\cite{Mat13b,Lea16} confirmed these findings and provided additional microscopic details of the process.
In particular, running the simulation for  several tens of picoseconds was sufficient to observe the AgHe exciplex formation as shown in Fig. \ref{fig28}.

\subsection{Rotational superfluidity}
\label{5.11}

\begin{figure}[t]
\vspace{42pt}
\begin{center}
\resizebox*{9cm}{!}{\includegraphics{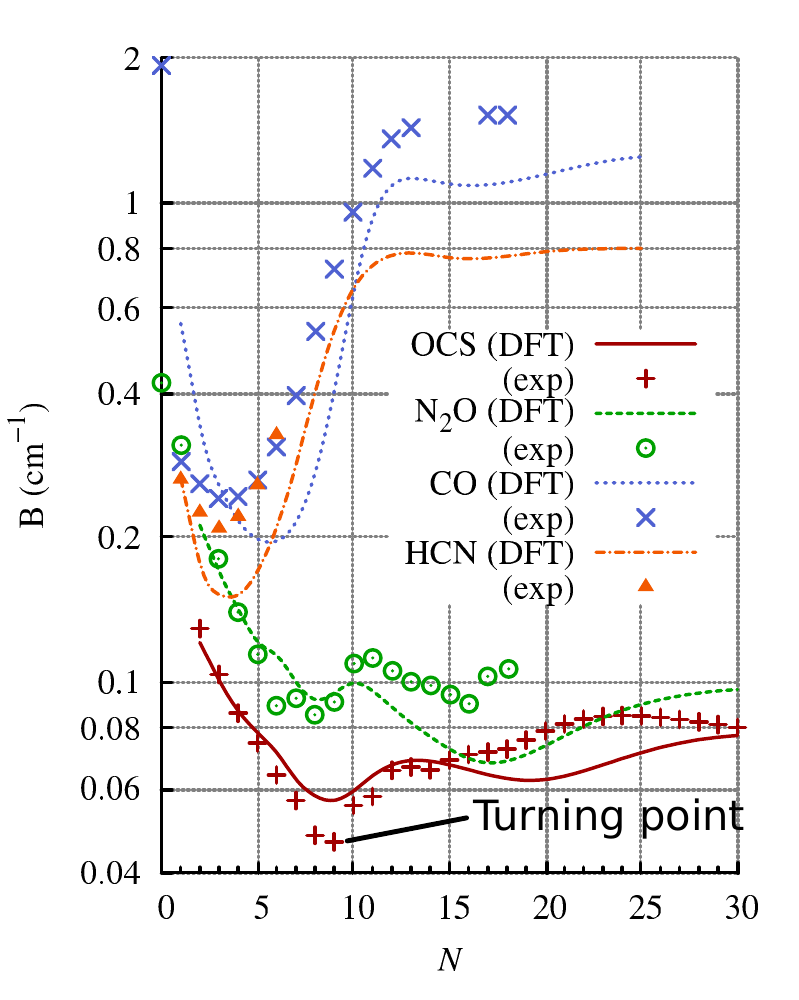}}
\caption{\label{fig29}
Experimental (exp) and calculated (DFT) effective rotational constants as a function of helium droplet size. The position of the non-classical turning
 point for OCS is indicated by a line.\cite{Mat15b}
}
\end{center}
\end{figure}

Superfluidity of helium droplets has been extensively studied by both experiments and theory.\cite{Bart96,Toe98,Gre00} As discussed earlier, previous
 experimental work has employed molecular probes to interrogate the droplet response to both radial and rotational  excitation.\cite{Har96,Jag02} 
 While these experiments have demonstrated the presence of the characteristic roton energy gap for larger droplets (i.e. the Landau criterion for superfluidity), microwave 
 spectroscopy experiments indicated that non-classical behavior already takes place in molecule-helium clusters with less than ten He atoms.\cite{Jag02} 
In these experiments, the rotational constant $B$ (proportional to the inverse of the moment of inertia) was determined as a function of the number of helium atoms  in the cluster.
$B$ was observed to initially decrease with $N_4$ as expected for a classical rotor, but then it started increasing again 
 from a given size on, which  depended on the probe,\cite{Jag03,Jag07,Jag08} as shown in Fig. \ref{fig29}.
This turning point has been interpreted as the onset of superfluidity in small helium droplets ($N_4 < 20$).\cite{Jag02}
It seemed to contradict the original Laudau criterion for superfluidity since previous QMC calculations had shown\cite{Sin89,Kro01}
 that there was no roton energy gap in droplets with less than \textit{ca.} 64 $^4$He atoms.

Recent OT-DFT calculations employing the rotational constraint of Eq. (\ref{eq31}) 
  have resolved this discrepancy by identifying the quantum mechanical origin of the non-classical reduction in rotational friction.\cite{Mat15b} 
  The only input to this model is the probe molecule-He interaction, i.e. the external potential for OT-DFT, taken from \textit{ab initio} electronic structure calculations.
The rotationally constrained OT-DFT equation was solved by the  ITM, which yielded the stationary order parameter $\Psi$ with the associated liquid
 density and velocity field rotating with the molecule. The added moment of inertia for the rotor ($I_{add}$), which is equivalent to $m_{add}$ for translation motion, can be computed from
\begin{equation}
I_{add} = \left<\Psi|L_z|\Psi\right>/\omega
\label{eq70}
\end{equation}
where $\hat{L}_z$ is the $z$-component of the liquid angular momentum operator and $\omega$ is the frequency of rotation -- typically less than 1 GHz; see Eq. (\ref{eq31}). 
The effective rotational constant of the molecule in superfluid helium, $B_{eff}$, is then given by
\begin{equation}
B_{eff} = \frac{\hbar}{4\pi c\left(I_{gas} + I_{add}\right)}
\label{eq71}
\end{equation}
where $I_{gas}$ is the moment of inertia of the molecule in the gas phase.

Four different probe molecules, which can be classified as `heavy'  
or `light' rotors based on their gas phase moments of inertia, were studied\cite{Mat15b} (see Fig. \ref{fig29}). 
The results showed that the experimentally observed turning points in the $B(N_4)$ curve
correlated with helium coverage of the probe molecule. When a connected path of helium forms around the molecule, 
helium attempts to remain irrotational by introducing  negative angular momentum to decouple from the rotational motion. 
This explains why the position of the turning point depends on the probe molecule itself.
In addition, this turning point correlates with the appearance of a continuous helium coverage around the probe rather than the completion of the first solvation shell. 
Secondary oscillations in $B(N_4)$ were related to the complete coverage of the subsequent helium layers. 

The appearance of global phase coherence around the 
probe molecule produces a Landau-type energy gap between the droplet rotational ground and first excited states. This gap plays a similar role 
for rotational motion as the Landau roton gap in traditional superfluidity. When analysed in the co-rotating frame of reference, the transition 
bears similarities to the Mott-1D superfluid quantum phase transition.\cite{Gre02} In order to distinguish this phenomenon from the traditional
 \textit{translational} superfluidity, we refer to it as \textit{rotational} superfluidity.\cite{Mat15b}

\subsection{Interaction of impurities with vortex lines}
\label{5.12}

Much of what is known on vortices in helium has been drawn using ions and electrons as probes in experiments. In this section we present
results for ions, electrons, and neutral impurities obtained by static DFT calculations and leave the discussion of dynamical
 capture of impurities by vortices in helium droplets\cite{Cop17} to Sec. \ref{5.15}. 
 The reader should note the different sign convention for the binding energy of an 
 impurity to the vortex line used in the following Secs. \ref{5.12.1} and \ref{5.12.2}. 
 We have kept the convention used in the original papers.

\subsubsection{Electrons}
\label{5.12.1}

DFT calculations can provide detailed information about electron trapping on quantised vortex lines.\cite{Pi07} The degenerate electronic states 
(e.g. p, d, f $\ldots$) of the bubble split if the bubble becomes non-spherical. For this reason, when the electron bubble becomes trapped on a vortex line,
 the resulting symmetry breaking leads to the splitting of the energy levels. Provided that this splitting is large enough, it could be observed by absorption 
 spectroscopy. 
 
\begin{figure}[t]
\vspace{42pt}
\begin{center}
\resizebox*{9cm}{!}{\includegraphics{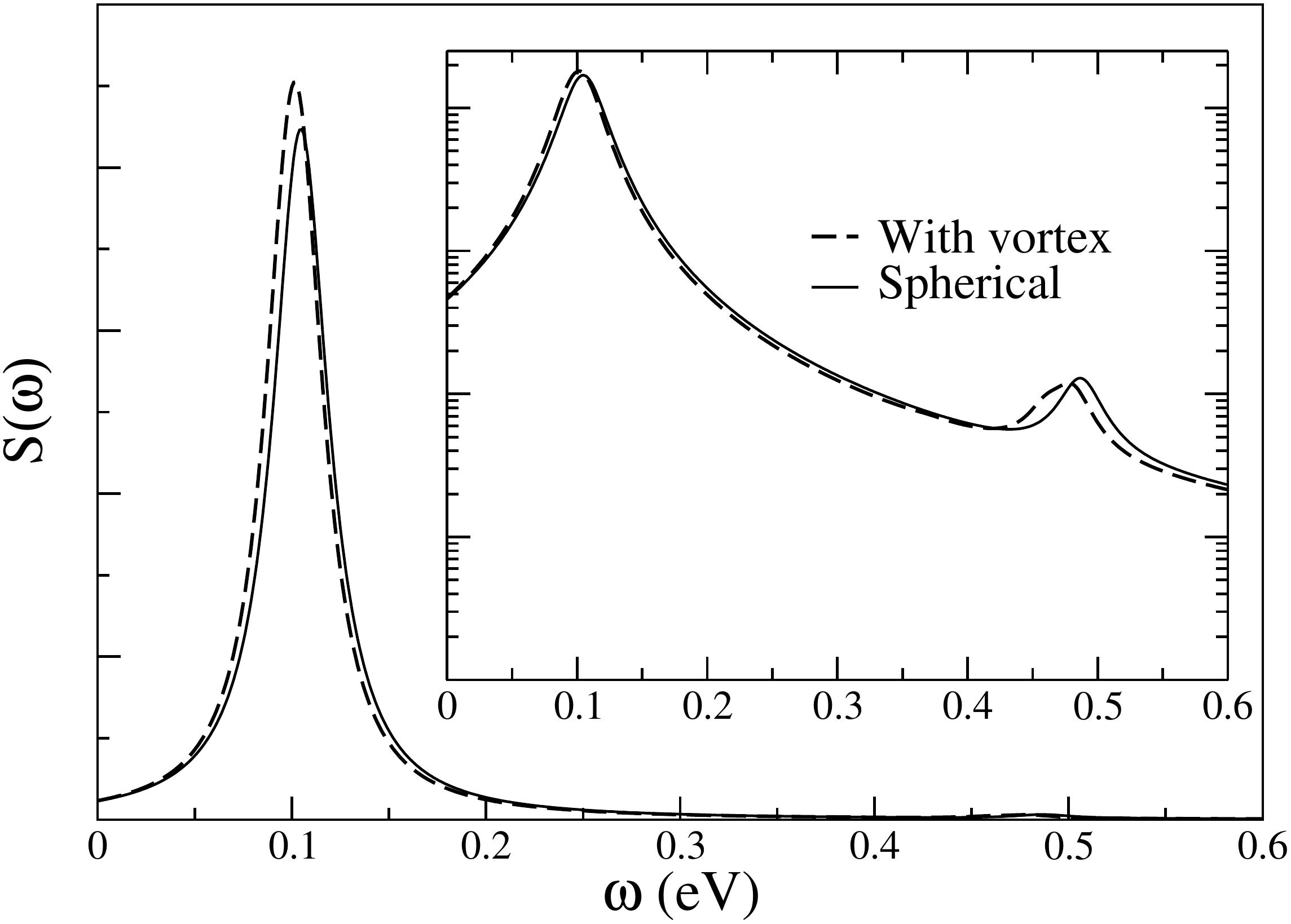}}
\caption{\label{fig30}
Dipole strength $S(\omega)$ (arbitrary scale) at $P=T=0$ for a spherical electron bubble (solid line) and for an electron bubble attached to a linear vortex (dashed line). 
The inset shows $S(\omega)$ in  logarithmic scale.\cite{Mat10b}}
\end{center}
\end{figure}

 The absorption spectrum of the electron bubble can be obtained from the dipole strength function $S(\omega)$ 
\begin{equation}
S(\omega)= \sum_{n\neq0} |\langle n | \mathbf{r} | 0 \rangle|^2 \delta (\omega- \omega_{n0})
\label{eq72}
\end{equation}
where $| 0 \rangle$ and $ | n\rangle$ correspond to the ground and excited states of the electron bubble, respectively. The infrared absorption spectrum of the e-bubble from DFT calculations is shown
 in Fig. \ref{fig30} in the range of 1s--1p and 1s--2p transitions. The function $S(\omega)$ displays peaks centered at the absorption energies $\omega_{n0}$. 
 Furthermore, the figure shows that the effect of vortex trapping on the absorption spectrum is very small, especially for the 1s--1p transition. Slightly larger changes 
  are observed for the 1s--2p transition, which originates from the more pronounced penetration of the 2p electron wave function into the liquid.
  However, this transition is much weaker than the 1s--1p one. Based on these results, it can be concluded that the infrared absorption spectrum of the e-bubble is 
  not very sensitive to the possible vortex trapping and hence, it is not generally suitable for detecting vorticity.

The 1p--1s emission energy has been calculated for both free and vortex trapped e-bubbles.\cite{Mat10b} This calculation assumes that the radiative lifetime
 of the excited state is longer than the time required for the liquid to equilibrate around the electron. Experimental data indicate that the radiative lifetime is 
 some tens of nanoseconds\cite{Gho05} or even tens of microseconds when calculated directly from the transition dipole moment.\cite{Leh07} Based on the 
 simulations, the equilibrium geometry around 1p is reached after several hundreds of picoseconds. 

The most important quantity that can be extracted from the calculations is the binding energy of the e-bubble to the vortex line. This can be obtained from the grand potential 
per unit volume, $\Omega \equiv (F - \mu_4 N_4)/{\cal V}$, for e-bubble configurations with and without a vortex line at the same temperature and pressure: 
\begin{equation}
B_{e-V}= [\Omega_e -\Omega] -[\Omega_{e+V} -\Omega_V] 
\label{eq73}
\end{equation}
where the subscripts refer to vortex-free e-bubble ($e$), bulk liquid (none), e-bubble attached to a vortex line ($e+V$), and
vortex line alone ($V$). Since all the terms above are evaluated under the same thermodynamic conditions, their chemical potential $\mu_4$ and the saturation liquid density are identical. 
The above expression is therefore well-defined and independent of the simulation box volume. Alternatively, one may use the substitution energy introduced in Sec. \ref{3.2}, 
which differs from the value of Eq. (\ref{eq73}) by \textit{ca.} 10\%. The calculations yield the electron binding energy as $B_{e-V} = 104.5$ K at $P=T=0$, which shows that trapping 
of electron bubbles on vortex lines is energetically very favored. A value of 109 K was found in Ref. \onlinecite{Mat15a};
the difference between the two values is attributed to the different electron-helium interactions used.
 
The attractive interaction between an electron and a vortex line can be understood in terms of the loss of kinetic energy due to the liquid displaced by the approaching 
electron bubble (i.e. the classical Bernoulli force).\cite{Bor07} The experimentally obtained values for electron binding to vortex lines vary between 55 and 59 K at 
$T=1.6$ K,\cite{Pra69,DeC74,Don91} which is in clear disagreement with the value   obtained from OT-DFT calculations at $T=0$. Since this rather deep binding 
value is consistent with the large radius of the electron bubble,\cite{Gri92,Gra06,Cla98,Elo02} the difference has been attributed to  the $T$  dependence of the vortex core parameter 
and local thermal deformations of the vortex line rather than to a deficiency in the OT-DFT-based model itself.\cite{Mat15a,Pi07,Pad72} Furthermore, the same OT-DFT model
 is able to reproduce the experimental binding energy of positive charges (i.e. He$^+_3$, Ref. \onlinecite{Mat15c}) to vortex lines: 16 K (calculated at 0 K) vs. 17.5 K 
 (experiment at 0.3 K).\cite{Mat15a} Once the thermal effects for the electron are taken into account, the OT-DFT binding energy is lowered down to 61 K, 
 which is very close to the experimental estimates.\cite{Mat15a} Thus, vortex lines present very deep traps for electrons in superfluid helium. 

\subsubsection{Atomic and molecular impurities}
\label{5.12.2}

Just as electrons trap on vortex lines, any impurity should be attracted towards them due to the Bernoulli force. Given a sufficiently high concentration of neutral impurities 
in the liquid, they can accumulate on vortex lines as a consequence of this attraction. Subsequent  diffusion along the vortex line may then lead to the assembly 
of nanowires, which are reminiscent of the original vortex line geometry. The final products from this process have been observed in metal doped bulk superfluid 
helium\cite{Gor14,Gor15,Pop13} and superfluid helium droplets.\cite{Gom12,Lat14,Spe14} One of the main factors influencing the initial stage of the nanowire assembly 
is the impurity-vortex line interaction, which has been modelled by OT-DFT calculations.\cite{Mat15a}

\begin{table}[t]
\vspace{0.1cm}
 {\begin{tabular}{ccccccc}
\hline\hline
Impurity &\;\;\;& $R$ &\;\; \;&$a_F$ & OT-DFT $V_{\textnormal{im-vortex}}$ & Exp.
$V_{\textnormal{im-vortex}}$\\
& &(\AA) & &(\AA) &  (K) & (K)\\
\hline
H$_2$ ($X$ $^1\Sigma_g$) & &3.1 && 0.38 & $-9.4$ & --\\
Ag$_2$ ($X$ $^1\Sigma_g$) & &3.9 & &0.51 & $-10.9$ & --\\
Cu$_2$ ($X$ $^1\Sigma_g$) & &4.0 & &0.52 & $-11.4$ & --\\
Ag ($^2S$) & &4.4 & &0.63 & $-12.0$ & --\\
Cu ($^2S$) & &4.5 & &0.65 & $-12.5$ & --\\
He$_3^+$ ($X$ $^2\Sigma_g$) & &$\approx 5.7$ & &$\approx 0.75$ & $-16.0$ & $-17.5$ at 0.28-0.6 K, Refs. \onlinecite{Ost75,Wil75}\\
Li ($^2S$) & &6.8 & &0.75 & $-21.5$ & --\\
He$^*$ ($2s$ $^3S$) & &7.1 & &0.77 & $-22.6$ & --\\
He$^*_2$ ($a$ $^3\Sigma_u$) & &8.6 & &0.80 & $-29.8$ & --\\
$e^-$ ($1s$) & &22.2 & &0.76 & $-109^{\;\rm a}$ & $-55$ to $-59$ at 1.6 K,  Refs. \onlinecite{Pra69,DeC74}\\
\hline\hline
\end{tabular}}
\caption{\label{table3}
Summary of the impurity-vortex interaction parameters at $P = T = 0$ based on Eq. (\ref{eq75}). $R$ represents the classical bubble radius 
for the impurity, $a_F$ is the healing length, and $V_{\textnormal{im-vortex}}(0)$ denotes the total binding energy.\cite{Mat15a} 
$^{\rm a}$ See discussion in Sec. \ref{5.6} regarding the   value for the electron.
}
\end{table}

The actual impurity trapping event is clearly a dynamic process, see Sec. \ref{5.15}. Assuming that the impurity impact velocity remains small and the vortex line geometry 
does not deviate from linear geometry, static interaction energy potentials for the vortex-impurity interaction can be obtained. This calculation can be carried out in
 imaginary-time where the vortex line structure is imposed by the initial guess given by Eq. (\ref{eq28}). The position of the impurity must remain fixed during the 
 calculation, which can be imposed by including the following penalty term in the external potential\cite{Mat15a} (see also Eq. (\ref{eq24}))
\begin{equation}
V[\Psi,\mathbf{r}] = 2\, \lambda_C\, (z - z_0)\int d\mathbf{r}'\Psi^*(\mathbf{r}')\,(z' - z_0) \Psi(\mathbf{r}')
\label{eq74}
\end{equation}
where 
$\lambda_C \sim 10^{-5}$ a.u., and the impurity is constrained along the $z$-axis (perpendicular to the vortex line) at position $z_0$. Calculation of the total energy of the system as a function of the distance between the vortex line and the impurity yields the static interaction potential.

\begin{figure}[t]
\vspace{42pt}
\begin{center}
\resizebox*{9cm}{!}{\includegraphics{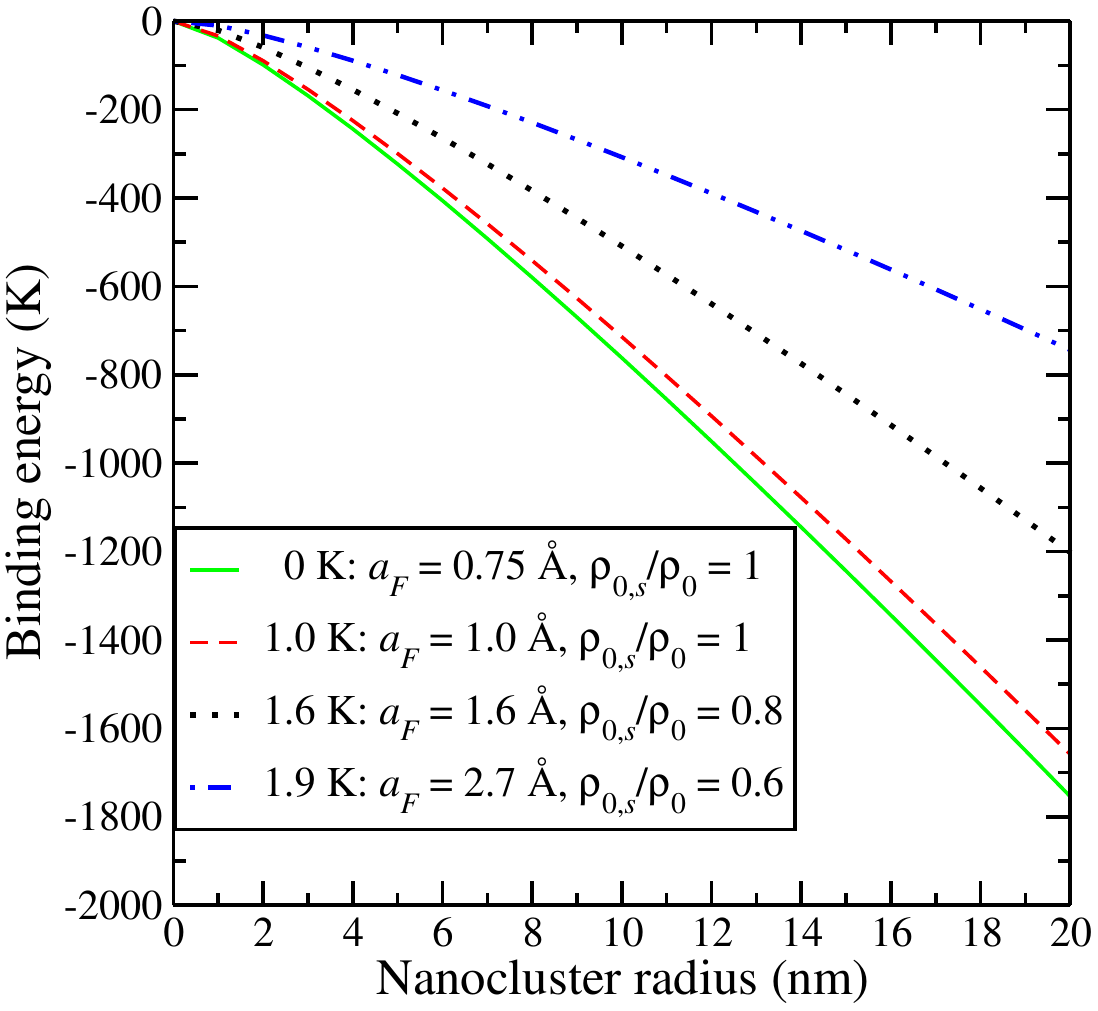}}
\caption{\label{fig31}
Nanocluster-vortex binding energies according to Eq. (\ref{eq75}) as a function of the cluster size at indicated temperatures.\cite{Mat15a}}
\end{center}
\end{figure}

In addition to providing the total binding energies of impurities to vortex lines, the interaction energies can be rationalised by the Donnelly-Parks potential function\cite{Par66}
 (see also Eq. (\ref{eq29}))
\begin{equation}
V_{\textnormal{im-vortex}}(r) = -2\pi\rho_{0,s}\left(\frac{\hbar^2}{m_4}\right)^2\int_0^{R}d\xi\frac{\left(R^2 - \xi^2\right)^{1/2}\xi}{\left[\left(\xi^2+r^2+a_F^2\right)^2 - 4r^2\xi^2\right]^{1/2}}
\label{eq75}
\end{equation}
where $R$ represents the classical radius for the spherical cavity containing the impurity, $r$ is the vortex core-impurity distance, $\rho_{0,s}$ is the bulk superfluid 
helium density, and $a_F$ is the  effective healing length. Non-linear least squares fit of this expression to the interaction potential energy curves from OT-DFT
 provides  effective size  estimates for  both $R$ and $a_F$.  Some of the obtained results are shown in Table \ref{table3}.

Nanowires  in liquid helium are believed to form through recombination of metal nanoparticles in the vortex rather than through building up from individual 
atoms.\cite{Gor15,Vol16} The size of such particles is too large for any practical OT-DFT calculation. Therefore, the binding energy estimates for nanoparticles 
bound to rectilinear vortex lines can only be obtained from Eq. (\ref{eq75}). Note that this assumes that the vortex lines are linear and longer than the diameter 
of the approaching nanoparticle. The binding energy data as a function of the nanoparticle radius is shown in Fig. \ref{fig31}. The effect of temperature was included 
in the estimate by varying the superfluid density ($\rho_{0,s}$) and the healing length ($a_F$) accordingly. While the binding energies quickly exceed 1000 K, under 
real experimental conditions this may rather be limited by the dimensions of the vortex line itself.

\subsection{Vortex arrays in $^4$He droplets}
\label{5.13}

Together with the frictionless motion of impurities at velocities below
the Landau critical velocity in superfluid $^4$He, the appearance of quantised vortices is another signature of superfluidity. Helium remains at rest when its container is rotated, 
until a critical angular velocity is reached. This leads to the appearance of vortices with quantised velocity circulation in units of $h/m_4$, where $h$ is the Planck constant. 

The vortex line distributions in superfluid $^4$He were first imaged by Williams and  Packard\cite{Wil74} by means of light scattered by electrons attached to the vortex lines; 
 quantised vortices have also been visualised  by suspending micron-sized solid particles of hydrogen in bulk superfluid $^4$He.\cite{Zha05,Bew06}
More recently,
femtosecond single-shot x-ray diffraction imaging of Xe doped $^4$He droplets employing a free electron laser,  revealed Bragg spots confirming the existence of 
quantum vortex arrays in helium droplets.\cite{Gom14,Jon16,Ber17} This result shows that large $^4$He droplets containing about 10$^{10}$ atoms are superfluid. 

\begin{figure}[t]
\vspace{42pt}
\begin{center}
\resizebox*{11cm}{!}{\includegraphics{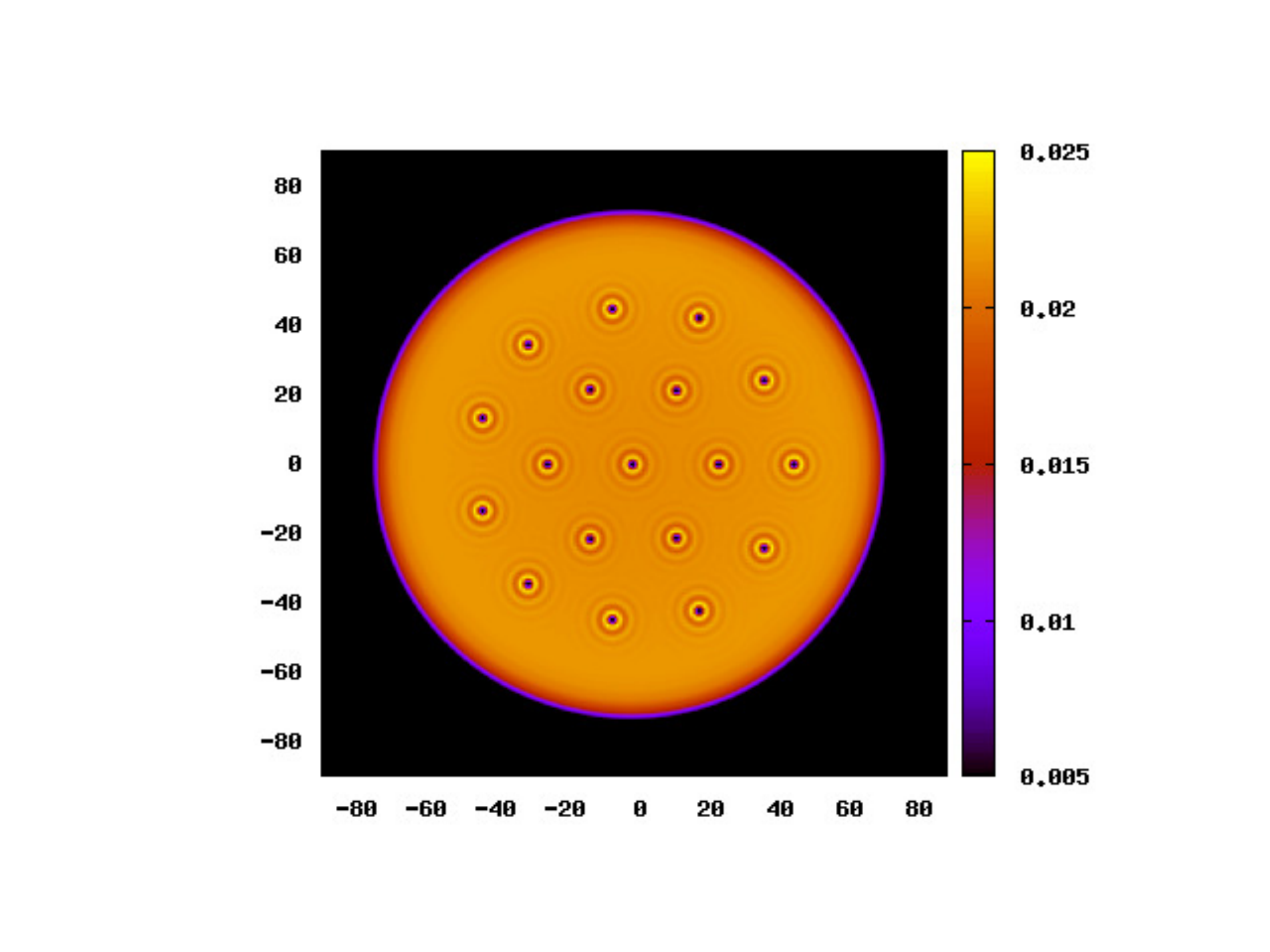}}
\caption{\label{fig32}
Lowest energy stationary 18-vortex configuration  in a nanocylinder of  $R=71.4$ \AA{} radius at $\Omega = 29.6$ (defined in Eq.~(\ref{eq76})). Distances are specified in \AA. 
The contour colors correspond to density  values between $\rho = 0$ and $\rho = 0.03$ \AA$^{-3}$.\cite{Anc14}}
\end{center}
\end{figure}

As discussed earlier, DFT has proven to be a very useful theoretical tool to study vortices in liquid $^4$He. In most recent applications, vortex arrays in a rotating $^4$He
 nanocylinder\cite{Anc14} and in $^4$He nanodroplets\cite{Anc15} were studied by DFT. In the former case, the $n_v$-vortex stability diagram was computed and compared 
 with that of classical vortex lines in an inviscid, incompressible fluid. Vortex array configurations in a rotating cylinder -- as that shown in Fig. \ref{fig32} --  can be completely characterised 
 within the Onsager-Feynman model by the dimensionless energy per unit length
${\cal E}\equiv (m_4/\rho_0 \pi \hbar ^2)E$, the dimensionless angular velocity $\Omega \equiv R^2 m_4\, \omega / \hbar$, and the scaled radial positions of the vortices $r_i/R$. 
Here $\rho_0 = 0.0218$ \AA$^{-3}$ is the bulk density and $R$ is the radius of the cylinder.\cite{Hes67} By scaling the calculated values to millimeter-scale, the nanoscale 
results agree with the experimental data on vortex arrays observed in the bulk liquid.\cite{Wil74}

\begin{figure}[t]
\vspace{42pt}
\begin{center}
\resizebox*{8cm}{!}{\includegraphics{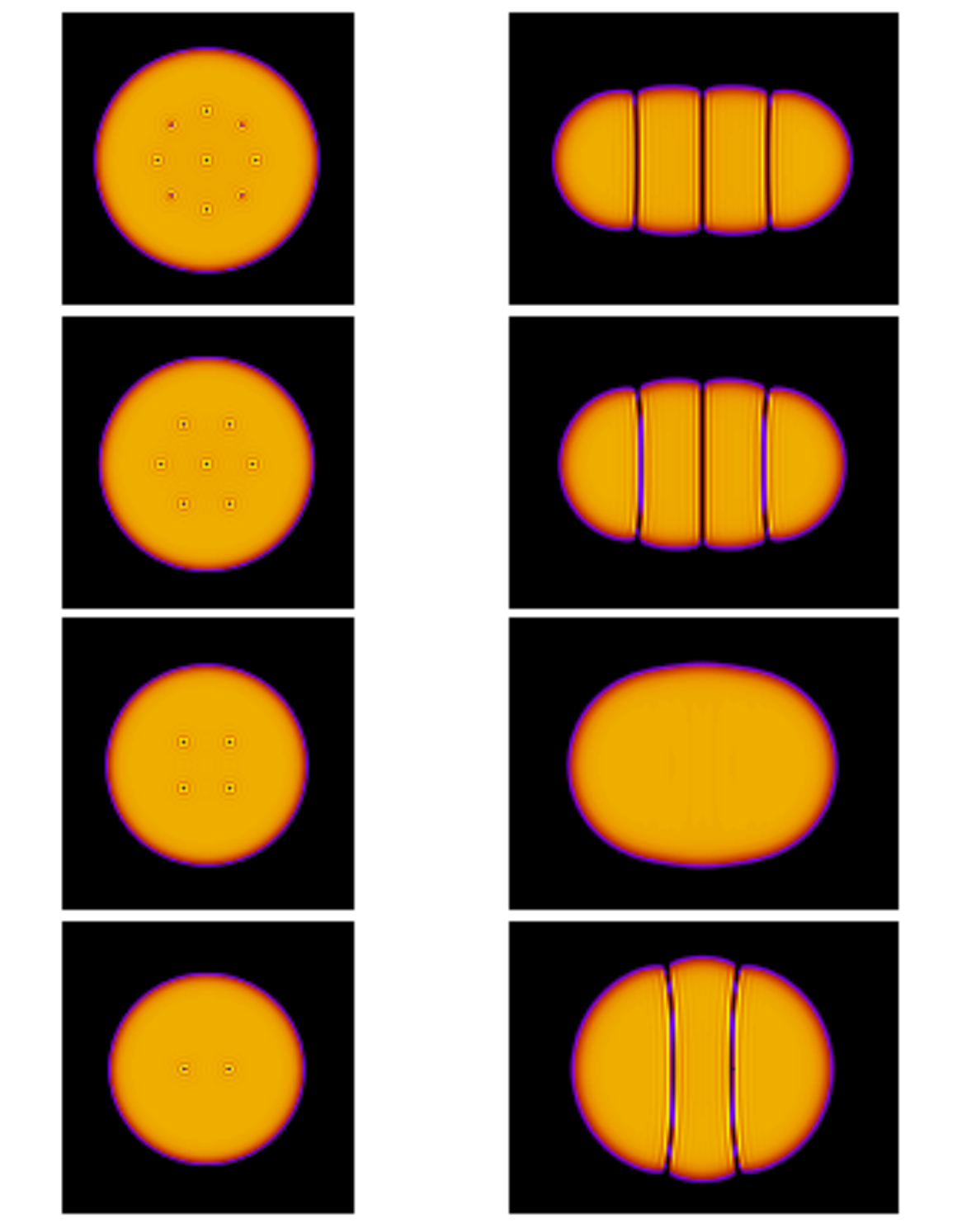}}
\caption{\label{fig33}
From bottom to top,  helium droplet configurations hosting  $n_v=$ 2, 4, 7 and 9 vortex lines. The left column shows the helium density on the $z=0$ 
symmetry plane (top view) and the right column on the $x=0$ plane (side view).\cite{Anc15}}
\end{center}
\end{figure}

The appearance of vortex arrays in rotating $^4$He nanodroplets at  $T=0$  was recently investigated by DFT.\cite{Anc15} The results
were compared with the theory developed for rotating classical fluid spheres, which was earlier used to analyze the shape and vorticity in helium droplet 
experiments.\cite{Gom14} In agreement with the experimental data, the droplets remain stable well above the stability limit predicted by classical theories despite 
their large shape deformations due to rotation.\cite{Cha65,Bro80} Vorticity inside the droplets changes their appearance from ellipsoidal to oblate and `wheel'-shaped with 
small and large vortex densities, respectively. In agreement with the experiments, the latter shape exhibits nearly flat upper and lower surfaces. Selected vortex array 
configurations for a $N_4=15000$ droplet are shown in Fig. \ref{fig33}.

The above results can be compared with the experimental data obtained for much larger droplets once they are scaled by a dimensionless characteristic rotational velocity $\Omega$ 
\begin{equation}
\Omega = \sqrt{\frac{m_4 \rho_0 R^3}{8\, \gamma}} \, \omega 
\label{eq76}
\end{equation}
where $\gamma= 0.274$ K \AA$^{-2}$ is the surface tension of the liquid. 
For a $N_4=15000$ droplet, $\Omega = 1$ corresponds to $\omega = 1.13 \times 10^{10}$ s$^{-1}$.

\begin{figure}[t]
\vspace{42pt}
\begin{center}
\resizebox*{10cm}{!}{\includegraphics{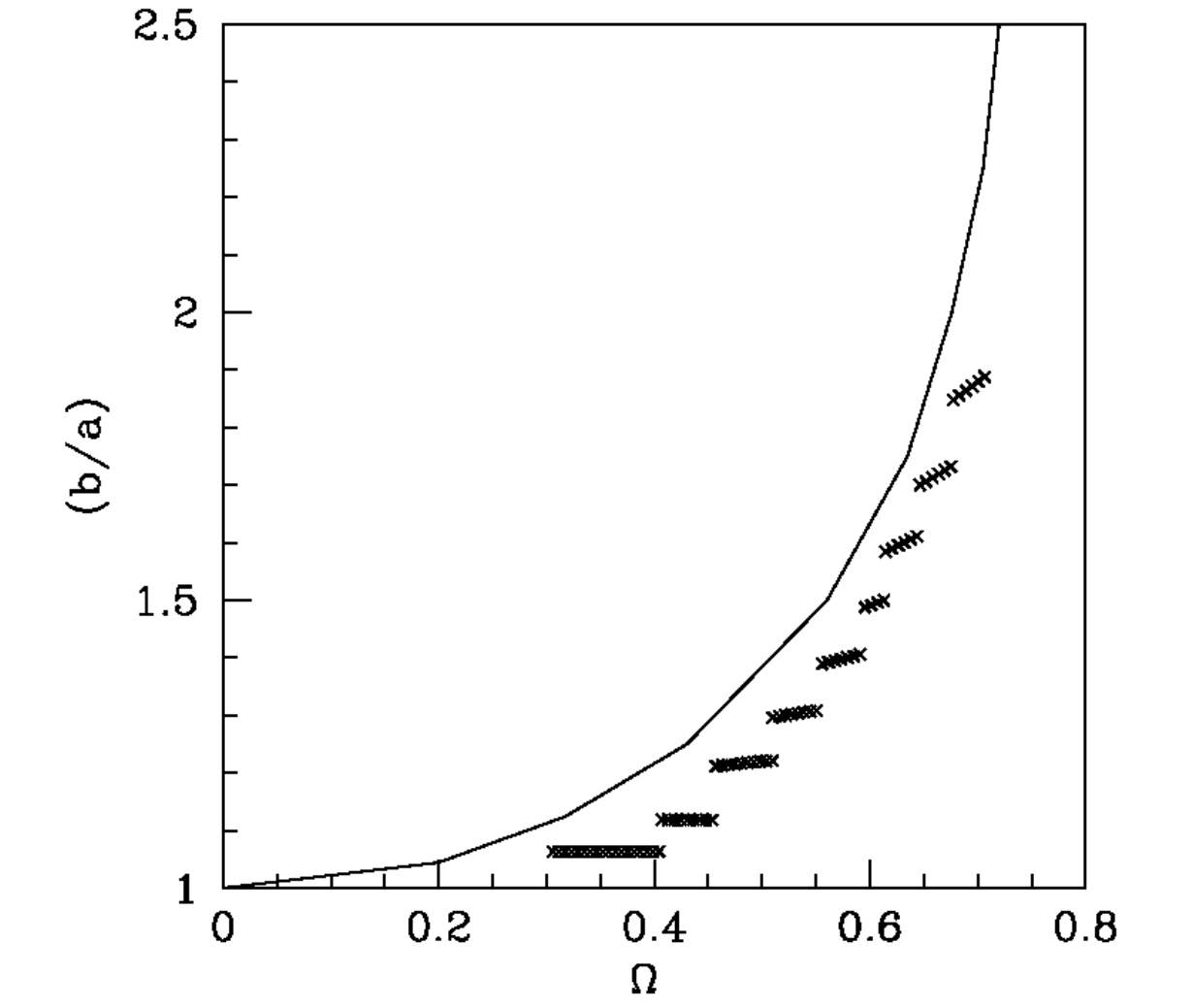}}
\caption{\label{fig34}
Aspect ratio $b/a$ as a function of the dimensionless angular velocity $\Omega$ [defined in Eq.~(\ref{eq76})] for a $N_4=15000$ droplet. The solid line shows the experimental
 curve obtained by using the classical model. Discontinuities in $b/a$ vs. $\Omega$ appear at the values  corresponding to phase transitions between 
 configurations with different number of vortices (from 1 to 9).\cite{Anc15}
}
\end{center}
\end{figure}

The data shown in Fig. \ref{fig33} demonstrate that the rotating droplet aspect ratio --  defined as $b/a$ where $a$ is the short half-axis and $b$ the long half-axis length --
depends on the angular frequency.   The calculated $b/a$ vs. $\Omega$ is plotted in Fig. \ref{fig34} 
 together with the classical model prediction, which was used in Ref. \onlinecite{Gom14} to interpret the experimental observations. Notice the gaps that appear in the calculated data, 
 which reflect the presence of forbidden values of the angular momentum per atom. Similar gaps have also been observed in calculations modelling trapped rotating 
 BEC.\cite{But99} Despite the apparent differences between rotating classical and superfluid droplets, the relationship between the aspect ratio and the angular 
 frequency looks very similar. The classical model underestimates the angular frequency only by less than 10\% for large vortex arrays. 

The experimental diffraction images of Xe doped He droplets ($\sim$200 nm diameter) have revealed configurations made of symmetrically arranged 
vortex arrays that are decorated with Xe clusters at unexpectedly large distances from the centre of the droplet.\cite{Jon16} These observations have been 
explained in terms of  angular momentum conservation. When the Xe atoms are drawn to the vortex cores, they start rotating along with the vortex array. 
The increased moment of inertia due to the additional Xe atoms must decrease the rotational angular velocity of the vortex array causing it to expand such that the 
inter-vortex distances increase. Note that the DFT results discussed above refer to pure helium droplets. Results for vortex arrays in helium nanocylinders that are 
decorated with Xe atoms are discussed in Ref. \onlinecite{Anc14}.

The above system was recently modelled by DFT using Ar doped helium droplets because the calculations are technically simpler than for Xe.\cite{Cop17} 
The system considered consisted of an array of six vortex lines filled with Ar atoms 
inside a $N_4 = 15000$ droplet ($\sim$11 nm diameter). 
In qualitative agreement with the experimental observations, 
the calculations show that the doping of the vortex cores 
substantially increases the rigidity of the system. 
This, in turn, makes the droplet stable at lower angular 
velocities and increases the inter-vortex distances. 
In contrast, a pure helium droplet with the same vortex array 
would have been unstable and the vortices would have been expelled off the droplet.
Moreover, the solvation potential effect  -- which tends to attract the Ar atoms
towards the center of the droplet -- becomes
apparent since, below some critical
value of the angular velocity, the vortices
cease to displace towards the surface and the
system reaches an equilibrium maximum distance
of the vortices from the droplet centre.

\subsection{Dynamics of alkali atoms excited on the surface of $^4$He droplets}
\label{5.14}

In a quest to understand how chemical reactions proceed in ultra-cold helium droplets, real time dynamics of photoexcited and photoionised atoms and molecules 
have been studied extensively by TDDFT. 
These processes share some elements with condensed phase chemical reactions, namely the dynamic liquid rearrangement and strong coupling of the electronic degrees 
of freedom to the surrounding liquid.

Photodissociation of Cl$_2$ and the following relaxation dynamics in $^4$He droplets has been recently addressed.\cite{Vil15a,Vil16a,Vil15b} 
These studies constituted the first application of TDDFT to describe photodissociation of a homonuclear diatomic molecule embedded in superfluid helium. 
Related  processes experimentally studied include photodissociation of alkyl iodides\cite{Bra07a,Bra07b,Bra07c} and Cr$_2$ molecules.\cite{Kau15} 
From the theory point of view, a major technical problem in modelling such systems is that a large amount of energy is deposited into the liquid. Unless the number 
of atoms in the droplet is very large (millions of atoms), the droplet is expected to disintegrate on a sub-picosecond time scale. Since large helium droplets can be 
approximated by the bulk liquid, theoretical calculations could be carried out in the bulk to avoid helium evaporation from the droplet surface. Irrespective of the geometry of the helium sample, 
these calculations require accurate dimer/molecule-helium interaction potentials, which poseses a challenge even to modern electronic structure methods.

The dynamics following photoexcitation of alkali atoms attached to helium droplets has been investigated in a series of joint experimental and theoretical 
works.\cite{Her12a,Van14} Photoelectron spectroscopy revealed that, upon excitation from the ground to the first excited s-state, alkali atoms desorb from
 the droplet surface. The mean kinetic energy of these atoms, which can be detected by ion imaging, shows a linear dependence on the excitation energy. 
 TDDFT calculations on these systems revealed that the desorption process is accompanied by the creation of highly non-linear liquid density waves in the droplet 
 that propagate at supersonic velocities.

\begin{figure}[t]
\vspace{42pt}
\begin{center}
\resizebox*{11cm}{!}{\includegraphics{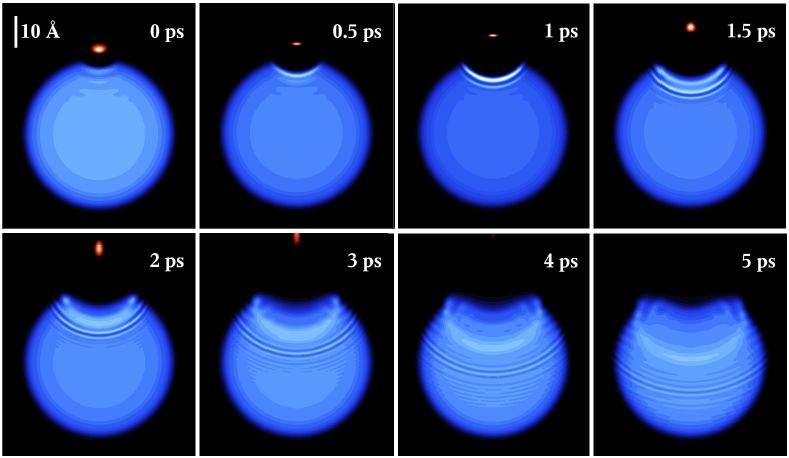}}
\caption{\label{fig35}
Time evolution of  the Na@$^4$He$_{1000}$ complex upon 4s $\leftarrow$ 3s excitation.\cite{Her12a}}
\end{center}
\end{figure}

The  test-particle method described in Sec. \ref{4.2} was introduced in the context of optical excitation of Li and Na atoms -- sitting on the surface of helium droplets   -- 
to their first excited  s-state.\cite{Her12a} The ejected alkali metal atoms acquire high  velocities, which makes the direct 
numerical solution of the Schr\"odinger equation difficult in the TTDFT context. An overview of the time evolution of the Na@$^4$He$_{1000}$ 
complex following the 4s $\leftarrow$ 3s excitation is shown in Fig. \ref{fig35}.\cite{Her12a} The sudden repulsive interaction between the excited Na 
atom and the droplet creates a series of supersonic shock waves in the droplet, which indicates that a significant fraction of the energy introduced
 by the optical excitation is transferred directly into the droplet. These  waves were observed to travel at velocities ranging from 370 to 890 m/s, 
 which are of similar magnitude as recently observed in laser ablation experiments in the bulk  liquid.\cite{Gar16} Furthermore, the moving high density peak,
  which originates from the first solvation shell, exhibits (bright) solitonic features such as constant propagation velocity (\textit{ca.} 590 m/s) and no spatial dispersion. 
  A similar result was found for the ejection of Rb and Cs atoms from helium droplets when excited from 5s to 6s and from 6s to 7s states, respectively.\cite{Van14} 

\begin{figure}[t]
\vspace{42pt}
\begin{center}
\resizebox*{9cm}{!}{\includegraphics{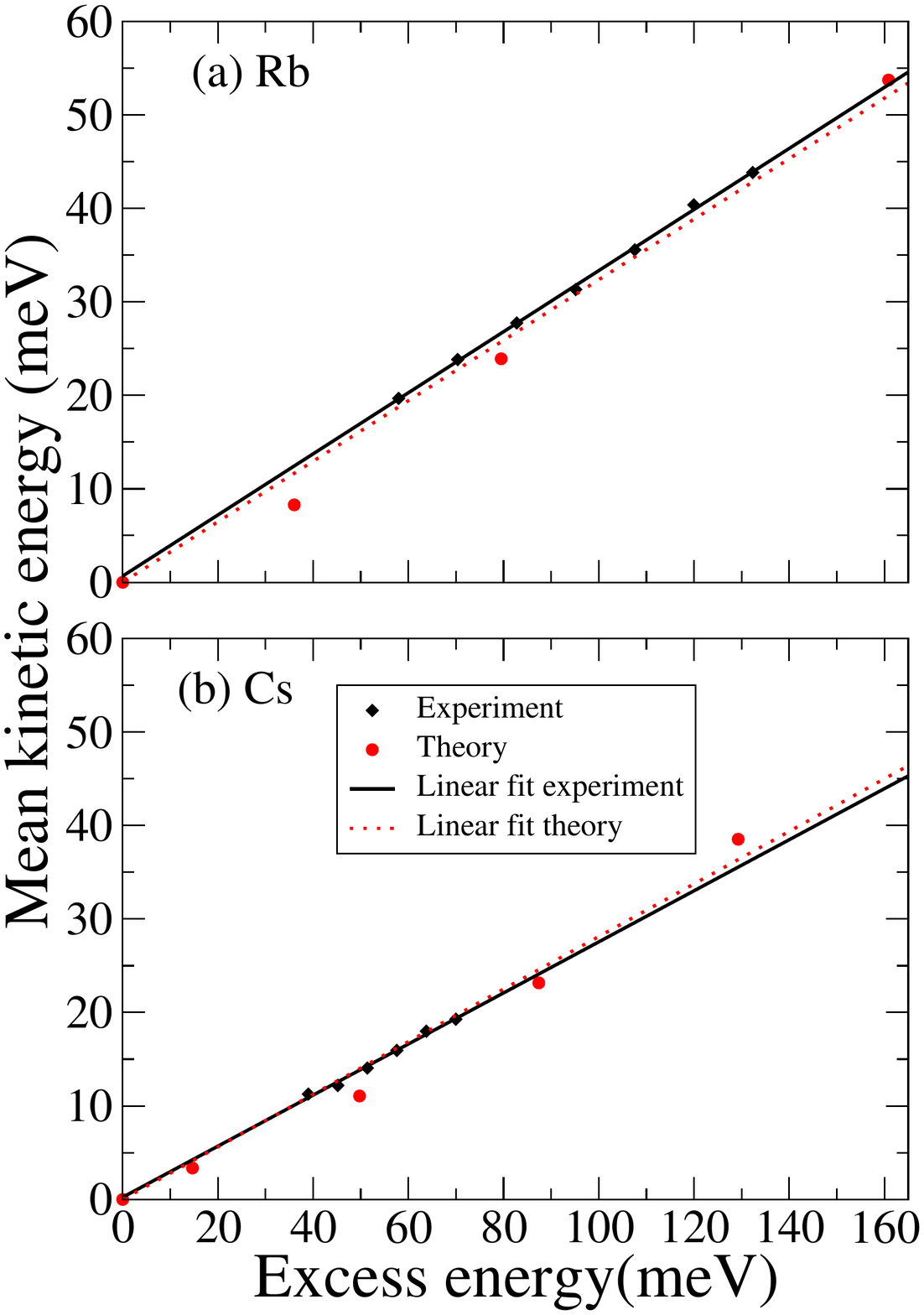}}
\caption{\label{fig36}
Mean kinetic energies of Rb (a) and Cs (b) atoms desorbing from helium droplets upon  6s$\Sigma$ and 7s$\Sigma$ state excitation, respectively. Straight and dotted 
lines: linear fits to the experimental and theoretical data, respectively.\cite{Van14}
}
\end{center}
\end{figure}

Detailed information about the kinematics of the process can be obtained by monitoring the kinetic energy of the atoms desorbed from the droplets 
as a function of the excitation energy.\cite{Her12a,Van14} The experimental results for Cs and Rb\cite{Van14} are shown in Fig. \ref{fig36}. The calculated 
points in that figure were obtained by starting TDDFT simulations from various impurity positions determined by a constrained minimization of the
 total energy of the complex. Whereas the light alkali metal atoms (Li and Na) require the use of the test-particle method, heavy alkali metals (Rb and Cs) can be treated classically.

For all alkali metals, the kinetic energy of the ejected atom exhibits a linear dependence on the excess excitation energy of the $(n+1)$s $\leftarrow$ $n$s transition. This indicates that, despite its apparent complexity, the ejection process can be well represented by a `pseudo-diatomic model'\cite{Bus72} in which the droplet is represented by one big atom bound to the alkali. By imposing energy and linear momentum conservation during the instantaneous ejection of the alkali atom from the droplet, the relative kinetic energy can be written as
\begin{equation}
E_{kin} =\eta (\hbar\omega-\hbar\omega_0) \; ,
\label{eq77}
\end{equation}
where $\omega$ denotes the excitation and $\omega_0$ the atomic transition frequencies. The slope $\eta$ is related to the effective mass of the helium droplet  
(the mass of the helium atoms effectively participating in the interaction with the alkali atom), $m_{\mathrm{eff}}$, by
\begin{equation}
\eta=  \frac{m_{\mathrm{eff}}}{m_{\mathrm{eff}}+m_{\mathrm{Ak}}}
\Longrightarrow
m_{\mathrm{eff}}=\frac{\eta}{1-\eta}  \, m_{\mathrm{Ak}} \; .
\label{eq78}
\end{equation}
Fitting the experimental and simulation data to Eq. (\ref{eq77}) yields the results summarised in Table \ref{table4}. It can be  seen that $m_{\mathrm{eff}}$ increases with the mass
 of the alkali atom as indicated by Eq. (\ref{eq78}). The variation of the corresponding number of helium atoms  reflect the differences in the dimple structure and
  the excited state interaction with the droplet.\cite{Van14} Note the lack of data for the 5s $\leftarrow$ 4s transition for the K atom. While there is no difficulty in simulating it with 
  TDDFT, this transition may overlap with 3d $\leftarrow$ 4s and this complicates the analysis of the experimental results.

\begin{table}[t]
\vspace{0.1cm}
{\begin{tabular}{cccccc}
\hline\hline
  Ak    & m$_{Ak}$  (exp) & $\eta$ (exp)         & $\eta$ (th)        & m$_{eff}$ (exp) & m$_{eff}$ (th)  \\
             & [amu]  &         &         & [amu] & [amu]  \\       
\hline
 Li     &   6.94 & 0.687 & 0.756 & 15.2     & 21.5  \\
   Na & 23.0 & 0.516 & 0.583 & 24.6        & 32.2  \\
   Rb & 85.5 & 0.327& 0.324 & 41.9        & 41.0  \\
   Cs & 132.9 & 0.281 & 0.273 & 51.8        & 50.5  \\
   \hline\hline
  \end{tabular}}
   \caption {\label{table4}
Some characteristics of the experimental and theoretical kinetic energy distributions of the desorbed alkali atoms;\cite{Her12a,Van14} see text for details.
}  
\end{table}

In addition to the  $(n+1)$s $\leftarrow$ $n$s transitions  discussed above (see also Refs. \onlinecite{Log11a,Log11b,Pif10,Lac11}), 
the lower energy $n$p $\leftarrow$ $n$s transitions have also been addressed in
 a series of experimental and theoretical studies.\cite{Sti96,Reh97,Bue07,Aub08,Lei08,Her10,Nak01} The first expectation was that alkali metal atoms would always detach from 
 the droplets upon $n$p $\leftarrow$ $n$s excitation. However, only the light alkali metals such as Li, Na, and K 
appear to detach;  the heavier alkalis (Rb and Cs) may remain attached  if they are excited with energies close to that of the D1 line in the gas-phase.\cite{Aub08,The11}
  
Photoexcitation and photoionisation of Rb atoms attached to helium droplets has been studied in real-time dynamics experiments.\cite{Van15}
It was shown that excitation of Rb atoms from the 5s to the 6p states leads to their detachment. 
Upon subsequent ionisation of the  excited Rb atom (Rb$^*$), the interaction with helium becomes attractive.
Hence, depending on the  time delay $\tau_D$ between the excitation and ionisation laser pulses, the resulting ion may be ejected as a bare  Rb$^+$ cation
 or as a Rb$^+$He$_n$ complex, or it can be drawn into the droplet. The critical time $\tau_c$ separating these processes is called the fall-back time.\cite{Van15} 

Recently, this study has been extended to the Rb 5p $\leftarrow$ 5s transition.\cite{Van17} 
Both transitions have been simulated with TDDFT as described in Sec. \ref{4.1}. The Rb atom located on the droplet surface is first photoexcited to 
either the 5p or the 6p state ($^2\Sigma_{1/2}$,  $^2\Pi_{1/2}$ or $^2\Pi_{1/2}$) and the system evolves on the excited state potential energy surface. 
After a fixed delay, Rb$^*$ is photoionised, which is simulated by suddenly switching the interaction potential to Rb$^+$-He. It was found that the desorption 
process for the 6p state is impulsive whereas the behavior of the 5p state is intermediate between impulsive and complex dissociation. The desorption time 
scales are also very different for the two states: $\sim 1$ ps for 6p and $\sim 100$ ps for 5p.

\begin{figure}[t]
\vspace{42pt}
\begin{center}
\resizebox*{10cm}{!}{\includegraphics{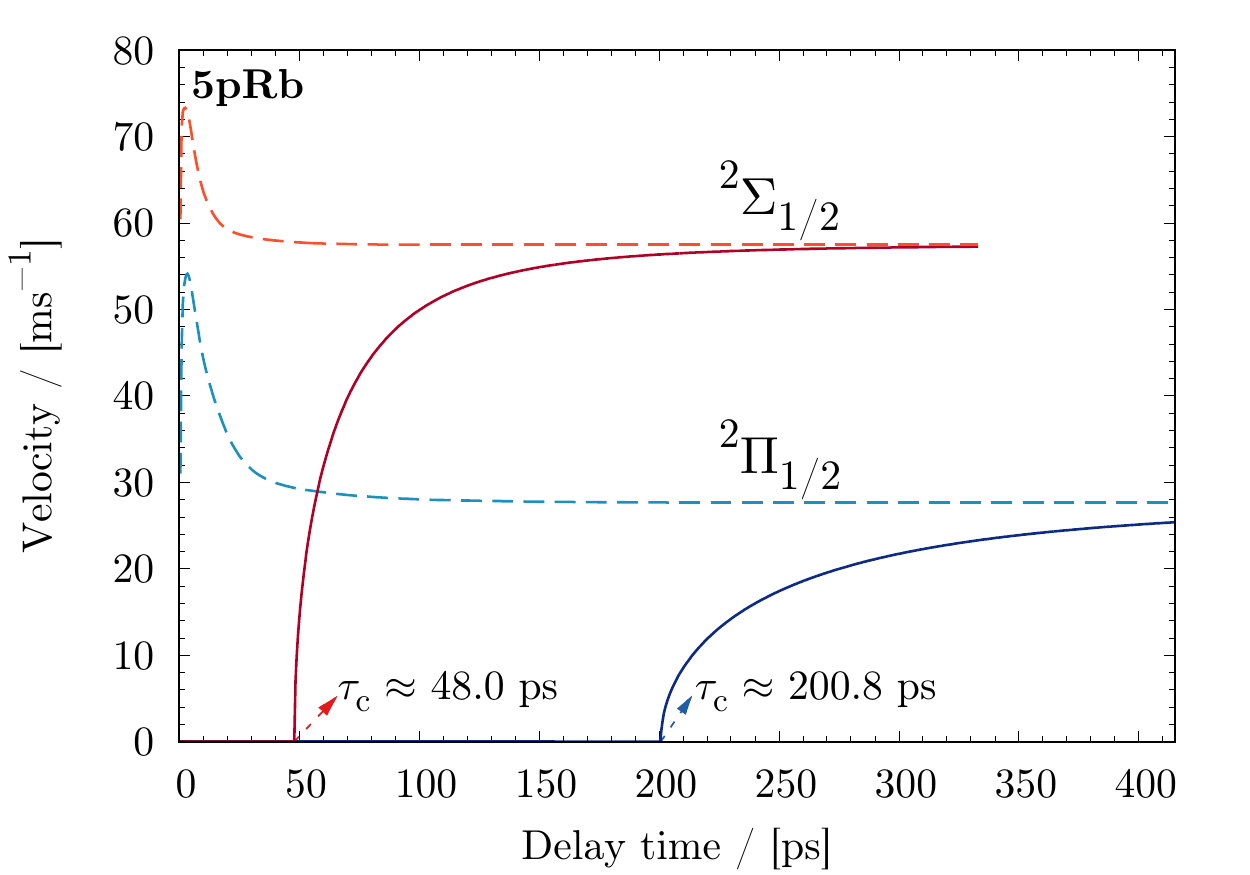}}
\caption{\label{fig37}
Velocity of the Rb$^*$ atom in the 5p\,$^2\Sigma_{1/2}$ and 5p\,$^2\Pi_{1/2}$ states (dashed lines) and  the Rb$^+$ cation produced by 
 photoionisation  (solid lines) after a given delay time vs. delay time. The fall-back times $\tau_c$ are indicated by arrows.}
\end{center}
\end{figure}

The velocities of Rb$^*$ and the Rb$^+$ as a function of the delay time are shown in Fig. \ref{fig37} for the 5p $\leftarrow$ 5s transition. Note that the 
largest (asymptotic) velocity of Rb$^+$ is simply given by the corresponding asymptotic velocity of Rb$^*$ because 
the ion escape velocity is not affected by the droplet at large distances.
TDDFT simulations have also been extended to the desorption of Cs following the 6p $\leftarrow$ 6s excitation.\cite{Cop17b} The general features of the dynamics 
appear very similar to Rb 5p $\leftarrow$ 5s. 

The TDDFT simulations for Rb and Cs atoms excited from their $n$s ground to their $n$p excited state 
can be summarised  as follows: i) excitation to the $^2\Sigma_{1/2}$ or $^2\Pi_{1/2}$ state leads to desorption of the excited alkali atom (Ak$^*$); ii) excitation to 
the $^2\Pi_{3/2}$ state produces an exciplex within $\sim$ 10 ps, which remains attached to the droplet surface.

Experiments and TDDFT calculations agree on the ejection of Ak$^*$ $^2\Sigma_{1/2}$ state from helium droplets. At first sight, the results for $^2\Pi_{3/2}$ and $^2\Pi_{1/2}$ 
states seem to disagree with the experiments,\cite{Aub08,The10,The11} but this may be explained as follows.\cite{Bar17} 

$\bullet$
In the case of the $^2\Pi_{1/2}$ state, the
 experiments have explored the low energy region of the D1 line whereas the TDDFT dynamics was initiated using a configuration that corresponds to the D1 resonance. 
Hence, the initial energy in the TDDFT simulation is larger than in experiments. 
 
$\bullet$ 
In the case of the $^2\Pi_{3/2}$ state, it is worth stressing that the electronic state of the ejected Ak$^*$ was not determined in the experiments. 
Only the state to which the Ak atom was excited is determined (by the excitation laser wavelength). 
In the case of Rb$^*$, experimental indications point to a non-radiative relaxation of the 5p$^2\Pi_{3/2}$ to the 5p$^2\Pi_{1/2}$ state.\cite{Van17} 
If this happens, the relaxed state may be a RbHe $^2\Pi_{1/2}$ exciplex. The energy available from this relaxation process is about 190 cm$^{-1}$. 
If even only one third of this energy is given to the RbHe ${}^2\Pi_{1/2}$ exciplex as additional initial kinetic energy, it is ejected from the droplet according to the TDDFT simulations. 
This process is compatible with the possibility of forming exciplexes in the ${}^2\Pi_{3/2}$ state which could remain attached to the droplet\cite{Gie12}
if no electronic relaxation occurs.
Note that no exciplex was produced upon direct ${}^2\Pi_{1/2}$ excitation because of a barrier preventing its formation.
A similar  relaxation process is also fully compatible with the experimental observations for Ag.\cite{Mat13b}

\begin{figure}[t]
\vspace{42pt}
\begin{center}
\resizebox*{8cm}{!}{\includegraphics{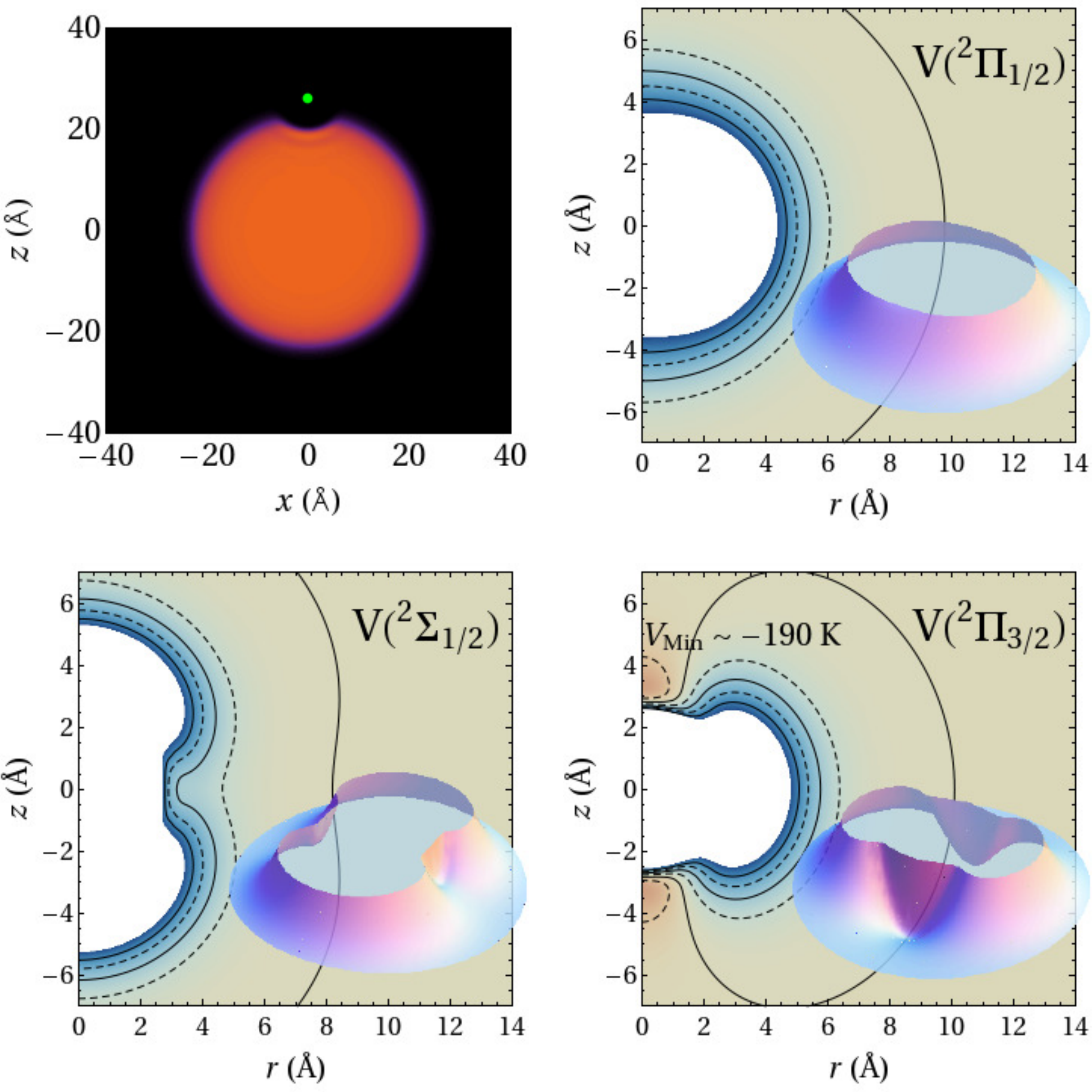}}
\caption{\label{fig38}
Top left panel: Equilibrium dimple configuration of Rb taken as starting point for the dynamics; the Rb atom is at 26.3 \AA{} from the COM of the droplet.
The other panels display the $^2$P $V_{\lambda}$ 5p Rb-He  potentials (Eq. (\ref{eq44}))
corresponding to this configuration. 
Regions where the potentials are attractive (repulsive) are represented in brown (blue).
The outermost equidensity line corresponds to zero potential.
The 5p\,$^2\Sigma_{1/2}$ and 5p\,$^2\Pi_{1/2}$ potentials have a shallow attractive minimum of about 1 K depth at 
a distance of $\sim $10 \AA{}.
}
\end{center}
\end{figure}

In a related study, DMC calculations have been carried out for Rb$^*$ in a small cluster or on a helium film, which can be considered as a simplified model of a 
large droplet surface.\cite{Lei08} This study found that Rb$^*$ stabilizes as a weakly bound metastable Rb $^2\Pi_{1/2}$  (not an exciplex) that forms a shallow dimple structure 
on the surface. There is no contradiction between the DMC and the {\it dynamic} TDDFT calculations in this respect: The DIM potentials indeed display a shallow minimum for this state
so that the imaginary-time DFT relaxation would also yield a weakly bound Rb$^*$. However, the energy available  in the real time dynamics  hinders the formation of this 
relaxed, weakly bound state. The same group later investigated Rb excited to the $^2\Pi_{3/2}$ state,\cite{Lei11} showing the appearance of a linear exciplex to which 
more helium atoms are attached, preferentially on one end of the linear exciplex. The TDDFT simulations are in agreement with these findings that, to a large extent, can be understood
by inspecting  the $V_{\lambda}$ DIM potentials, Eq. (\ref{eq44}), as plotted in Fig. \ref{fig38}.  Notice for example the appearance of two deep wells in the $^2\Pi_{3/2}$ potential; the 
 filling of these wells by helium atoms yields the Rb$_2$$^2\Pi_{3/2}$ linear exciplex.

\begin{figure}[t]
\vspace{42pt}
\begin{center}
\subfigure{
\resizebox*{5cm}{!}{\includegraphics{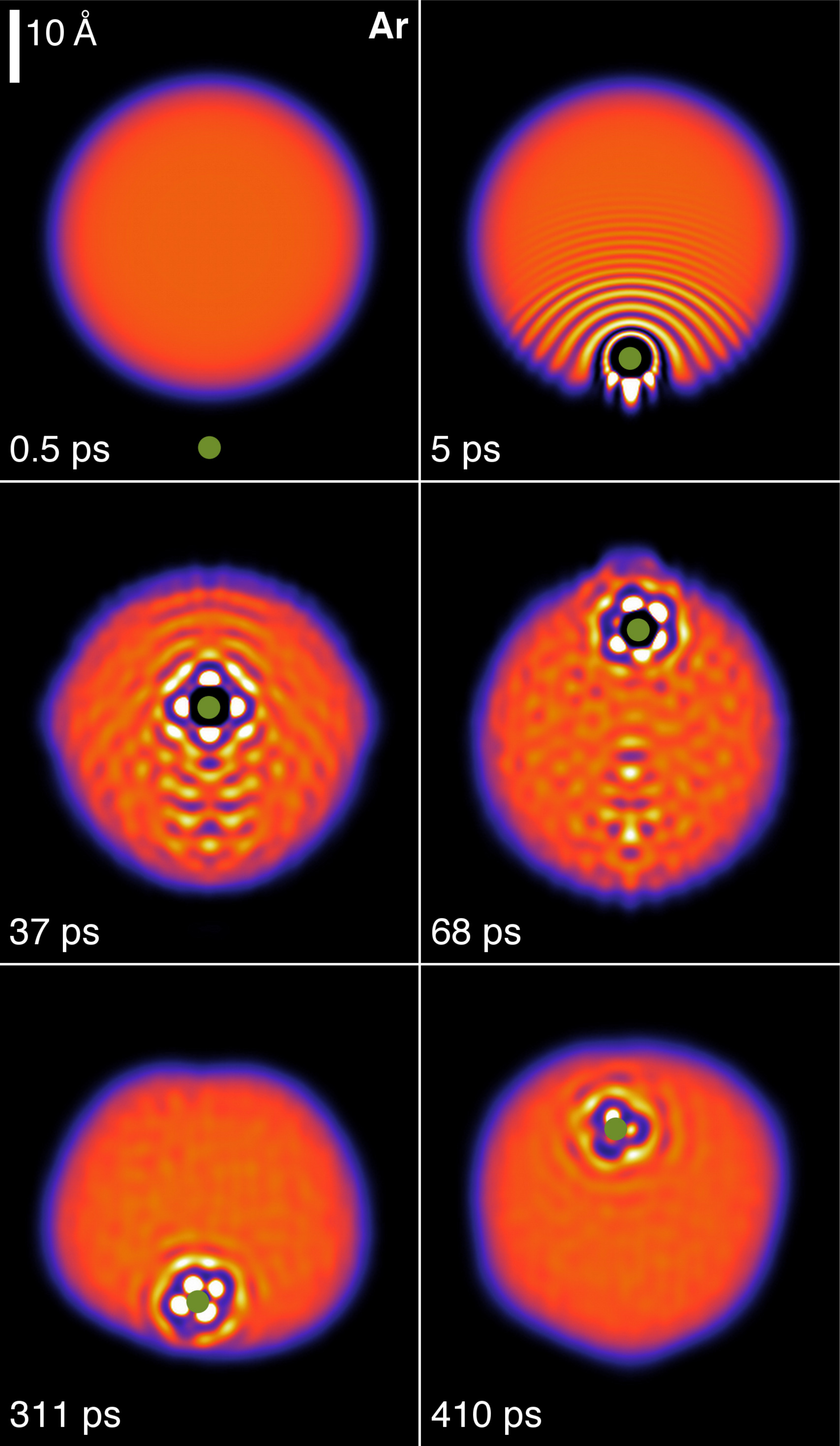}}}%
\subfigure{
\resizebox*{5cm}{!}{\includegraphics{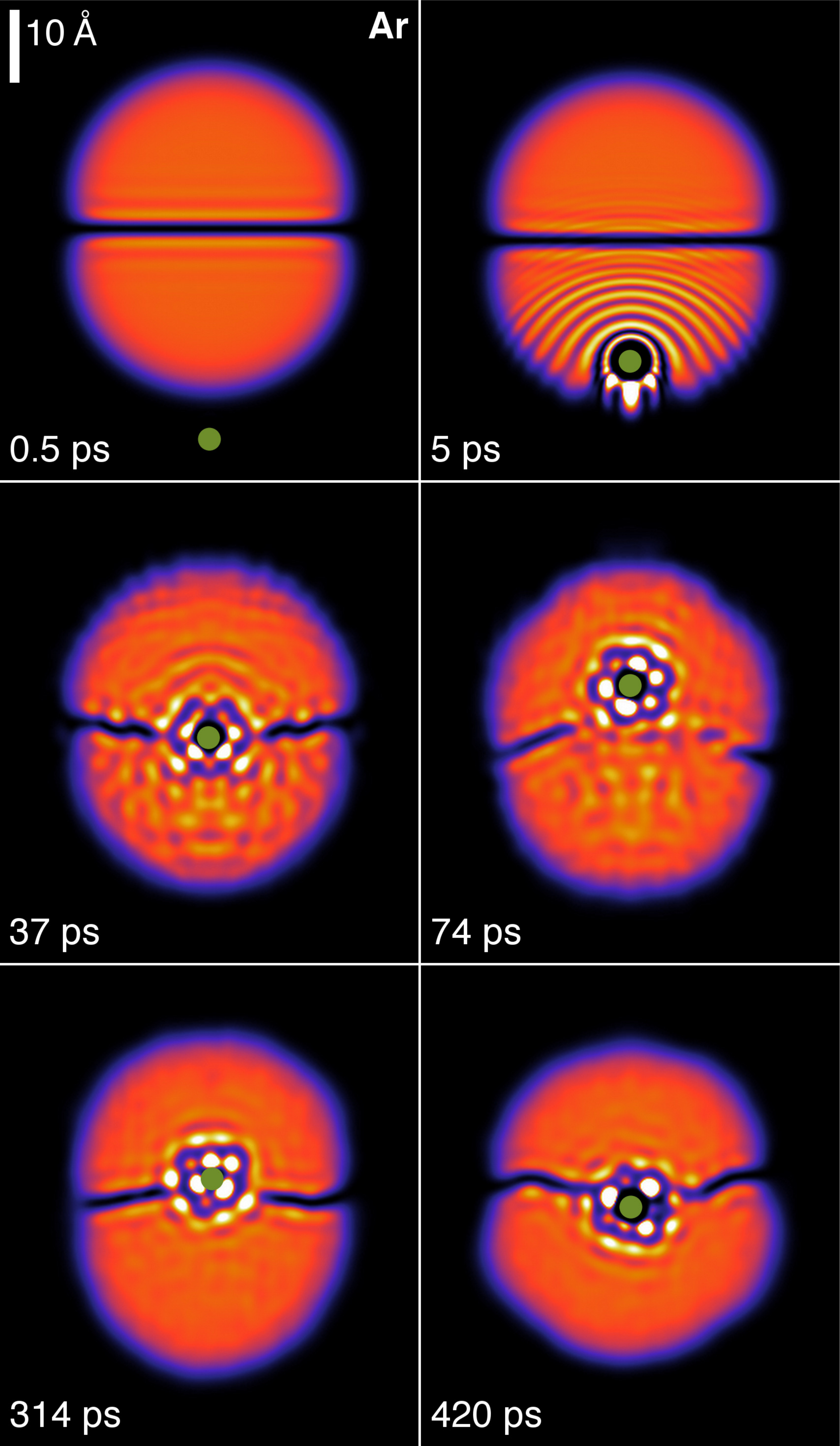}}}%
\caption{\label{fig39}
(a) Simulation of an  Ar atom (green dot) approaching a $^4$He$_{1000}$ droplet from below at $v_0 = 360$ m/s. The corresponding time is indicated in each frame. 
(b) Same as (a) but the droplet hosts a vortex line. 
We have included mp4 movies as Supplemental Material (on line) that show the complete simulations corresponding to this figure.
}
\end{center}
\end{figure}

\begin{figure}[t]
\vspace{42pt}
\begin{center}
\resizebox*{9cm}{!}{\includegraphics{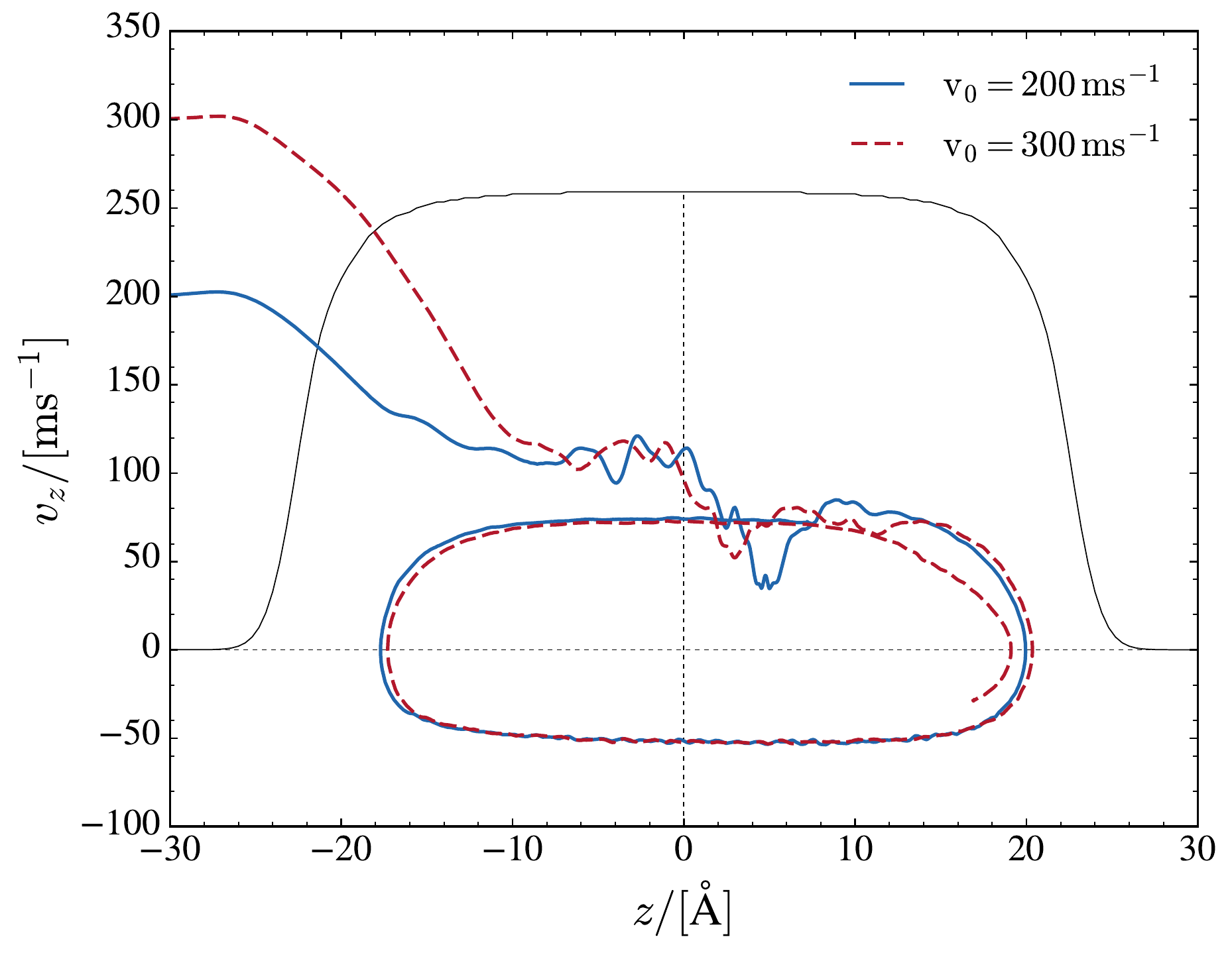}}
\caption{\label{fig40}
Phase space evolution of the Xe atom for two values of the initial velocity $v_0$ (200 and 300 m/s) during a head-on collision with a 
$^4$He$_{1000}$ droplet. The initial droplet density profile is also shown in arbitrary density scale.
}
\end{center}
\end{figure}

\begin{figure}[t]
\vspace{42pt}
\begin{center}
\resizebox*{10cm}{!}{\includegraphics{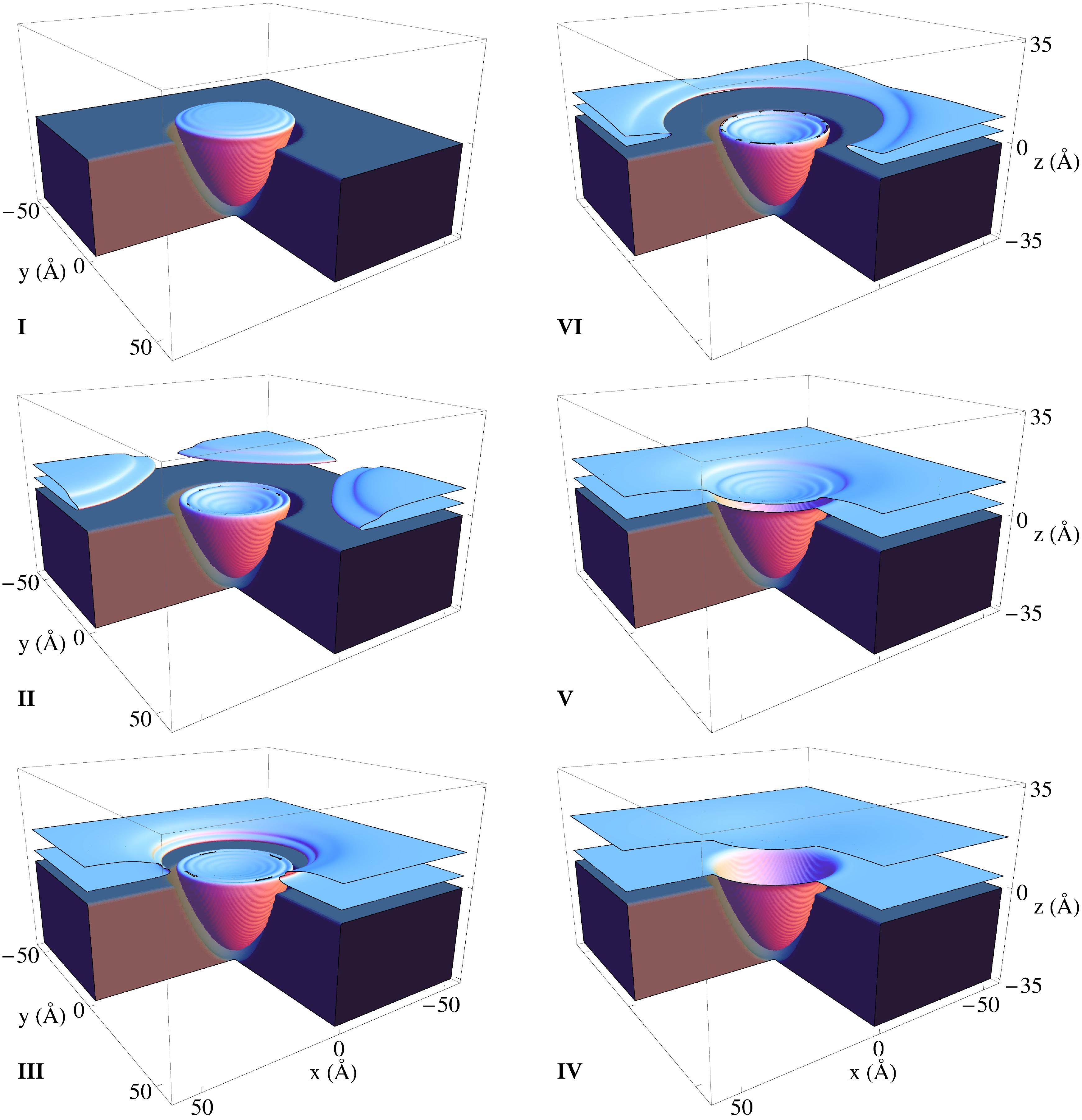}}
\caption{\label{fig41}
Liquid $^4$He  on a Na patterned surface.\cite{Anc09b}
The panels show some illustrative equilibrium
configurations for  different coverages;
the panels display the isodensity surfaces drawn at a value which is
half the bulk liquid density at $T=0$ ($\rho_0= 0.0218$ \AA$^{-3}$).
The dark area represents the Na planar surface. 
} 
\end{center}
\end{figure}

\subsection{Capture of impurities by $^4$He droplets}
\label{5.15}

The previous sections frequently consider situations that appear after an impurity has been captured by a helium droplet or after it has been injected into liquid helium.
 It is well known that helium droplets are able to capture atoms and molecules, as first shown for Ne atoms.\cite{Sch90} This finding has had a tremendous impact 
 on low temperature experiments as the technique allows to use helium droplets as an ultra-cold matrix.\cite{Toe04}

The pickup process of gas phase Ar, Kr, and Xe atoms by $^4$He$_N$ droplets ($N_4 > 10^3$ atoms) produced in nozzle-beam expansions was first
 studied  by Toennies and co-workers.\cite{Lew95} These experiments characterised the deflection of a helium droplet beam 
 by a secondary crossed beam made of rare gas atoms. This cross-beam technique was later used to characterize the helium droplet density in the beam 
 by comparing the measured integral cross-section with the helium droplet density profiles predicted by DFT calculations.\cite{Har98,Har01}

Theoretical work addressing the capture process is scarce. The earliest  work on the scattering of $^4$He  atoms by $^4$He 
droplets was largely inspired by the nuclear physics optical model.\cite{Eic88} More recently, scattering of helium atoms from inhomogeneous quantum 
liquids has also  been considered.\cite{Kro07,Kro08}

\subsubsection{Pure droplets}
\label{5.15.1}

The first TDDFT calculations modelling the capture of impurities by helium droplets were recently carried out for heliophobic (Cs)\cite{Lea14a} and 
several heliophilic atoms (Ne, Xe and Ar).\cite{Vil16b,Cop16,Cop17} The heavy impurities were treated classically whereas the lighter Ne was treated quantum mechanically; 
however, the collision process considered was strictly 1D (see Sec. \ref{4.1}).

Depending on the energy and the impact parameter of the impinging atom, a rich variety of dynamical phenomena may be observed.\cite{Lea14a,Cop17} 
DFT calculations have shown that for a Cs atom  to be trapped on the droplet surface, its excess kinetic energy must be transferred to the droplet very efficiently 
because the Cs-droplet binding energy is only 10.5 K.  In a head-on collision
with a heliophilic Xe atom, whose binding energy is 316.3 K,  if the impact velocity is sufficiently high ($v_0 > 600$ m/s), the Xe atom may pass through the droplet;\cite{Cop16}
 otherwise it remains trapped inside. As an example, 
Fig. \ref{fig39} shows the head-on collision of an Ar atom at $v_0 = 360$ m/s with a $^4$He$_{1000}$ droplet. 

Most of the excess kinetic energy  of the impurity is deposited into the droplet, which results in the ejection of He atoms and the emission of sound and shock waves. 
Contrary  to the naive expectation that the average energy per ejected He atom simply corresponds to its binding energy ($\sim$ 7 K),
the atoms ejected at early times (prompt-emitted atoms)  carry significant amounts of kinetic energy -- see Table \ref{table5}. Whether the impurity is heliophilic or 
heliophobic plays a role in the process. For example, for heliophilic Xe,  18 He atoms are ejected after 200 ps for $v_0$ = 200 m/s  
  whereas only 6 He atoms are ejected in the case of heliophobic  Cs during the same period of time.\cite{Lea14a}

\begin{table}[!]
\vspace{0.1 cm}
\begin{tabular}{c c cc c c  c}
\hline 
\hline
Species &\hspace{0.5 cm} & $v_0$ (m/s) &\hspace{0.5 cm}  & $N_e$ & \hspace{0.5 cm} & $E_e$  (K) \\
\hline
Xe  & &200 &  &18  & &19    \\
   && 300&  &28  &  &23   \\
   & &400 & &37 & &30 \\
   \hline
   Ar& & 360& & 16 & & 22 \\
\hline 
\hline
\end{tabular}
\caption{\label{table5}
Number of  ejected helium atoms ($N_e$)  and average energy per ejected atom ($E_e$) 
 for  the indicated head-on collisions  during the first 200 ps.\cite{Cop17}
}
\end{table}

For an impurity to be captured by a helium droplet, the excess kinetic energy of the impurity must be dissipated such that it becomes less than the impurity-droplet 
binding energy.   Fig. \ref{fig40} shows the
trajectory in phase space of a  Xe atom captured in a $^4$He$_{1000}$ droplet for a head-on collision at $v_0 = 200$ and 300 m/s.\cite{Cop17} It can be seen that  for these collisions 
-- corresponding to thermal velocities -- the motion of the impurity inside the droplet is independent of $v_0$ to some extent. We attribute this to the fact that 
dissipation occurs  mostly  during the very first stages of the process.\cite{Cop16,Cop17}   

In grazing collisions not only excess energy but also angular momentum is deposited into the droplet. This allows to visualize the resulting irrotational superfluid flow inside the droplet and to 
calculate the capture cross-section. At low energies and small impact parameters, the impurity is captured by the droplet and may even orbit around the droplet COM.

A simple expression for the capture cross-section of classical dopants can be obtained provided that the reduced de Broglie wavelength of the impurity is much smaller
 than the droplet\cite{Lea14a} 
\begin{equation}
\sigma(E)= \frac{\pi}{\kappa^2 } \sum_{\ell=0}^{\ell_{cr}} (2 \ell +1)= \frac{\pi}{\kappa^2 } (\ell_{cr} +1)^2 
\label{eq79}
\end{equation}
where $E$ is the energy available  in the COM frame, $\ell_{cr}$ is the critical relative angular momentum leading to impurity capture, and $\kappa=(2 \mu E/\hbar^2)^{1/2}$ with $\mu$ being the reduced mass of the system.  For a given energy, $\ell_{cr}$ is determined by carrying out a series of simulations with varying impact parameters. This procedure was implemented for
 Cs\cite{Lea14a} and recently also for Xe.\cite{Cop17}

For a Xe atom at $v_0 = 200$ m/s, the impact parameter leading to its capture is approximately 20.5 \AA{}, which can be compared with the sharp-density radius of the $^4$He$_{1000}$ 
droplet, 22.2 \AA{}. Hence, at thermal velocities the calculated cross-section for Xe capture is  close to the geometrical cross-section of the droplet itself. 
The angular momentum of the impinging Xe at $v_0=200$ m/s with an impact parameter of 22.2 \AA{} is 917 $\hbar$. 
This collision was simulated for 200 ps\cite{Cop17} and it was found that
15 He atoms were ejected during this time period, of which 5 remained attached to the Xe atom. 
After the collision, the Xe+$^4$He$_5$   complex  carries away  522 $\hbar$ angular momentum units, while some 95 $\hbar$  units are deposited into the droplet  as vortex loops and
capillary waves. The remaining angular momentum is taken away by the  promptly emitted He atoms.
 
\subsubsection{Droplets hosting vortices}
\label{5.15.2}

Recently, an experimental technique for determining the size of large He droplets ($N_4> 10^5$) has been introduced\cite{Gom11} 
that is based on the attenuation of a continuous droplet beam through collisions with Ar atoms at room temperature. 
The pickup chamber of the droplet beam apparatus is filled with argon gas and the helium droplets are subjected to multiple isotropic collisions with Ar atoms 
on their way to the detection chamber; large helium droplets could also be doped by impurities using this approach.
The experimental situation for large superfluid He droplets is discussed in Ref. \onlinecite{Tan17}.

  This method has been instrumental for visualizing quantised vortex arrays in large helium droplets ($10^8-10^{11}$ atoms) doped with Xe atoms  and clusters.\cite{Gom14,Jon16,Ber17} 
  Although experimental data for Ar was also recorded, the analysis has been limited so far to Xe  because of the higher sensitivity in coherent 
  x-ray diffractive imaging. These experiments have motivated a series of TDDFT simulations on the impurity capture process by vortex lines at impact 
  velocities relevant to the experimental conditions.

The capture of thermal Ar and Xe atoms by a linear vortex line hosted inside a $^4$He$_{1000}$ droplet has been recently studied by TDDFT.\cite{Cop17} 
The vortex line was generated by the imprinting method described in Sec. \ref{3.2} and the perpendicular impurity-vortex impact took place on the 
equatorial plane of the droplet at 240 (Xe) or 360 m/s (Ar). In both cases, the impurity is `captured'  by the vortex line in the sense that, after a few hundred ps, it orbits around 
the vortex line and remains at a close distance from it.\cite{Psh16,Cop17} 

The right panel of Fig. \ref{fig39} shows snapshots of the collision process for Ar ($v_0 = 360 $ m/s) with a $^4$He$_{1000}$ droplet 
hosting a single vortex line. These data, together with those in the left panel, show the corresponding Ar atom trajectories with and without a vortex line. 
The first turning point of Ar in a vortex-free droplet is reached at 68 ps, and is located close to the droplet surface. The equivalent configuration in the presence 
of a vortex line is shown by the snapshot at 74 ps. The vortex-free configurations at 311 and 410 ps correspond to the 4\textsuperscript{th} and 5\textsuperscript{th} 
turning points of the Ar trajectory. Similar configurations hosting a vortex line are shown in the right panel, which demonstrates that the Ar atom trajectory
 becomes localised in the immediate neighborhood of the vortex line due to the impurity-vortex binding. Note that additional sources of dissipation 
 (e.g. viscosity), which are not included in TDDFT, may only enhance this localisation process.
 
\subsection{Liquid helium on nanostructured surfaces}
\label{5.16}

As a consequence of the extremely weak He-He interaction, it is expected that liquid helium   interacts strongly  with almost any substrate and 
wet the surface such that the vapor and the substrate are not in contact. However, since the interaction of helium  with alkali atoms is even weaker than the He-He interaction, 
they  might represent a notable exception to this rule.  Indeed,  while liquid $^3$He wets any substrate, it has been shown that
liquid $^4$He does not wet surfaces made of  heavy alkali metals such as Cs  at $T=0$.\cite{Che91,Tab92} 
 
 The wetting properties of $^4$He on the surface of heavy alkali metals have been studied in the past by using a $T$-dependent free-energy density functional, 
 which  describes the surface properties of liquid $^4$He  accurately in the  $0<T<3$ K temperature range.\cite{Anc00}
  The resulting liquid structure on the Cs surface was elucidated, providing both the $T$-dependence of the
   contact angle and the wetting temperature, which are in good agreement with experiments. 

The most recent research employing DFT to study the adsorption of 
 helium samples on various substrates has been 
reviewed in Ref.  \onlinecite{Anc09c}. The issues addressed 
in this review include the deposition and spreading of 
helium droplets on flat alkali metal surfaces; the 
determination of isotherms; the construction of the phase diagram 
of helium on such substrates; the adsorption of helium on 
spherical and cylindrical surfaces; the filling of wedges and the filling/emptying transitions at $T=0$ taking place at fixed values 
of the wedge opening angle;\cite{Her06} the filling of  
infinite polygonal pores, and the adsorption on 
planar surfaces structured with an array of parabolic nanocavities.
In particular, the prewetting line and isotherms for helium-Cs
adsorbed on nano-patterned surfaces with  parabolic 
cavities were studied by the finite temperature DFT approach.\cite{Anc00,Anc09b} 
The results obtained for Cs surfaces (non-wettable) were 
compared with the corresponding planar Na (wettable at $T=0$) 
and nano-patterned Na surfaces. 

To illustrate how wetting of a patterned surface
proceeds, we show  some 
configurations in Fig. \ref{fig41} for 
the wetting sequence of a Na surface at $T=0.5$ K, from low to higher coverage.
The Na surface is
patterned with an array of periodically repeated 
parabolic cavities of nanoscopic size, and the He coverage
is increased continuously.
Panel I shows the low coverage phase up to the complete filling of each
heliophilic parabolic cavity; in panel II,  
droplets grow on the flat region
between the periodically repeated cavities as coverage increases;
in panel III the drops on the flat region
have merged together and only a ring-shaped region around the edge of the
cavity remains covered by a thin helium film. Finally,
complete filling of the  
annular region
occurs (panels IV and V);
at relatively higher coverages, 
a very thick film made of several monolayers covers the surface 
and grows continuously with increasing coverage (panel VI).

\subsection{Soft-landing of helium droplets}\label{5.17}

\begin{figure}[t]
\vspace{42pt}
\begin{center}
\subfigure{
\resizebox*{8.0cm}{!}{\includegraphics{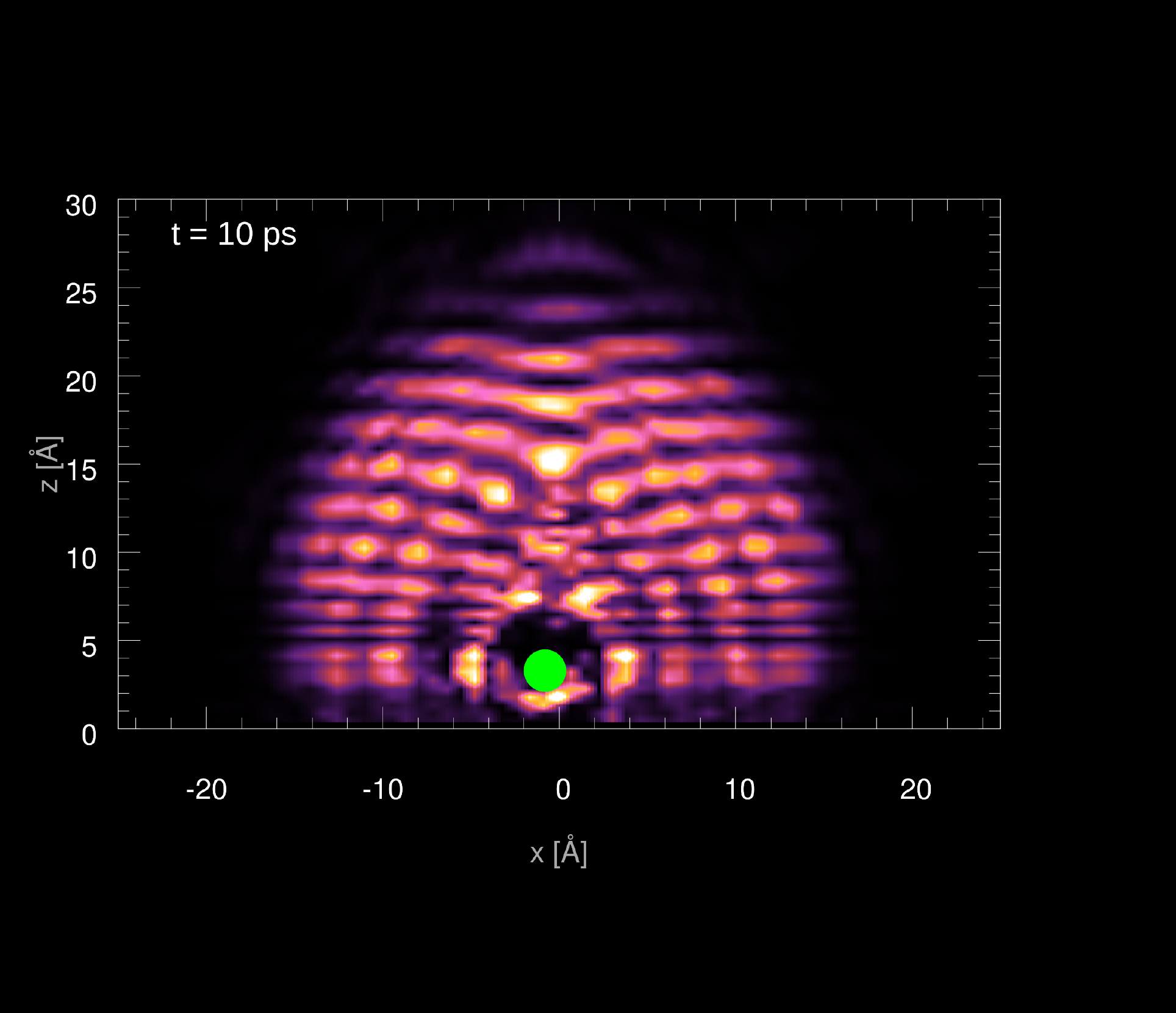}}}%
\subfigure{
\resizebox*{7.0cm}{!}{\includegraphics{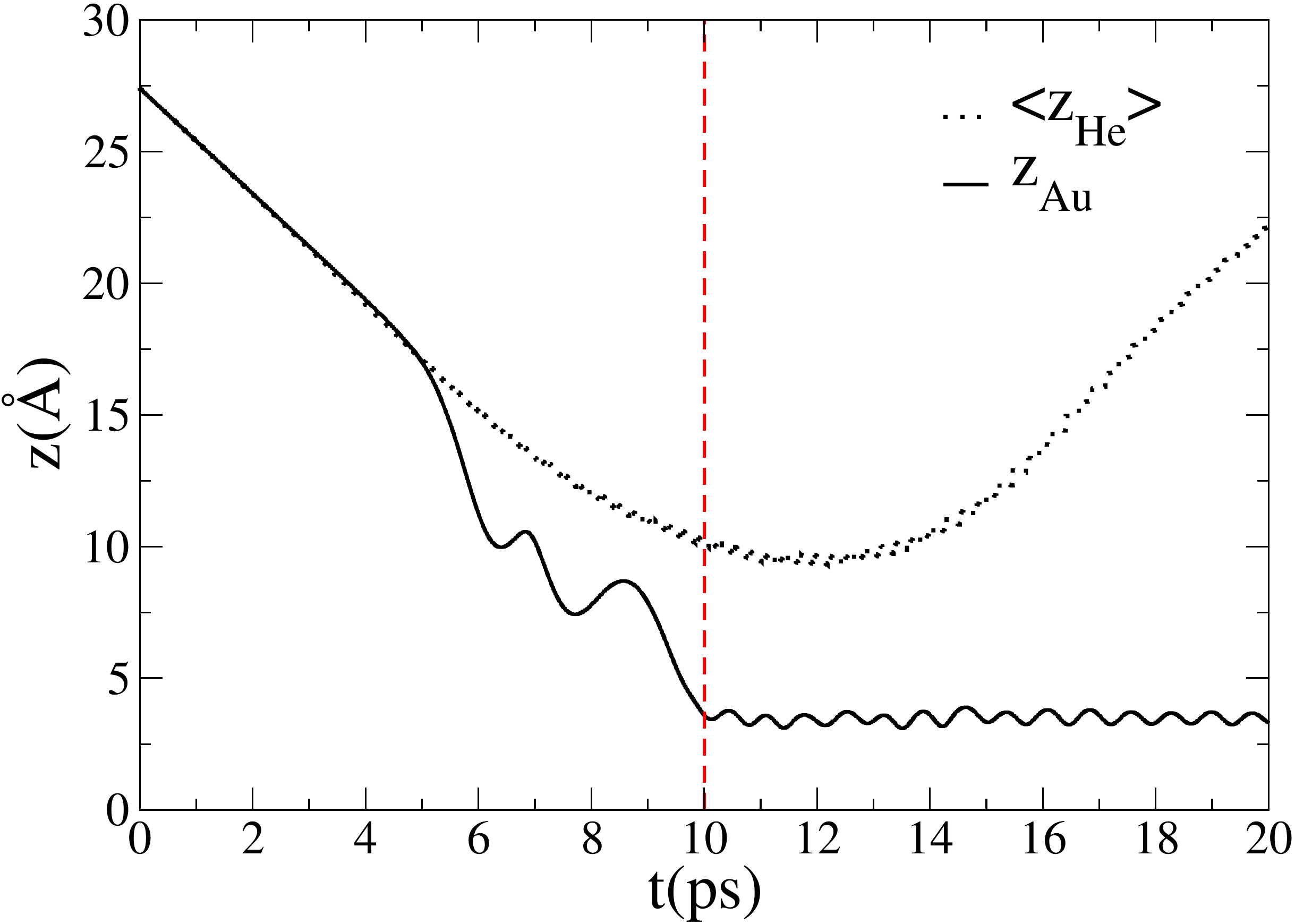}}}%
\caption{\label{fig42}
Au@$^4$He$_{300}$ landing on rutile TiO$_2 (110)$-surface at 200 m/s. (a) Droplet density on a plane perpendicular to the surface at $t = 10$ ps. 
Bright spots correspond to  high helium densities and the Au atom is represented by a green dot. (b) Position of the Au atom (solid line) and the COM of the helium 
droplet (dotted line) with respect to the surface plane. The vertical dashed line indicates the time corresponding to the snapshot shown in (a).\cite{deL15}  
}
\end{center}
\end{figure}

The time-dependent process of helium droplet landing on solid surfaces, which bears similarities with the surface wetting process discussed in the previous section, has been studied by TDDFT. 
In addition to contributing to the understanding of the basic physics governing such a collision event, the species formed inside the droplets can be gently deposited onto the surface (soft-landing). 
With the recent general interest in nano-sciences and nano-technology, this approach has received significant attention because it can potentially be used to extract the nanostructures
 formed inside helium droplets for practical applications.\cite{Moz07,Log11d} For example, metallic nanoclusters and nanowires, which are predicted to be excellent catalysts 
 due to their large surface area, could be produced by using this technique. For a review on soft-landing, see  Ref. \onlinecite{Joh11}. 

Soft-landing of metal (e.g. Ag) doped large helium droplets has also been instrumental for the discovery of vortex lines inside the droplets.\cite{Gom12,Lat14,Tha14} 
The experiments revealed the presence of linear structures on the deposition surface along which a series of nanoparticles were distributed. The observed linear geometries 
have been suggested to arise either from vortex mediated nanowire assembly (and subsequent decomposition) or from direct trapping of multiple metal nanoparticles on a 
vortex line. Both mechanisms require the presence of quantized vortices inside the droplets.

The first TDDFT studies for the softlanding of pure helium droplets on surfaces with potential technological interest have been 
carried recently.\cite{Agu12,deL14}  The model system considered consisted of a $^4$He$_{300}$ droplet traveling at 200 m/s towards a TiO$_2$ (110)-surface. To identify the possible 
quantum effects, both classical molecular dynamics and TDDFT simulations were carried out. In contrast to the classical results, which show the helium droplet splashing 
off the surface on impact, the TDDFT evolution leads to the spreading of the liquid on the surface. This thin film formation is a process similar to the surface wetting described in the 
previous section. In addition to TiO$_2$, a graphite sheet was also considered as a target.\cite{deL14}  Despite the omission of 
 thermal effects and the small droplet size considered, these studies have provided a solid starting point for simulating helium droplet mediated deposition of metallic clusters on substrates. 

Deposition of an Au atom embedded inside a $^4$He$_{300}$ droplet on a TiO$_2$ (110)-surface was addressed in Ref. \onlinecite{deL15}. This was the first theoretical 
study that considered the experimentally studied landing process as described e.g. in Refs. \onlinecite{Moz07,Log11d,Vol13}. The outcome of the simulation for an Au@$^4$He$_{300}$ 
complex  at 200 m/s, with COM initially  located at 27.4 \AA{} from the surface, is shown in Fig. \ref{fig42}. 
As shown in the figure, the Au atom initially follows the droplet, then begins to oscillate back and forth inside it, and finally becomes trapped inside the Au-TiO$_2$ 
surface potential minimum. After \text{ca.} 10 ps, the atom keeps oscillating about the potential energy minimum until the end of the simulation (zero average acceleration). 
The spreading of the droplet on the surface was observed, but to a lesser degree than for the pure helium droplet.\cite{Agu12} By comparing the results from TDDFT and classical 
calculations, it was concluded that the proper description of this $^4$He droplet-assisted process must be carried out using quantum mechanical simulations.

\section{Summary and outlook}
\label{6.0}

The density functional approach offers a unique method to study both static and dynamic response of superfluid helium. In addition to atomic and molecular impurities, 
it can also be applied in its present form to model the interaction of nanometer-scale objects with the liquid. From the computational resource perspective, the method is easily applied to 
systems up to 100 nm in size in 3D, the main limitations being both computer time and memory requirements. The formulation allows for the description of both helium droplets and bulk
 liquid through suitable boundary conditions. With the recent improvements to the OT-DFT functional, strongly inhomogeneous snowball systems can now be modelled. Unlike QMC-based 
 methods, DFT can yield real time quantum dynamics. The examples summarised in this review provide extensive evidence that the TDDFT approach is capable of reproducing the 
 results from a wide range of time-resolved experiments, especially in superfluid helium droplets. It is the only method that allows this close interplay between theory and experiment in this field. 

Despite the enormous success of OT-DFT, and DFT in general, there is still room for improvement in both accuracy and functionality. For example, just like any other DFT-based method, it is not straightforward to couple DFT to any degrees of freedom that follow traditional quantum mechanics. The often employed coupled quantum and OT-DFT equations presented in this review all either ignore the quantum correlations between the two subsystems or incorporate them in a phenomenological way into the interaction potential. In many systems, such as the electron bubble, this correlation may not be significant due to the large mass difference between the electron and helium. But it will arise, for example, in the treatment of molecular rotation in superfluid helium. Up to now the latter problem has only been addressed in terms of classical rotation and the associated moment of inertia, which is the likely origin of the remaining discrepancy between the OT-DFT calculations and experimental results (Sec. \ref{5.11}). A similar issue should arise in the treatment of molecular vibrations of impurities solvated in superfluid helium. 

While PIMC calculations can elegantly model superfluid helium at finite temperatures, OT-DFT in its basic formulation is restricted to 0 K. The static liquid response up to 
3 K temperature has been introduced into OT-DFT, but it is still missing dynamic contributions such as the viscous response. The viscous response can be included from continuum 
fluid mechanics (Navier-Stokes) into OT-DFT,  but it is not \textit{a priori} clear how it should behave at the typically observed rather wide gas-liquid interfaces around solvated impurities. 
In general, one possible strategy may be to follow the very successful Landau's two-fluid model and treat the superfluid and normal fractions in DFT separately. 
 
Many elegant experiments have been carried out in bulk superfluid helium over the years, which would require at least a mesoscopic-size description of the system. Therefore, they are currently not accessible to the OT-DFT approach due to the limitations in current computational resources. In order to employ the DFT approach for such systems, new strategies are needed to reduce especially the memory requirements of the calculations. An obvious approach is to use any symmetry present in the system and formulate the problem in 1D or 2D rather than the full 3D. However, often such symmetry is not present and furthermore, the numerical implementation of OT-DFT in reduced dimensions is sometimes  far from trivial. 

If a distant part of the system can be treated with a limited accuracy, multi-scale-type models could be developed to expand the spatial domain considered. Another option would be to vary the resolution of the spatial grid, which allows allocating more points to the regions of interest. However, all current numerical implementations of OT-DFT are restricted to uniform grids because they employ the finite difference approximation and the FFT algorithm for evaluating the non-linear potential. While the well-known finite element method could replace the finite difference approach, the heavily used FFT would still not be applicable for non-uniform grids.

Notwithstanding the need for improvements, the applications of the current OT-DFT approach, especially in its time-dependent version, are noteworthy. 
 Among the projects that are being conducted during the completion of this review, let us mention e.g. the  desorption of intrinsic and extrinsic impurities from 
 helium droplets; the description of soft-landing
 processes under conditions closer to the experimental situation;
 the multiple capture of impurities by droplets hosting vortex arrays, and the propagation of shock waves and solitons, and gas bubble dynamics and vortex nucleation in liquid helium.
 
Finally, let us indicate that the CSU at Northridge and Barcelona-Toulouse helium-DFT codes are available at the following repositories:

$\bullet$ CSU at Northridge He-DFT code:

https://sourceforge.net/projects/libgrid/

https://sourceforge.net/projects/libdft/

$\bullet$ BCN-TLS DFT He-code:

https://github.com/bcntls2016/4hedft

https://github.com/bcntls2016/4hedft-vortex

https://github.com/bcntls2016/4hetddft-isotropic

https://github.com/bcntls2016/4hetddft-anisotropic

\begin{acknowledgements}
We would like to thank the many colleagues whose collaboration in the development of the DFT method, triggered in most cases by their experimental work, 
has made possible that the DFT approach has reached the level of sophistication and applicability this review is based on. This work has been performed under 
Grants No. FIS2014-52285-C2-1-P from DGI, Spain, and  2014SGR401 from Generalitat de Catalunya. MB thanks the Universit\'e F\'ed\'erale Toulouse 
Midi-Pyr\'en\'ees for financial support during the completion of this work throughout the `Chaires d'Attractivit\'e 2014' Programme IMDYNHE. 
JE acknowledges financial support from NSF grant DMR-1205734.
\end{acknowledgements}

\end{document}